\documentclass{aa}

\usepackage{graphicx}
\usepackage{txfonts}
\usepackage{subfigure}
\usepackage{placeins}
\newcommand{\um}{\,$\mu$m}
\newcommand{\kms}{km\,s$^{-1}$}

\newcommand{\cii}{[C\,{\sc ii}]}

\newcommand{\thcii}{[$^{13}$C\,{\sc ii}]}
\newcommand{\ci}{[C\,{\sc i}]}
\newcommand{\oi}{[O\,{\sc i}]}
\newcommand{\sii}{[S\,{\sc ii}]}

\newcommand{\hii}{H\,{\sc ii}}
\newcommand{\hi}{H\,{\sc i}}
\newcommand{\hh}{H$_2$}

\newcommand{\twco}{$^{12}$CO}
\newcommand{\thco}{$^{13}$CO}
\newcommand{\ceio}{C$^{18}$O}

\begin{document}

   \title{Bright-rimmed clouds in IC~1396\\I. Dynamics}

   \titlerunning{}

   \subtitle{}

   \author{Yoko Okada \inst{1}
          \and
          Slawa Kabanovic \inst{1}
          \and
          Rolf G\"{u}sten \inst{2}
          \and
          Volker Ossenkopf-Okada \inst{1}
          \and
          Nicola Schneider \inst{1}
          \and
          Robert Simon \inst{1}
          \and
          Christof Buchbender \inst{1}
          \and
          Ronan Higgins \inst{1}
          \and
          Craig Yanitski \inst{1}
          \and
          Markus R\"{o}llig \inst{3,1}
          \and
          J\"{u}rgen Stutzki \inst{1}
          \and
          Daisuke Ishihara \inst{4}
          \and
          Kunihiko Tanaka \inst{5}
          \and
          Edward Chambers \inst{6,7}
          \and
          Netty Honingh \inst{1}
          \and
          Matthias Justen \inst{1}
          \and
          Denise Riquelme \inst{8,9,2}
          }

   \institute{I. Physikalisches Institut der Universit\"{a}t zu K\"{o}ln, Z\"{u}lpicher Stra{\ss}e 77, 50937 K\"{o}ln, Germany
              \email{okada@ph1.uni-koeln.de}
              \and
              Max-Planck-Institut f\"{u}r Radioastronomie, Auf dem H\"{u}gel 69, 53121 Bonn, Germany
              \and
             Physikalischer Verein - Gesellschaft für Bildung und Wissenschaft, Robert-Mayer-Str. 2, 60325 Frankfurt, Germany
            \and
              Institute of Space and Astronautical Science, Japan Aerospace Exploration Agency, 3-1-1 Yoshinodai, Chuo-ku, Sagamihara, Kanagawa 252-5210, Japan
              \and
              Department of Physics, Faculty of Science and Technology, Keio University, 3-14-1 Hiyoshi, Yokohama, Kanagawa 223--8522 Japan
              \and
              SOFIA Science Center, NASA Ames Research Center, Moffett Field, CA 94045, USA
              \and
              Space Science Institute, 4765 Walnut St, Suite B, Boulder, CO 80301, USA
              \and
              Instituto Multidisciplinario de Investigaci\'on y Postgrado, Universidad de La Serena, Ra\'ul Bitr\'an 1305, La Serena, Chile
              \and
              Departamento de Astronom\'ia, Universidad de La Serena, Ra\'ul Bitr\'an 1305, La Serena, Chile
             }

   \date{}

 
  \abstract
   {}
   {We investigate the dynamical and physical structures of bright-rimmed clouds (BRCs) in a nearby \hii\ region. We focused on carbon- and oxygen-bearing species that trace photon-dominated regions (PDRs) and warm molecular cloud surfaces in order to understand the effect of UV radiation from the exciting stars on the cloud structure.}
   {We mapped four regions around the most prominent BRCs at scales of 4--10 arcmin in the \hii\ region IC~1396 (IC~1396A, B, D, and E) in \cii\ 158\um\ with (up)GREAT on board SOFIA. IC~1396 is predominantly excited by an O6.5V star. Toward IC~1396A, we also observed \oi\ 63\um\ and 145\um. We combined these observations with JCMT archive data, which provide the low-J transitions of CO, \thco,\ and \ceio. All spectra are velocity-resolved.}
   {The line profiles in the four mapped regions show a variety of velocity structures, which we investigated in detail for all observed emission lines. IC~1396B and D show clearly distinct velocity components that overlap along the line of sight. We find no clear sign of photoevaporating flows in the \cii\ spectra, although the uncertainty in the location of the BRCs along the line of sight makes this interpretation inconclusive. Our analysis of the \thcii\ emission in IC1396~A, which has the best signal-to-noise ratio, suggests that the \cii\ is likely mostly optically thin. The heating efficiency, measured by the (\cii+\oi\ 63\um)/far-infrared intensity ratio, is higher in the northern part of IC~1396A than in the southern part, which may indicate a difference in the dust properties of the two areas.}
{The complex velocity structures identified in the BRCs of IC~1396, which is apparently a relatively simple \hii\ region, highlight the importance of velocity-resolved data for disentangling different components along the line of sight and thus facilitating a detailed study of the dynamics of the cloud. We also demonstrate that the optically thin \thcii\ and \oi\ 145\um\ emission lines are essential for a conclusive interpretation of the \cii\ 158\um\ and \oi\ 63\um\ line profiles.}


   \maketitle
%

\section{Introduction}\label{sec:intro}

\begin{figure*}
\centering
\includegraphics[width=\hsize]{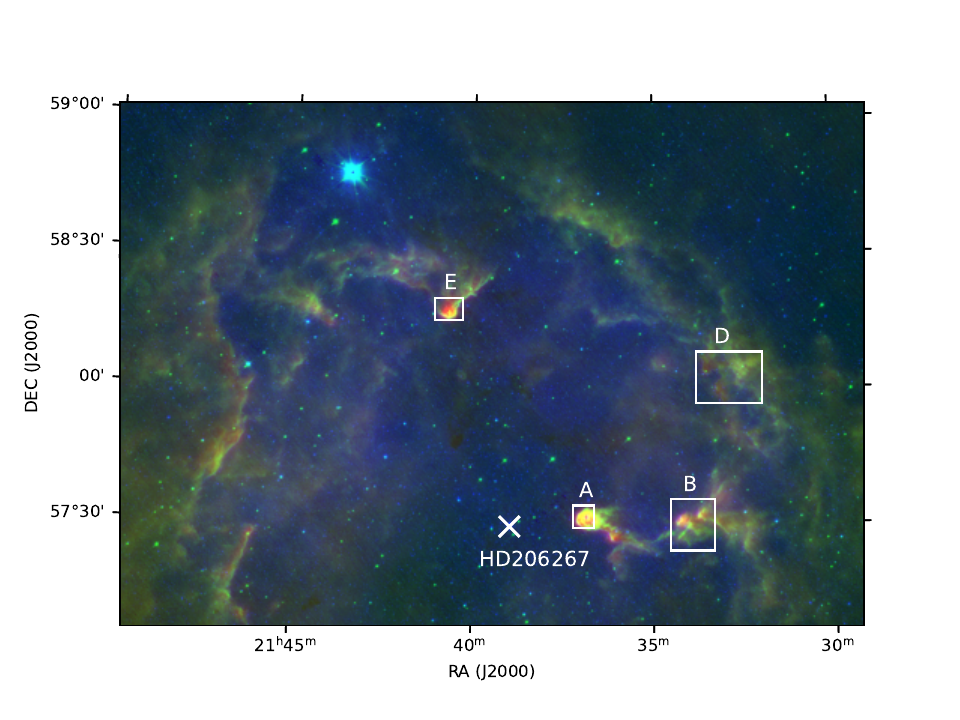}
\caption{Overview of IC~1396. In this three-color composite image, blue is DSS-red, green is the 9\um \ map from the MIR all-sky survey by the Infrared Camera \citep[IRC;][]{Onaka2007,Ishihara2010} on board AKARI \citep{Murakami2007}, and red is the WIDE-L (140\um) map from the FIR all-sky survey by the Far-Infrared Surveyor \citep[FIS;][]{Kawada2007,Doi2015} on board AKARI.  The maps are not convolved to the common spatial resolution, and the coarsest spatial resolution is 88\arcsec\ \citep{Takita2015} for FIS (red), which is about a third of the size of the box around at IC~1396A.  The fields observed with upGREAT in the \cii\ line emission are overlaid as white boxes.  The regions are labeled following the notation of \citet{Weikard1996}. The position of the exciting source, HD~206267, is marked.}
\label{figure:IC1396_overview}
\end{figure*}

Bright-rimmed clouds (BRCs) are located near the edges of evolved \hii\ regions and are exposed to the UV radiation from the exciting stars, creating a photon-dominated region (PDR) at their surface (the bright rim). Their evolution and morphology, which sometimes includes young stellar objects (YSOs), are often discussed in the context of triggered star formation from a radiation-driven implosion \citep{Bertoldi1989,Lefloch1994,Miao2009}. Early analytical studies \citep{OortSpitzer1955} and more sophisticated 3D radiation-magnetohydrodynamic simulations \citep{Henney2009} predict a photoevaporation flow from the illuminated surface and a rocket effect to the cloud as its back reaction. \citet{Sugitani1991} classified BRCs into three types based on the morphology (rim curvature), which is explained by the combination of the initial self-gravitational state and the evolutionary sequence \citep{Miao2009}. The term ``BRC'' has been used in the literature to refer to a surface illuminated by exciting sources, while the term ``pillars'' (a column-like shape) and ``globules'' (isolated and having a head-tail structure) are used to specify their morphology \citep{Schneider2016}.

IC~1396 is an \hii\ region with a diameter of a few degrees on the sky \citep[Fig.~\ref{figure:IC1396_overview}; 1 degree corresponds to 16~pc at IC~1396's distance of 925~pc;][]{Pelayo-Baldarrago2023}, located at the edge of the Cepheus Bubble. The whole region is predominantly excited by the O6.5V star in the binary system HD~206267 \citep{Abt1986,Sota2011}, which is a member of the cluster Tr 37. An optical image of this region shows a number of cometary silhouettes, some of them with bright rims facing the exciting source. Most of them are bright in the mid-infrared (MIR) and far-infrared (FIR; Fig.~\ref{figure:IC1396_overview}). This rich selection of BRCs in an apparently simple geometry with a single dominant exciting star makes IC1396 a textbook example of an evolved \hii\ region and ideal for studying the evolution of BRCs \citep[e.g.,][]{Sugitani1991,Patel1995,Weikard1996,Sicilia-Aguilar2019}. 

The \cii\ 158\um\ and \oi\ 63\um\ FIR fine-structure lines are the dominant cooling lines in PDRs and thus are excellent tracers of the physical conditions of BRCs and, if observations are velocity-resolved, their kinematics. Recent studies demonstrate that large, velocity-resolved \cii\ 158\um\ maps are a powerful tool for investigating the radiative and mechanical feedback as well as the physical properties of the illuminated surface, together with other PDR and molecular gas tracers such as \oi\ and CO in nearby \hii\ regions \citep{Schneider2012,Mookerjea2019,Pabst2019,Luisi2021,Tiwari2021,Schneider2021,Kabanovic2022}.

So far, the \cii\ emission from IC~1396A has been analyzed using only an under-sampled map in \citet{Okada2012}. In the current study, we presented 4--10\arcmin\ scale fully sampled \cii\ maps around four BRCs in IC~1396 and a deep \oi\ 63\um\ map of IC~1396A, which allowed us to investigate the dynamics and physical properties of the regions. We focused on the detailed structure and dynamics of individual BRCs. The excitation and column densities of the carbon-bearing species and the modeling using PDR models will be presented in a follow-up paper. 

The organization of this paper is as follows. In Sect.~\ref{sec:intro_IC1396} we give an overview of IC~1396 based on previous studies, and in Sect.~\ref{sec:obs} we present the data used in this study. Section~\ref{sec:results} presents the results of our analysis. After providing an overview of IC~1396 as a whole (Sect.~\ref{subsec:global_picture}) and explaining the methods for calculating integrated line intensities (Sect.~\ref{subsec:integrated_intensities}), we reveal the details of the spatial distributions and dynamics of the four observed subregions (Sects.~\ref{subsec:result_IC1396A} to \ref{subsec:result_IC1396E}). The \thcii\ analysis of IC~1396A is presented in Sect.~\ref{subsec:13CII_IC1396A}, followed by a discussion of the heating efficiency and the \cii--mid-infrared correlation in Sects.~\ref{subsec:heating_eff} and \ref{subsec:CII-MIR_correlation}. Signs of photoevaporation flows, 3D kinematic structures, and a comparison of different BRCs are discussed in Sect.~\ref{sec:discussion}.  Our main results are summarized in Sect.~\ref{sec:summary}.

\section{IC1396} \label{sec:intro_IC1396}

The Cepheus Bubble is thought to have been created by a first generation of stars formed 13--18~Myr ago, followed by the formation of the Cep OB2 association in the compressed expanding shell \citep{Patel1998}. HD~206267 is estimated from its stellar properties to have been born about 4~Myr ago \citep[][ and references therein]{Patel1995}, creating an \hii\ region.  The BRCs are found at the interface between the \hii\ region and the associated molecular cloud. \citet{Patel1995} proposed an expanding ring model based on the different CO velocities in the surrounding BRCs. They show that the momentum of the BRCs cannot be explained by the stellar wind, and model the dynamical evolution with the ``rocket effect,'' suggesting a dynamical age of IC~1396 of 2--3 Myr. Using \textit{Gaia} data, \citet{Pelayo-Baldarrago2023} found that the stellar component in IC~1396 is composed of several substructures that are clearly distinct in proper motion and partly in age, suggesting a complex star formation history with multi-episode formation processes.

Figure~\ref{figure:IC1396_overview} shows a three-color composite image of optical, MIR, and FIR emission. As already shown in previous studies \citep{Weikard1996,Barentsen2011}, BRCs often appear as a silhouette with a bright rim in optical images. \citet{Weikard1996} made an extensive \twco/\thco(1-0) survey covering an area of more than 6 deg$^2$ in IC~1396, together with focused maps around a number of BRCs in the $J\!=$2--1 and 3--2 transitions.  They show the presence of warmer molecular gas in the BRCs and discuss the detailed kinematics of individual clouds. Whether or not the detected clouds are associated with IC~1396 has been investigated based on several criteria such as excitation conditions and kinematics. In general, clouds with central velocities of $-8$~\kms\ to $+7$~\kms\ associate with IC~1396. There are a number of clouds that are not associated with IC~1396 despite having similar central velocities. For example, there is a prominent cometary optical dark cloud between HD~206267 and IC~1396E (Fig.~\ref{figure:IC1396_overview}) that was identified in CO with a velocity of $-0.7$\,\kms\ \citep[B161, part of the broad dark lane Khav 161]{Weikard1996}. Despite its appearance, they found no indication that it is associated with IC~1396. This area is quite dark in the MIR image and at shorter wavelength in the FIR image (Fig.~\ref{figure:IC1396_overview}). Some structure, including B161, starts to become visible between 140\um\ and 160\um.

\citet{Carlsten2018} explained the observed positional offset of H$_2$ and Br$\gamma$ in some of the BRCs of IC~1396 with photoevaporative flow models. \citet{Sicilia-Aguilar2019} also suggested that the velocity of CN is indicative of photoevaporative flows. \citet{Soam2018} studied the magnetic field and the shape of BRCs in IC~1396 and confirmed the theoretical prediction \citep{Henney2009} that BRCs with small angles between the direction of the magnetic field and the ionizing radiation have broad heads, while BRCs with large angles of these have a more flattened and elongated shape. 

Of the four BRCs that we study in this paper, IC~1396A and IC~1396E have been particularly well studied, and we summarize them in the following two subsections. IC~1396B is connected to IC~1396A, and is located further away from the exciting source (Fig.~\ref{figure:IC1396_overview}). IC~1396D has the largest projected distance from the exciting source and is weaker in the infrared among the four BRC samples. It contains a cometary-shaped BRC with an embedded Infrared Astronomical Satellite (IRAS) source identified by \citet{Sugitani1991}, SFO34.

\subsection{IC~1396A}

IC~1396A is the prominent BRC closest in projection to the exciting source (Fig.~\ref{figure:IC1396_overview}). A cavity within the BRC is visible in $^{12}$CO and \thco\ maps \citep[][see also Fig.~\ref{figure:integmap_IC1396A} in Sect.~\ref{subsec:result_IC1396A}]{Wootten1983,Nakano1989}, and is associated with the YSOs LkH$\alpha$ 349a and c. The cavity is also clearly visible in the MIR \citep{Reach2004,Reach2009} and FIR \citep{Sicilia-Aguilar2014} images. \citet{Reach2009} showed that it is plausible that the stellar wind from LkH$\alpha$ 349a, more massive than LkH$\alpha$ 349c, created a bubble inside the BRC. LkH$\alpha$ 349a is likely a very young Herbig Ae/Be star and will be a late B- to early A-type star on the main sequence \citep{Reach2009}. It shows P Cygni profiles in H$\alpha$, indicating ejection of material at velocities above 300~\kms\ \citep{Hernandez2004}. The rim of the cavity is bright in shock-tracing optical \sii\ emission \citep{Sicilia-Aguilar2013}, reflecting interaction between the YSOs formed inside the cavity and the surrounding cloud. In addition to this most prominent cavity, the \textit{Spitzer} image shows that IC~1396A hosts a number of cavities and that all of the major cavities contain YSOs, suggesting that the BRC is being reshaped from the inside via outflows from protostars \citep{Reach2009}. IC~1396A must be located in front of the ionizing region based on its appearance as a dark cloud with a bright rim in optical wavelengths \citep{Barentsen2011}, a sharp edge of H$\alpha$ \citep{Barentsen2011} and \sii\ \citep{Sicilia-Aguilar2013} emission, and its bluest velocity among the clouds in IC~1396 \citep{Weikard1996}. \citet{Sicilia-Aguilar2014} found a Class 0 object (IC~1396A-PACS-1) with \textit{\textit{Herschel}}/Photodetector Array Camera and Spectrometer (PACS) observations. It is located inside the BRC, $\sim 0.1$~pc from the bright rim defined by a sharp boundary in the PACS map, surrounded by a cold region, and is consistent with its formation being triggered via radiation-driven implosion \citep{Sicilia-Aguilar2014}. Combining the Institut de radioastronomie millim\'{e}trique (IRAM) 30m observations and \textit{Gaia} data, \citet{Sicilia-Aguilar2019} confirm that IC~1396A has undergone several episodes of star formation. From the \textit{Gaia} proper motion of the stars and the radial velocity of the gas, they derived a total (3D) velocity of IC~1396A of $\sim 8$\,\kms\ away from Tr 37. This relative velocity is too low compared to the expected velocity due to rocket acceleration if IC~1396A had been much closer to HD206267 in the past. \citet{Moriarty-Schieven1996} found an \hi\ tail behind IC~1396A (opposite to the exciting stars) extending across IC~1396B and further west. The blueshifted \hi\ velocity relative to CO at the head of the BRC and the relatively constant velocity of the tail is consistent with a scenario where \hi\ traces ablated material from the BRCs, drifting into the shadow of the BRC so that it is not further accelerated. \citet{Okada2012} presented the first velocity-resolved \cii\ map in IC~1396A, revealing the prominent \cii\ wings at the northern and southern edges of the rim. They performed a simple KOSMA-$\tau$ PDR model fit to the \cii, \ci,\ and CO(4-3) data and suggest that the UV radiation is shielded at the rim.

\subsection{IC~1396E}

Another well-studied BRC is IC~1396E (also called IC~1396N), which hosts prominent protostellar outflows, located in the northern part of IC~1396 (Fig.~\ref{figure:IC1396_overview}). \citet{Sugitani1989} mapped the central outflow in CO(2-1) and show that it is associated with IRAS 21391+5802. The density profile derived from CS observations is steep and centered at this IRAS source \citep{Serabyn1993}, well away from the ionization front, suggesting a preexisting density enhancement rather than compression by external shocks. IRAS 21391+5802 was resolved into three sources in the millimeter continuum \citep[BIMA1, 2, and 3;][see also Fig.~\ref{figure:integmap_IC1396E} in Sect.~\ref{subsec:result_IC1396E}]{Beltran2002}, and the brightest, BIMA2, which drives the outflow, was further resolved into three sources in millimeter interferometric observations \citep{Neri2007}, forming an intermediate-mass protocluster. \citet{Fuente2009} used molecular line observations and suggested that the brightest identified source in BIMA2, IRAM 2A, is a low-mass or Herbig Ae star, or a Class 0/I transition object that has already formed a small PDR. \citet{Codella2001} found another outflow in the northern part of IC~1396E, and \citet{Beltran2012} showed that it collides with yet another outflow. The outflows have a clumpy morphology, suggesting that they are the result of episodic mass-loss events \citep{Codella2001,Beltran2012}. H$_2$ 2.12\um\ knots are observed in collimated stellar jets, excited by the shock produced by the outflows \citep{Nisini2001,Beltran2009}. The interaction of the outflows with the ambient gas is also seen in the distorted shape of the western lobe of the central outflow. \citet{Codella2001} suggested a significant contribution of swept-up material to the central outflow mass based on the low velocity component ($\pm 3$\kms\ with respect to the ambient velocity).  CO emission up to $J$=25 was detected with Infrared Space Observatory (ISO)/Long Wavelength Spectrometer (LWS) \citep{Saraceno1996}, likely tracing shock heated gas \citep{Molinari1998,Nisini1998}. \citet{Patel1995} revealed the dynamics of the CO gas using \thco(1-0); there is a pair of symmetrical extended features from the cometary head to the northeast and northwest, likely moving away from the cometary head. Their velocities differ by 1.7~\kms\ (also visible in the \thco(3-2) channel maps in Fig.~\ref{figure:channelmap_large_IC1396E}). They proposed that the velocity difference in the two tails is due to either the different inclinations or the different densities, the former resulting in different line-of-sight velocities and the latter causing different shock velocities.

\begin{table}
\caption{Line combinations and chopping parameters of the upGREAT observations in IC~1396A.}
\label{table:obs_summary_IC1396A}
\centering
\begin{tabular}{ccccc}
\hline\hline
Date (UT) & Flight ID & Chop$^\mathrm{a}$ & receivers and lines\\
\hline
2016-11-04 & F345 & 200\arcsec/0$^\circ$ & HFA: \oi\ 63\um\\
&&& L2: \cii\ 158\um\\
\hline
2017-06-09 & F403 & 240\arcsec/0$^\circ$ & HFA: \oi\ 63\um\\
&&& LFAH: \cii\ 158\um\\
\hline
2017-06-10 & F404 & 240\arcsec/0$^\circ$ & HFA: \oi\ 63\um\\
&&& LFAH: \cii\ 158\um\\
&&& LFAV: \cii\ 158\um\\
\hline
2018-12-04 & F533 & 300\arcsec/45$^\circ$ & HFA: \oi\ 63\um\\
&&& LFAH: \oi\ 145\um\\
&&& LFAV: \cii\ 158\um\\
\hline
2020-03-06 & F668 & 300\arcsec/45$^\circ$ & HFA: \oi\ 63\um\\
&&& LFAH: \cii\ 158\um\\
&&& LFAV: \oi\ 145\um\\
\hline
2020-03-07 & F669 & 300\arcsec/45$^\circ$ & HFA: \oi\ 63\um\\
&&& LFAH: \cii\ 158\um\\
&&& LFAV: \oi\ 145\um\\
\hline\hline
\end{tabular}
\begin{list}{}{\setlength{\itemsep}{0ex}}
\item[$^\mathrm{a}$] Chop throw and chop angle, which is defined in counter-clock direction from north.
\end{list}
\end{table}

\begin{table}
\caption{CO observations in the JCMT data archive. $\checkmark$ means detected, $\times$ means not detected, and -- means the line is either not observed or the calibrated data are not available in the archive.}
\label{table:co_jcmt}
\centering
\begin{tabular}{ccccc}
\hline\hline
Lines & IC~1396A & IC~1396E & IC~1396D & IC~1396B\\
\hline
CO(3-2) & $\checkmark$ & $\checkmark$ & $\checkmark$ & $\checkmark$\\
\thco(3-2) & $\checkmark$ & $\checkmark$ & $\checkmark$ & --\\
\ceio(3-2) & $\checkmark$ & $\checkmark$ & $\times$ & --\\
CO(2-1) & -- & -- &  $\checkmark$ & --\\
\thco(2-1) & -- & -- &  $\checkmark$ & --\\
\ceio(2-1) & -- & -- &  ($\checkmark$)$^a$ & --\\
\hline
\end{tabular}
\begin{list}{}{\setlength{\itemsep}{0ex}}
\item[$^\mathrm{a}$] Detected only in a small area.
\end{list}
\end{table}

\begin{figure*}
\centering
\includegraphics[width=0.48\hsize]{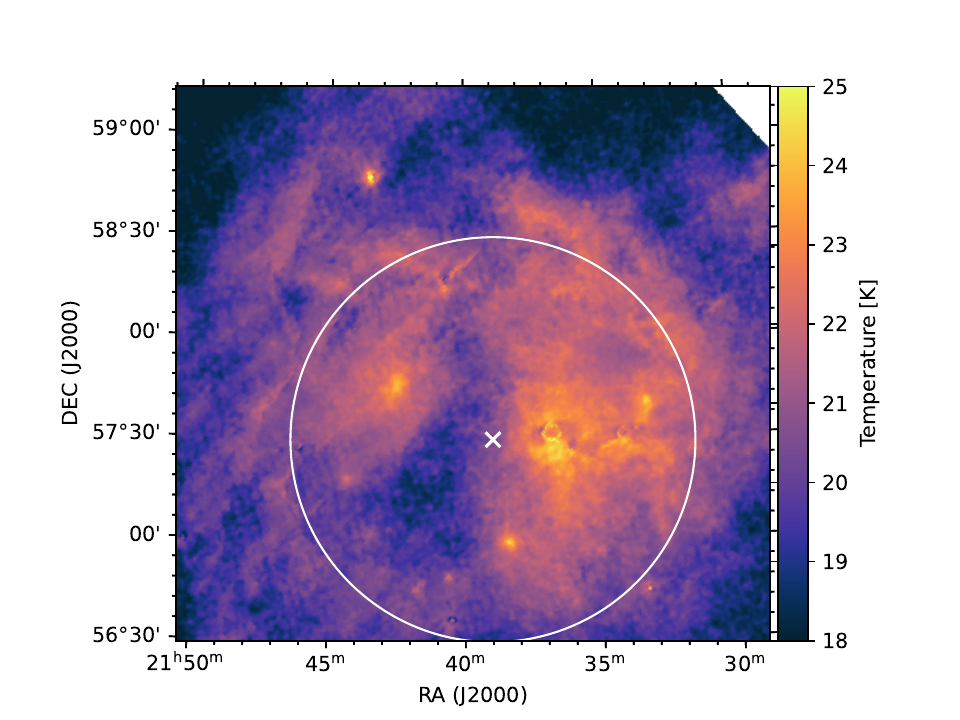}
\includegraphics[width=0.48\hsize]{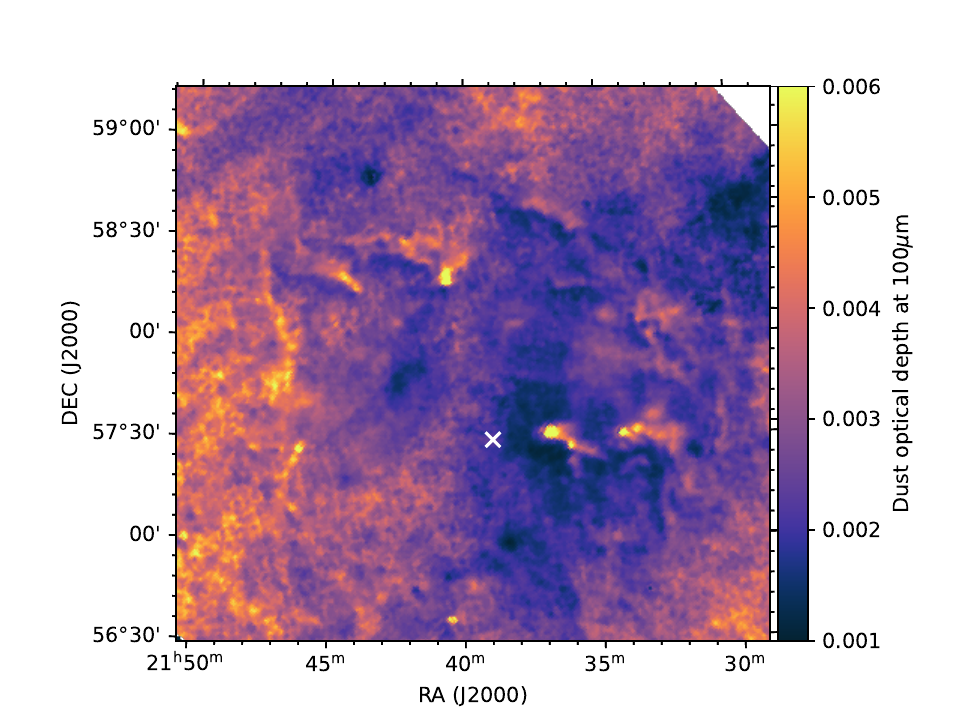}
\caption{Dust temperature (left) and the continuum optical depth at 100\um\ (right) fitted with a single temperature model to the AKARI/FIS data (see the main text). The white cross marks the position of HD~206267, and the white circle in the left panel indicates a 1 deg radius, where the dust temperature is 25~K from a purely geometric estimate (see the main text).}
\label{figure:FIS_tempfit}
\end{figure*}

\section{Observations and data reduction} \label{sec:obs}

\subsection{\oi\ 63\um, 145\um, and \cii\ 158\um\ observations with upGREAT on board SOFIA in IC~1396A} \label{subsec:obs_great_IC1396A}

We observed the \oi\ lines at 4744.77749~GHz (63\um) and 2060.06886~GHz (145\um) and the \cii\ line at 1900.53690~GHz (158\um) with the German REceiver for Astronomy at Terahertz Frequencies\footnote{GREAT is a development by the MPI f\"{u}r Radioastronomie and the KOSMA / Universit\"{a}t zu K\"{o}ln, in cooperation with the MPI f\"{u}r Sonnensystemforschung and the DLR Institut f\"{u}r Planetenforschung} \citep[upGREAT;][]{Heyminck2012,Risacher2018} on board the Stratospheric Observatory for Infrared Astronomy \citep[SOFIA;][]{Young2012} in IC~1396A, as a guaranteed time project. The High Frequency Array (HFA) with 7 pixels in a hexagonal configuration was tuned to \oi\ 63\um\ and all observations were made in single phase chopped on-the-fly (OTF) mode with 3\arcsec\ step size (see Table~\ref{table:obs_summary_IC1396A} for the chop parameters).  We observed the \cii\ emission in parallel, either with the 7 pixels of the Low Frequency Array (LFA) or, earlier on,  with the single pixel L2 channel. On some flights, we split the two polarizations (H and V) in the LFA and observed \oi\ 145\um\ in parallel (see Table~\ref{table:obs_summary_IC1396A} for receiver combinations).  We used several generations of the Fast Fourier Transform Spectrometer (FTS) as backends, which were continuously upgraded over several years of observations in this project.

The data were calibrated along the following steps. First, we performed a frequency correction on the HFA data. This is a standard procedure for the HFA, because the absolute frequency stability of the quantum cascade laser local oscillator is limited by the temperature and current stability of the laser \citep{Risacher2018}.  We corrected the frequency scale scan by scan using the known narrow telluric \oi\ line as a reference. We then corrected for a gain drift by scaling the raw counts of individual observations to those immediately after each load measurement for individual pixels. The obtained data were calibrated using the standard GREAT pipeline \citep{Guan2012}, which converts the observed counts to the main beam temperature scale ($T_\mathrm{mb}$) and corrects for atmospheric attenuation using an atmospheric radiative transfer model. We then resampled the spectra to a resolution of 0.5~\kms. To improve the baseline quality, we extracted principal components from the OFF spectra and subtracted these components from the ON spectra of \cii\ and \oi\ 63\um\ (see our Appendix~\ref{app:pca} as well as \citealt{Tiwari2021} and Buchbender et al. in prep. for details of the methods and evaluation of the results). We then convolved the \cii\ and \oi\ 63\um\ maps to a spatial resolution of 16\arcsec\ or 25\arcsec. We use 16\arcsec\ resolution to discuss the velocity profiles and spatial structure in channel maps, integrated intensity maps, velocity profiles in selected positions and position-velocity (p-v) diagrams. For the analysis of the line widths and the \thcii\ line, and the derivation of the heating efficiencies, we used 25\arcsec\ maps to have a better signal-to-noise ratio (S/N). The median rms noise of the spectra at a given pixel on the 16\arcsec\ resolution map is 0.4~K and 0.18~K for \cii\ and \oi, respectively. 

The \oi\ 145\um\ is located on a shoulder of a significant atmospheric ozone feature in the signal sideband (Fig.~\ref{figure:IC1396A_OI145_S-H}). The line is detected at only a few positions on the 16\arcsec\ map. Because the atmospheric transmission is smooth in the velocity range of interest, we consider the velocity profile of \oi\ 145\um\ as reliable as that of the other lines (Appendix \ref{app:atm_oi145}).

Since we obtained a fully sampled map for \oi\ (beam size of 6.3\arcsec), the parallel \cii\ map (beam size of 14.1\arcsec) has a S/N ten times higher than the data shown in \citet{Okada2012}, observed with a single pixel. The \cii\ map obtained in this study thus reveals much more detailed structures.

\subsection{\cii\ 158\um\ observations with (up)GREAT on board SOFIA in IC~1396E, D, and B}

\cii\ mapping observations were also made in three other regions: IC~1396E, D, and B, as part of the guaranteed time observations. IC~1396E was observed with the single pixel L2 channel in May 2014 and January 2015, and with the 7 pixel upGREAT LFA in May 2015 during the upGREAT commissioning flights. We also observed IC~1396B and D in May 2015. IC~1396D observations were extended in December 2015. All observations were made in total-power OTF mode with a 5--6\arcsec step size. The OFF positions are (RA, Dec.)(J2000) = (21:34:13.0, 57:23:18.0), (21:33:39.4, 57:58:20.0), and (21:40:39.0, 58:13:28.0) for IC~1396B, IC~1396D, and IC~1396E, respectively.

The data were calibrated using the standard GREAT pipeline as described above. We subtracted linear baselines and spectrally resampled to a resolution of 0.5~\kms. We did not apply the principal component analysis (PCA) to these data because the analysis with the chopped data in IC~1396A (Sect.~\ref{subsec:obs_great_IC1396A}) shows that the baseline improvement by using PCA is not significant in the LFA data and the total power observations do not provide a sufficient number of OFF measurements to identify the principal components. We convolved the \cii\ map to 16\arcsec\ or 25\arcsec\ resolution as for IC~1396A (except that we use 25\arcsec\ maps for p-v diagrams in IC~1396B and IC~1396D for better S/N). The median rms noise on the 16\arcsec\ resolution map is 1.2~K, 0.74~K, and 0.77~K for IC1396~B, IC1396~D, and IC1396~E, respectively.

\subsection{Complementary observations: CO, \thco, and \ceio\ from JCMT archive data} \label{subsec:obs_jcmt}

We obtained calibrated CO spectral maps from the Heterodyne Array Receiver Program \citep[HARP;][]{Buckle2009} on the\textit{ James Clerk Maxwell }Telescope (JCMT\footnote{Data before March 1, 2015: The JCMT has historically been operated by the Joint Astronomy Centre on behalf of the Science and Technology Facilities Council of the United Kingdom, the National Research Council of Canada and the Netherlands Organisation for Scientific Research.}) from their data archive\footnote{Project ID: M07AH24A and M08BU15, observed between 2007 and 2010.}. Table~\ref{table:co_jcmt} summarizes the available isotopes and transitions. The original beam size is 14\arcsec\ for CO(3-2) and 19.7\arcsec\ for CO(2-1). In this paper, we focus on CO(3-2) and its isotopologues.

We subtracted a zeroth-order baseline and spectrally resampled to a resolution of 0.5~\kms. We then convolved the maps to the same spatial resolution and grid as the upGREAT \cii\ data (16\arcsec\ or 25\arcsec\ resolution). To convert the intensity to the main beam scale, we utilized main beam and forward efficiencies of 0.63 and 0.88, respectively, for CO(3-2) and its isotopologues %
(following JCMT's heterodyne data reduction cookbook ver1.0).

\begin{figure*}
\centering
\includegraphics[width=\hsize]{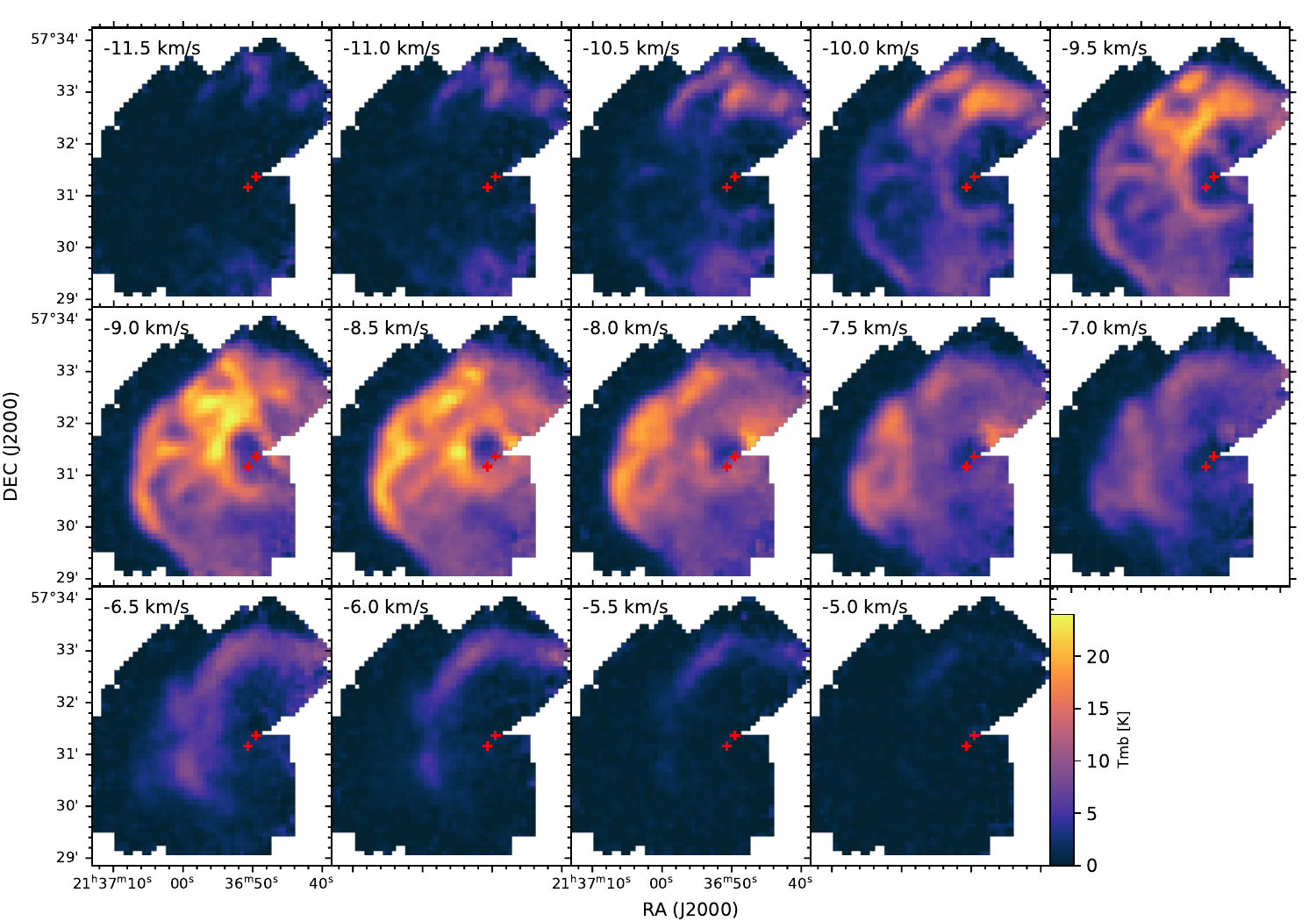}
\caption{Channel maps of \cii\ in IC~1396A (16\arcsec\ resolution). The color scale is $T_\mathrm{mb}$ averaged over a velocity bin centered at the value noted in each panel with a width of 0.5\kms. Red crosses mark LkH$\alpha$ 349a (southeast) and c (northwest). HD~206267 is east of the BRC at (RA, Dec.)(J2000) = (21:38:57.619, 57:29:20.54).}
\label{figure:channelmap_IC1396A_CII}
\end{figure*}

\subsection{Complementary observations: Far-infrared continuum} \label{subsec:obs_cont}

To derive the global distribution of the dust temperature and column density (Sect.~\ref{subsec:global_picture}), we obtained FIR maps from the FIR all-sky survey by the Far-Infrared Surveyor \citep[FIS;][]{Kawada2007,Doi2015} on board AKARI. They have four bands, N60, WIDE-S, WIDE-L, and N160, centered at 65\um, 90\um, 140\um, and 160\um. While the beam sizes along the in-scan and cross-scan directions are not the same \citep{Takita2015}, we use a simplified circular Gaussian with the mean full width at half maximum to convolve the N60 and WIDE-S maps (63\arcsec\ and 78\arcsec, respectively) to the resolution of WIDE-L and N160 (88\arcsec). We also applied the color correction tabulated in \citet{Shirahata2009} to take into account the discrepancy between the integrated flux in an imaging band and the intensity at the central wavelength of the band. We used the correction factor corresponding to a gray-body with $\beta=2$ and $T=30$~K.

In the analysis of the heating efficiencies in IC~1396A (Sect.~\ref{subsec:heating_eff}), we used higher spatial resolution data from \textit{Herschel}\footnote{\textit{Herschel} is an ESA space observatory with science instruments provided by European-led Principal Investigator consortia and with important participation from NASA.} and the JCMT\footnote{Data after March 1 2015: The JCMT is operated by the East Asian Observatory on behalf of The National Astronomical Observatory of Japan; Academia Sinica Institute of Astronomy and Astrophysics; the Korea Astronomy and Space Science Institute; Center for Astronomical Mega-Science (as well as the National Key R\&D Program of China with No. 2017YFA0402700). Additional funding support is provided by the Science and Technology Facilities Council of the United Kingdom and participating universities in the United Kingdom and Canada. Additional funds for the construction of SCUBA-2 were provided by the Canada Foundation for Innovation.}. We downloaded the following data from their data archives; PACS 70\um\ and 160\um\ data presented in \citet{Sicilia-Aguilar2014}\footnote{Proposal ID: OT1\_asicilia\_1, ObsID: 1342259791, 1342259792}, and SCUBA-2 850\um\ data\footnote{Project ID: M17BD002 and M18AN001, observed in 2018} \citep{Holland2013,Chapin2013,Dempsey2013}%
. For the SCUBA-2 data, we selected maps observed on four different nights and combined them with equal weight. We convolve the PACS and SCUBA-2 maps to 25\arcsec\ resolution with the same spatial grid as the line emission data, with which we performed the heating efficiency analysis. In the SCUBA-2 850\um\ map, the cavity around LkH$\alpha$ 349 appears to have negative intensities. The most negative intensity in the convolved map is $-26$\,MJy\,sr$^{-1}$, which was added as an offset to the whole map and is considered as a systematic uncertainty in the absolute flux.

\section{Results} \label{sec:results}

\subsection{Global picture from dust emission}\label{subsec:global_picture}

\begin{figure*}
\centering
\includegraphics[width=\hsize]{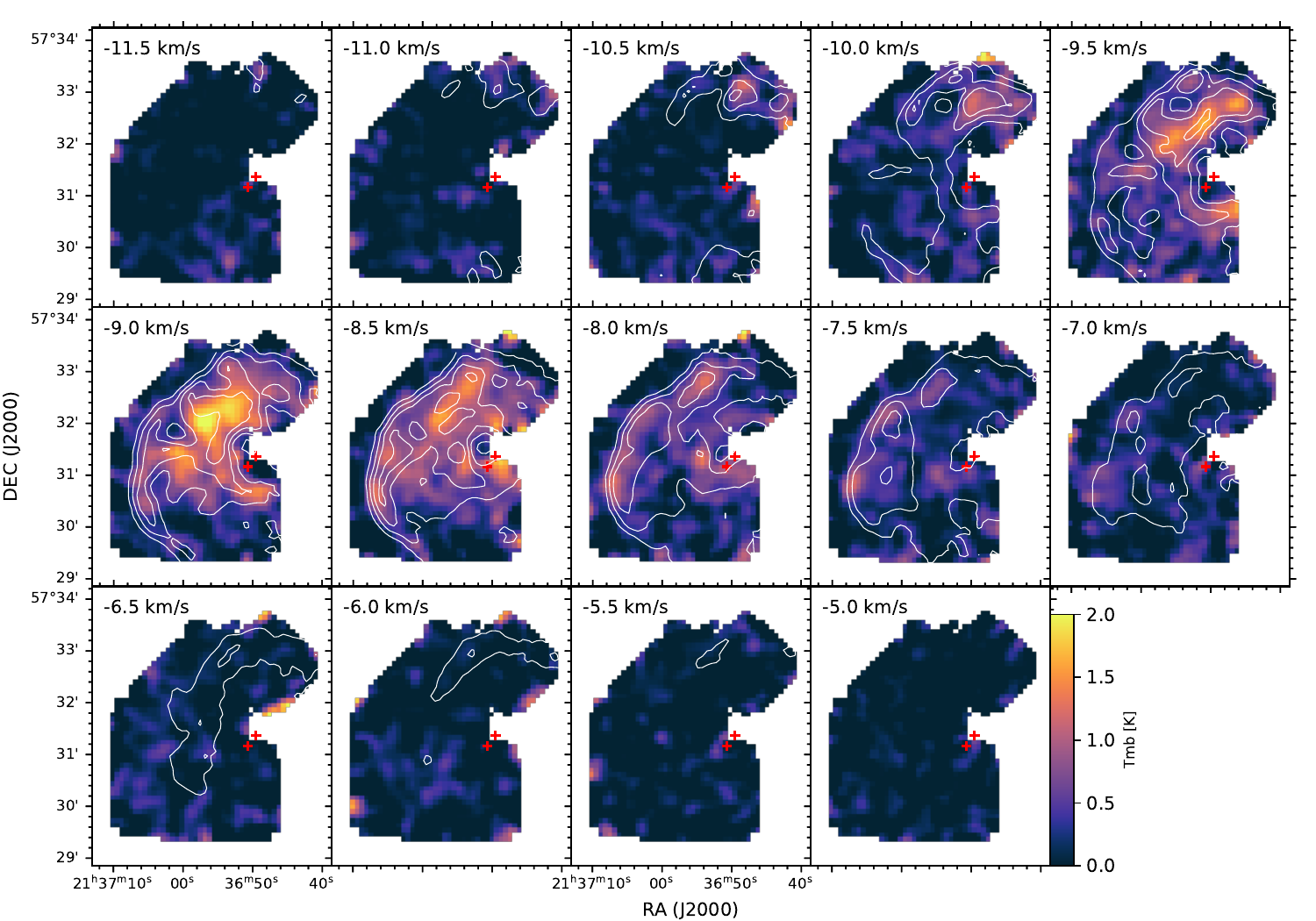}
\caption{Same as Fig.~\ref{figure:channelmap_IC1396A_CII} but for the \oi\ 63\um\ emission (colors) overlaid with the \cii\ 158\um\ channel map contours (white with a spacing of 5~K) of IC~1396A.}
\label{figure:channelmap_IC1396A_OI63}
\end{figure*}

We fit the spectral energy distribution (SED) from the four FIS bands with a single temperature modified gray-body with $\beta=2$. This corresponds to a simplified dust model in which the wavelength dependence of the absorption coefficient follows $\lambda^{-\beta}$ in the infrared (for the dust emission) and is constant in the optical (for the stellar light absorption). $\beta=2$ falls in the range of the dust models of \citet{Jones2013}, \citet{DraineLi2007} or \citet[see Fig. 16 therein]{Hensley2023}. Dense molecular clouds show a shallower spectral index $\beta$. We fit the dust temperature ($T$) and the optical depth at a reference wavelength (100\um, $\tau_{100}$) in the equation $\tau_{100}(\lambda/100\mu\textrm{m})^{-\beta}B_\nu(T)$. Due to an insufficient number of data points to differentiate between different dust models, here we focus on the relative spatial distribution of the dust temperature and optical depth.
Figure~\ref{figure:FIS_tempfit} shows the resulting temperature and 
dust optical depth at 100\um\ for the entire IC~1396 region. The area with $T_\mathrm{dust} \geq 20$~K has a more or less round distribution centered at HD~206267. It does not have a clear gradient toward the center (cross in Fig.~\ref{figure:FIS_tempfit}) and there is no significant structures except for a colder dust lane running from the southeast to the north. This cold dust lane includes Khav 161, and probably traces foreground gas that is not heated by HD~206267. We calculated the expected dust temperature as a function of distance from HD~206267, assuming a luminosity of $10^{5.2} L_\odot$ for an O6.5V star \citep{Martins2005} and the dust model described above. The derived dust temperature 16~pc away from the source (1 deg at the distance of 925~pc) is 25~K, which is shown as a white circle in Fig.~\ref{figure:FIS_tempfit}. This is a few kelvins higher than the temperature obtained from the SED fit, corresponding to a factor of 1.9 in energy ($\propto T^6$), but considering that the fit model is very simplified and the theoretical estimate ignores energy loss within the \hii\ regions and considers only the projected distance, we conclude that they are in good agreement and confirm that HD~206267 is the dominant heating source. In the dust optical depth map, the BRCs are clearly peaking out, but otherwise the dynamic range of the dust column density is not high. The small variation of the dust temperature and the dust optical depth, except for the BRCs, and the fact that the region is well visible in H$\alpha$ \citep{Barentsen2011} give the picture of a spherical \hii\ region breaking out only in the direction toward the observer. 

\subsection{Integrated line intensities}\label{subsec:integrated_intensities}

\begin{figure*}
\centering
\includegraphics[width=0.355\hsize]{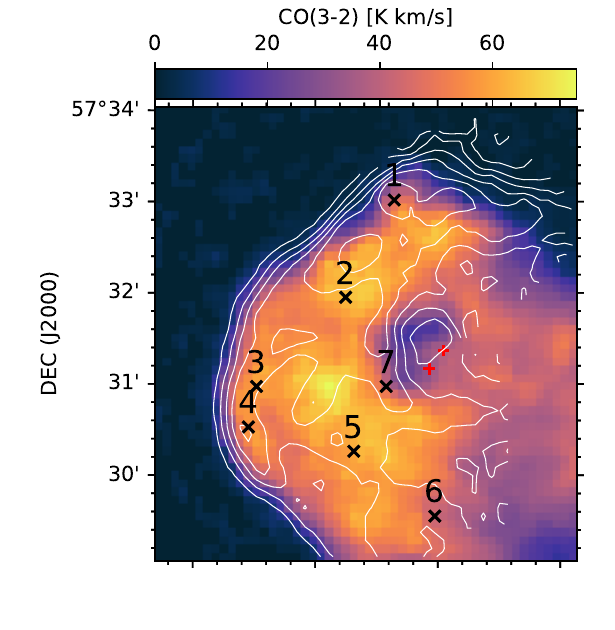}
\includegraphics[width=0.29\hsize]{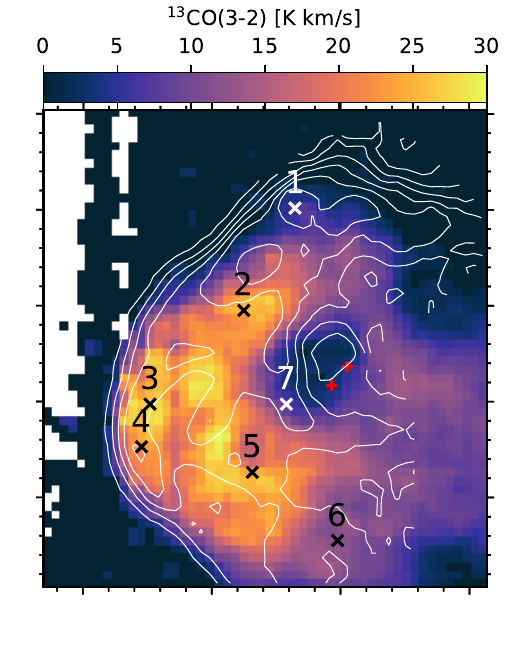}
\vspace{-0.5cm}
\includegraphics[width=0.29\hsize]{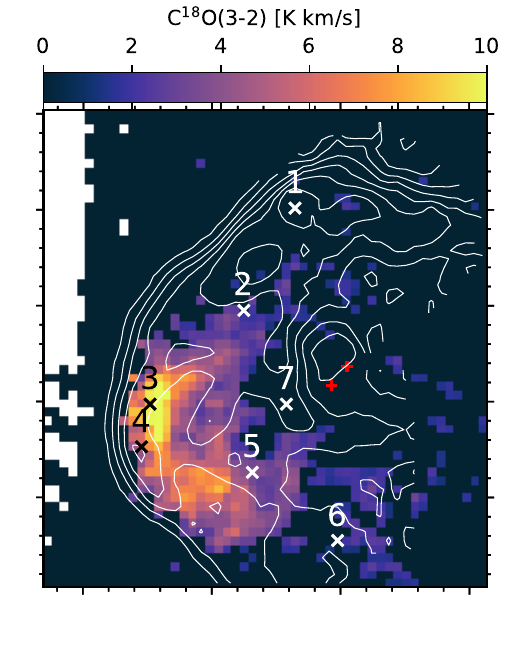}
\includegraphics[width=0.355\hsize]{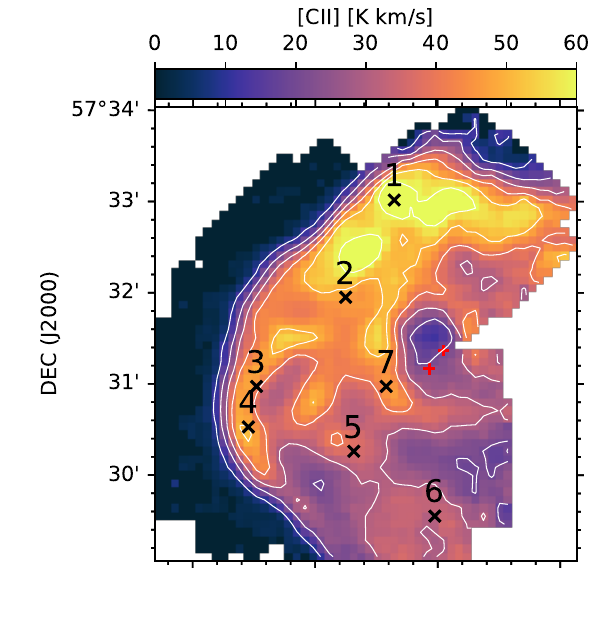}
\includegraphics[width=0.29\hsize]{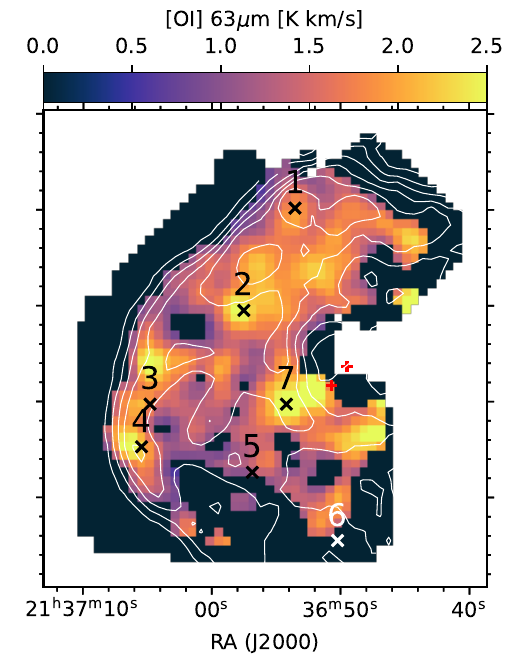}
\vspace{-0.4cm}
\includegraphics[width=0.29\hsize]{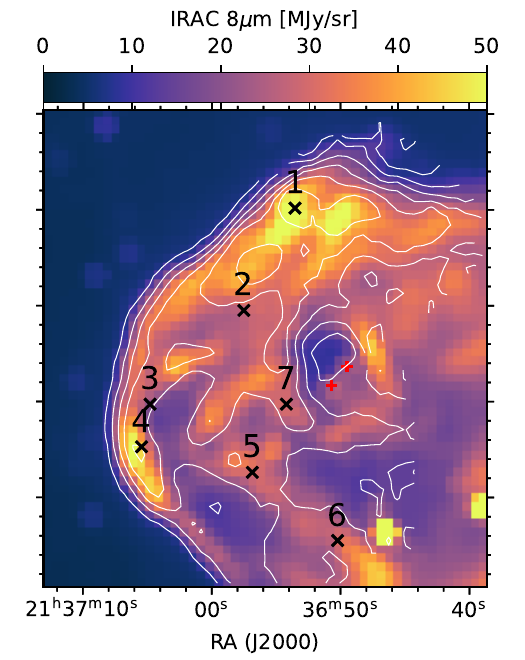}
\includegraphics[width=0.355\hsize]{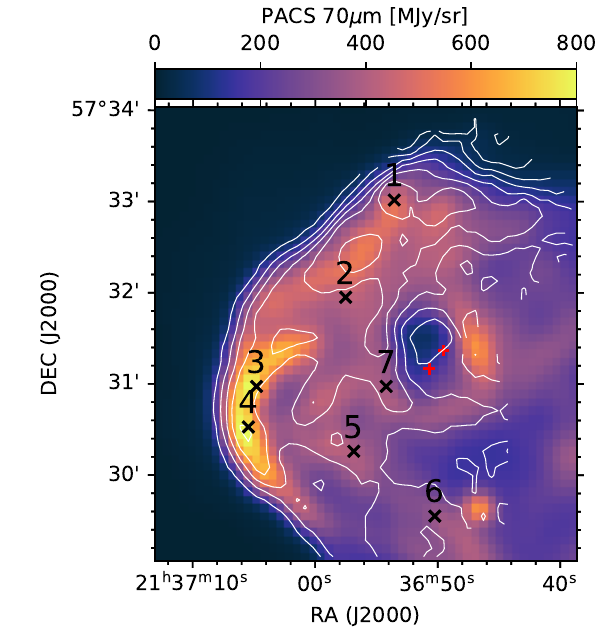}
\hspace{2.7cm}
\includegraphics[width=0.43\hsize]{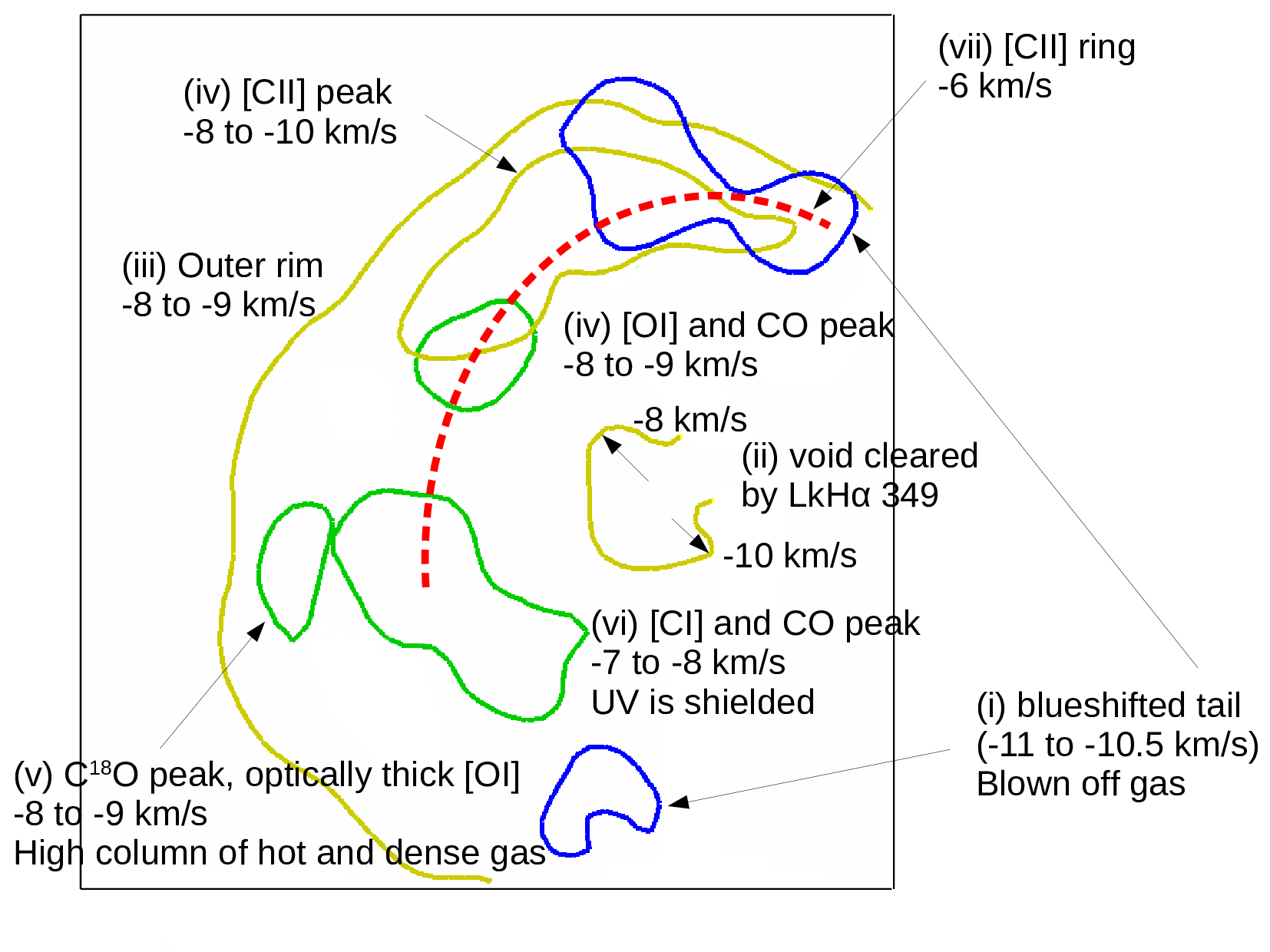}
\caption{Integrated intensity maps (colors, 16\arcsec\ resolution) of the different tracers (denoted on top of each panel) and overlaid with contours of the \cii\ integrated intensity (solid white lines, 16\arcsec\ resolution) in IC~1396A.  Integrated intensities were calculated via a direct integration over the velocity range ($-13$, $-4$)~\kms. White areas indicate that a region was not observed, and black areas indicate that no line was detected (positions where the integrated intensities are below $3\sigma$ of the baseline are masked). Red crosses mark LkH$\alpha$ 349a (southeast) and c (northwest). Black crosses with numbers mark the positions of the spectra shown in Fig.~\ref{figure:spec_selected_IC1396A}. The bottom-right panel shows the schematic decomposition. Different colors indicate a difference in velocity (red represents the most redshifted component and blue is the most blueshifted), and the Roman numerals correspond to the description in the text. The black box border indicates the size of the images in the other panels.}
\label{figure:integmap_IC1396A}
\end{figure*}

\begin{figure*}
\centering
\includegraphics[width=\hsize]{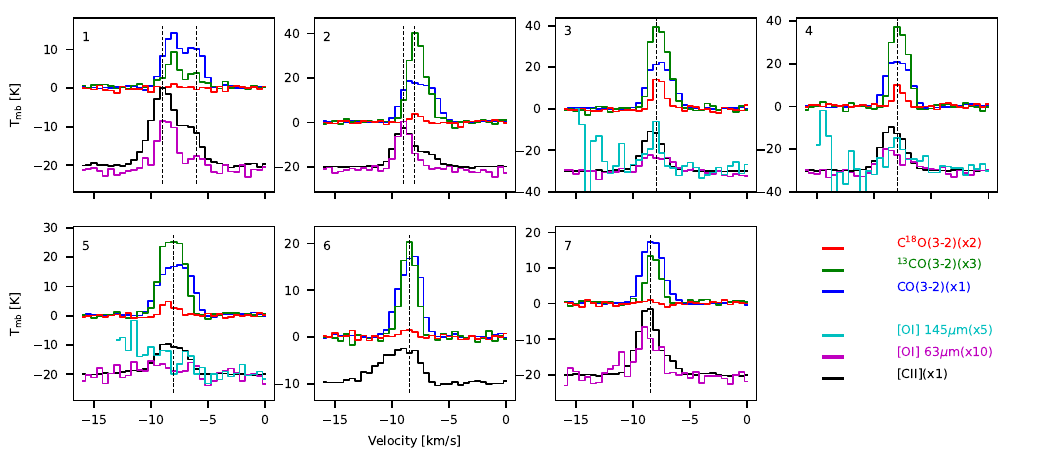}
\caption{Spectra at selected positions in IC~1396A, as marked in Fig.~\ref{figure:integmap_IC1396A} (16\arcsec\ resolution). The \cii\ and \oi\ spectral baselines are offset for a better visibility of the individual lines, and the vertical dotted lines are for guidance when comparing the peaks between different lines. \oi\ 145\um\ spectra are affected by the atmospheric ozone feature at velocities of $< -10$~\kms\ (see the main text).}
\label{figure:spec_selected_IC1396A}
\end{figure*}

In IC~1396A and IC~1396E, the integrated line intensities were calculated by integration over a fixed velocity range in each region ($-13$ to $-4$~\kms\ and $-5$ to $8$~\kms, respectively). On the other hand in IC~1396B and IC~1396D, the velocity structure is more complex and different velocity components coexist in the observed regions, which requires a broader velocity range to integrate ($-15$ to $10$~\kms\ and $-12$ to $10$~\kms, respectively) and leads to significantly higher noise (see the lowest panel of Fig.~\ref{figure:integmap_IC1396D_CII_compare}). For these two subregions, we derived the integrated intensities using dynamic integration windows based on a 3D dendrogram analysis.  Dendrograms are a hierarchical structure representation \citep{Rosolowsky2008}.  There are three control parameters: minimum intensity, minimum number of pixels, and minimum significance. We limited our analysis to the area with a baseline noise of 2 times the median noise over the map ($\sigma_\mathrm{med}$) for each subregion, so that the dendrogram routine runs over a cube with a relatively homogeneous noise distribution. We use the minimum intensity limit of $2\sigma_\mathrm{med}$ at each velocity channel and at each position. We set the minimum significance for a leaf to be considered an independent entity to $4\sigma_\mathrm{med}$. The number of minimum pixels for an independent leaf is set to $3^3=27$. Three spatial pixels correspond to the spatial resolution of the map. Then we created a 3D mask using all detected leaves, branches, and trunks (we call them clusters in the following). Each spatial and spectral pixel belonging to one of the detected clusters has a value of 1 in this 3D mask. We then derived the integrated intensity map by integrating the 3D spectral cube over this mask. This is equivalent to defining a matching velocity range for integration pixel by pixel, thus avoiding integration over a noisy baseline range without signal. However, this systematically underestimates the integrated intensities by ignoring wing contributions (spatially and spectrally) that fall under the minimum intensity (see Appendix~\ref{app:integrated_intensity} for more discussion and comparisons of the dendrogram-based methods and the masked moment methods with different threshold parameters).

\subsection{Spatial distributions and dynamics}\label{subsec:spatial_dist_dynamics}
\subsubsection{IC~1396A}\label{subsec:result_IC1396A}

\begin{figure*}[ht]
\centering
\includegraphics[width=0.355\hsize]{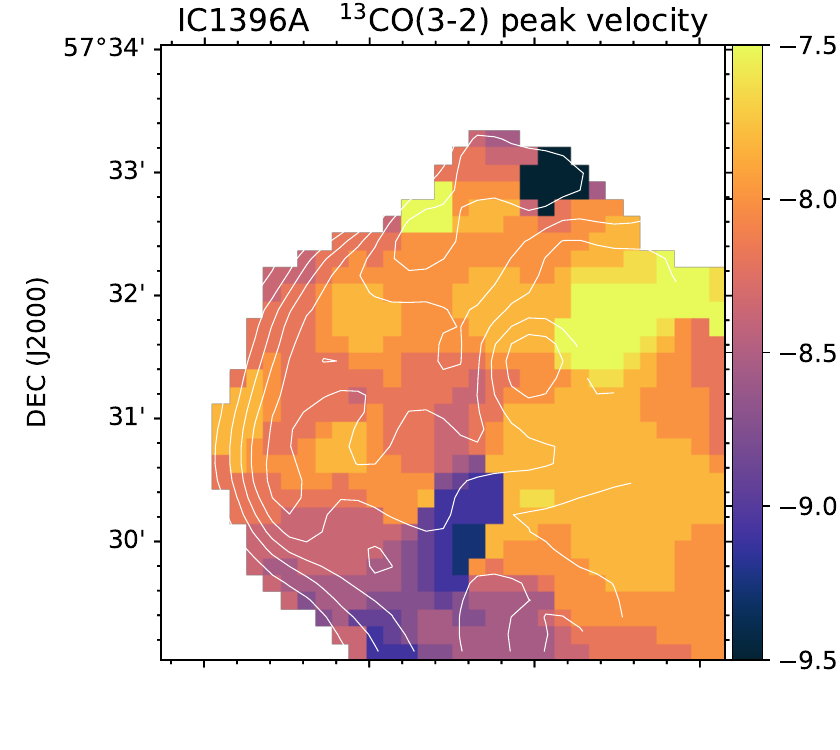}
\includegraphics[width=0.315\hsize]{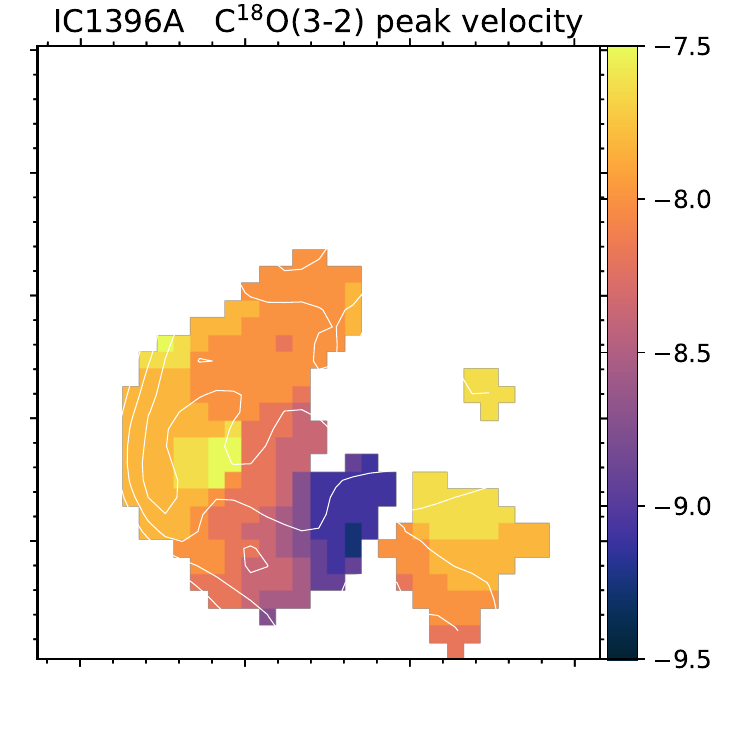}
\includegraphics[width=0.315\hsize]{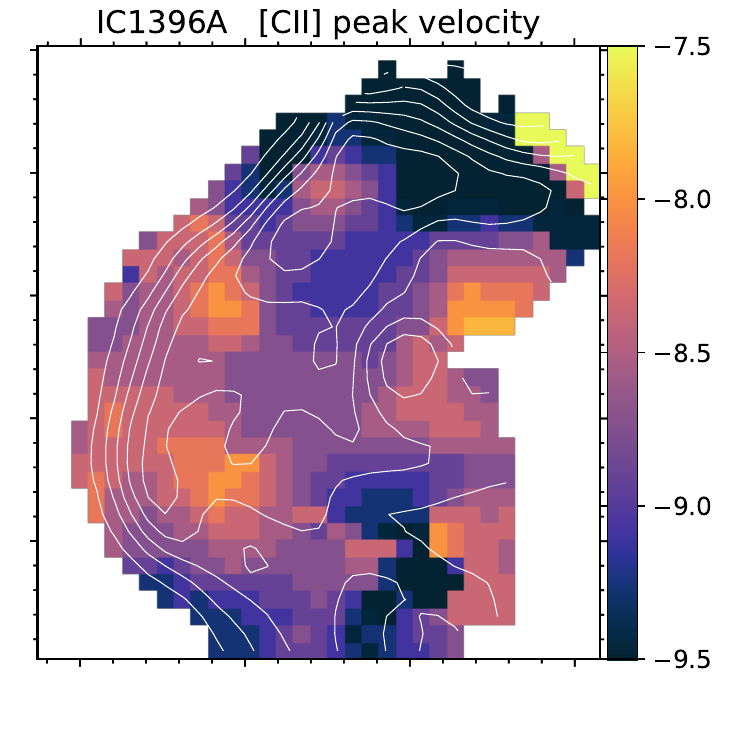}
\includegraphics[width=0.355\hsize]{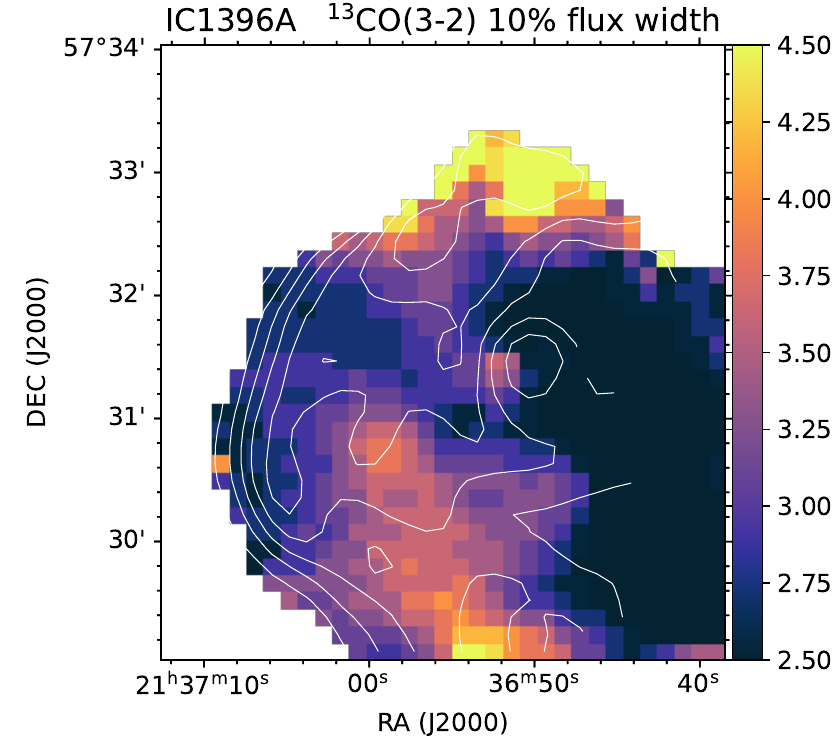}
\includegraphics[width=0.315\hsize]{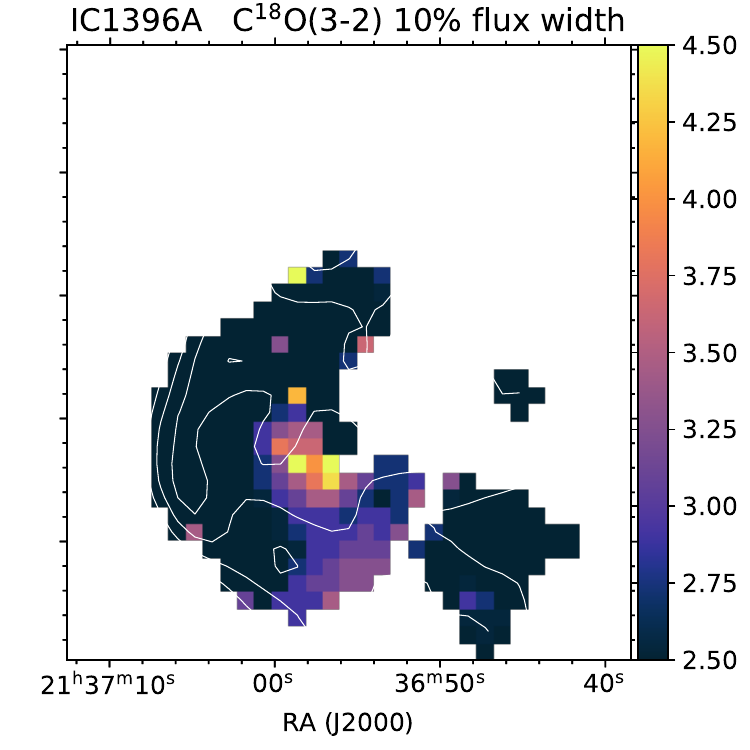}
\includegraphics[width=0.315\hsize]{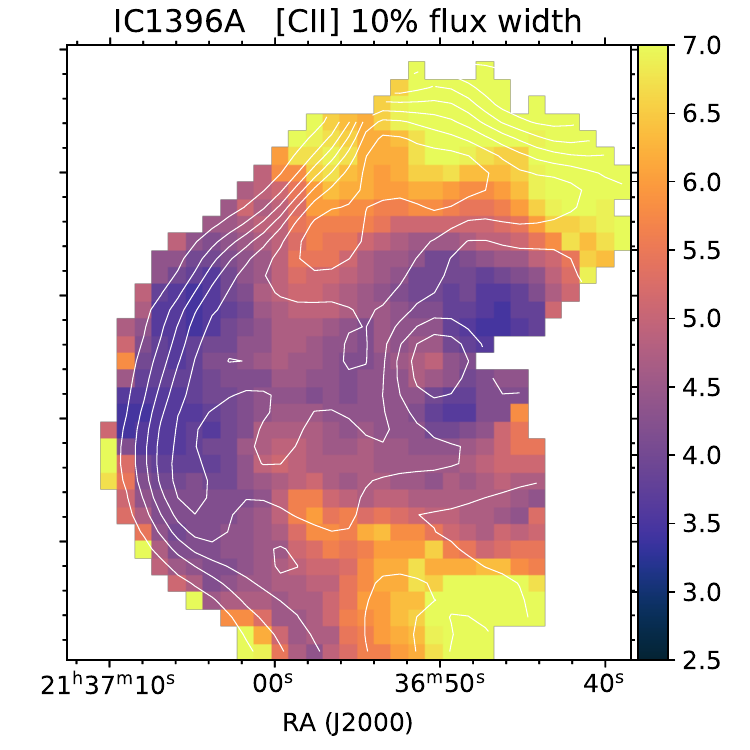}
\caption{Peak velocity (upper panels) and line width at 10\% of the peak intensity (lower panels) in units of \kms\ for \thco(3-2), \ceio(3-2), and \cii\ (from left to right) in IC~1396A at 25\arcsec\ resolution. The contours are the integrated intensity of \cii\ at 25\arcsec\ resolution. Note the different scale of the color wedge used for the \cii\ line width, as the spatial distribution of the \cii\ width would not be visible at the color scale used for the other two lines.}
\label{figure:width_IC1396A}
\end{figure*}

Figures~\ref{figure:channelmap_IC1396A_CII} and \ref{figure:channelmap_IC1396A_OI63} show the channel maps of \cii\ and \oi, Fig.~\ref{figure:integmap_IC1396A} shows the integrated intensity maps for each observed species, Fig.~\ref{figure:spec_selected_IC1396A} plots the spectra of selected positions marked in Fig.~\ref{figure:integmap_IC1396A}, and Fig.~\ref{figure:pvdiagram_IC1396A} shows the p-v diagrams for two cuts through IC~1396A. \oi\ 145\um\ is detected at only a few positions, and thus shown only in Fig.~\ref{figure:spec_selected_IC1396A}. Position 1 is at the \cii\ peak, and position 2 is at the \oi\ peak as well as the \thco(3-2) local peak (Fig.~\ref{figure:integmap_IC1396A}). Position 3 is the position of the Class 0 object PACS-1 \citep{Sicilia-Aguilar2014}, position 4 is the \textit{Spitzer} InfraRed Array Camera (IRAC) 8\um\ peak at the tip of the BRC. Position 5 is inside the BRC deeper than the \thco\ peak and \ci\ peak (Fig.~\ref{figure:integmap_IC1396A_CI492}), and position 6 is where \cii\ shows a strong blue wing in the spectrum (Fig.~\ref{figure:spec_selected_IC1396A}).

At the bottom of Fig.~\ref{figure:integmap_IC1396A}, we sketch the structures identified in IC~1396A. In the following, we explain these structures in detail.

(i) At $-11$ and $-10.5$\,\kms \ in the channel maps of \cii\ (Fig.~\ref{figure:channelmap_IC1396A_CII}), blueshifted components are visible in the northern and southern regions, appearing as a clear blueshifted tail in the spectra (position 6 in Figs.~\ref{figure:integmap_IC1396A} and \ref{figure:spec_selected_IC1396A}).

(ii) At $-10$ to $-8$~\kms, a shell around the cavity with LkH$\alpha$ 349a/c is visible with a velocity gradient: at $-10$~\kms, the southwest of LkH$\alpha$ 349 shows a clearer void, while at $-8$~\kms\ the void is more northeast of LkH$\alpha$ 349. \citet{Sicilia-Aguilar2013} detected \sii\ emission at the rim of this void, with particularly strong emission from the northwest region, suggesting that low-mass stars contribute to shaping of the surrounding cloud. The velocity gradient in \cii\ can be explained by expansion of the gas, likely also caused by the low-mass stars located inside the cavity. At these velocities, more void-like structures are observed in IC~1396A, with thin bridges and shells surrounding them.

(iii) The outer rim that is prominent in the IRAC 8\um\ and PACS 70\um\ maps (Fig.~\ref{figure:integmap_IC1396A}) is the interface to the ionized region illuminated by the UV radiation from HD~206267. The \cii\ peak velocity at the tip of the rim (the eastern side) is at $-8$ to $-9$\,\kms.

(iv) As shown in \citet{Okada2012} with lower S/N \cii\ data, the integrated intensity map of \cii\ correlates well with the \textit{Spitzer} IRAC 8\um\ map, with a peak in the northern part of the BRC (position 1).  Inside the northern part of the BRC (position 2), the velocity profile of CO(3-2) shows a clear flat-top shape in contrast to that of \thco(3-2), probably due to its high optical depth. In the northern part of the BRC, \cii\ has its peaks closer to the rim, and \oi\ 63\um\ is stronger inside the BRC, similar to CO(3-2). This indicates a clumpy structure in the northern part to allow the UV radiation to reach the middle of the BRC and excite the \oi.

(v) Position 3 is the position of the Class 0 object PACS-1 with cold dense gas on its western side \citep{Sicilia-Aguilar2019}. We also see that \ceio(3-2) is concentrated west of the illuminated rim, while \thco(3-2) and CO(3-2) are more widely distributed in the BRC. The \oi\ 63\um\ spectra at positions 3 and 4 show a flat top shape with a slope (the apparent peak of the \oi\ 63\um\ is blueshifted compared to the \oi\ 145\um\ peak), and the \oi\ 145\um\ has the same peak velocity as those of \thco(3-2) and \ceio(3-2), clearly indicating the self-absorption of \oi\ 63\um.  The peak velocity of \ci\ at lower spatial resolution \citep{Okada2012} also matches those of \thco(3-2) and \ceio(3-2). We conclude that there is a high column of hot and dense gas at these positions. This is consistent with the strong 70\um\ emission there and the temperature distribution derived by \citet{Sicilia-Aguilar2019}.

(vi) At position 5, the flat-top spectra of CO(3-2) and even \thco(3-2) indicate a high CO column density. While the \cii\ spectrum also appears to have a flat-top, the \thcii\ analysis shows no evidence of significant optical thickness (Sect.~\ref{subsec:13CII_IC1396A}). Inside the southern part of the BRC, including position 5, \cii\ and \oi\ 63\um\ are generally weak. The \oi\ 145\um\ is not detected at position 5 (as mentioned above, the baseline around $-15$~\kms\ is affected by the atmospheric absorption line, but there is no sign of a strong emission line at $-8$~\kms), and the \oi\ 63\um\ is also weak, although self-absorption is not fully excluded due to the limited S/N of \oi\ 145\um. Position 5 is near the peak of \ci\ 492 GHz \citep[Fig.~\ref{figure:integmap_IC1396A_CI492};][]{Okada2012}, which is located deeper in the BRC than the \ceio(3-2) peak, close to the CO(3-2) peak (roughly along the lines of positions 3--5). All these are consistent with the scenario that the UV radiation in the southern part is strongly shielded by the dense molecular gas west of the rim, and the PDR emissions are weak in the southern part of the BRC. The cold temperature derived by \citet{Sicilia-Aguilar2019} behind the rim in the southern part also supports our interpretation.

(vii) At a redshifted velocity of $-6$\,\kms\ there is a ring-like structure tracing the northern edge of the BRC and continuing inside the BRC toward the south (west of the \ceio\ integrated intensity peak). This redshifted ring exists only in \cii, not in the \oi\ channel maps (Fig.~\ref{figure:channelmap_IC1396A_OI63}).

The spectra in Fig.~\ref{figure:spec_selected_IC1396A} show that the peak velocity of \cii\ is often blueshifted compared to \thco(3-2) and \ceio(3-2) (position 1--4 and 7). The displacement of the \cii\ and \thco(3-2) peak velocity is also clearly visible in the p-v diagrams (Fig.~\ref{figure:pvdiagram_IC1396A}). The difference is small, while it exists, near the edge of the rim (cut a), but it is significant on the backside of the BRC (cut b). Cut (b) also clearly traces the blueshifted tails on the northern and southern edges of the BRC that are visible in the \cii\ channel maps (Fig.~\ref{figure:channelmap_IC1396A_CII}), and \thco(3-2) also follows this curve to a lesser extent.  We also found that the velocity profile of \ci\ 492~GHz \citep{Okada2012} is overall very close to that of \ceio(3-2). As mentioned above, \oi\ 145\um\ at positions 3 and 4 has the same peak velocity as that of \thco(3-2) and \ceio(3-2). This confirms that there is no peak velocity shift between \oi\ and CO at these positions. Positions 1 and 2 were not covered by the \oi\ 145\um\ observations, but a steep decrease in the \oi\ 63\um\ at the peak velocity of \thco(3-2) at position 2 is also consistent with the \oi\ 63\um\ being self-absorbed. We discuss the interpretation of the blueshifted \cii\ spectra compared to other emission lines in Sect.~\ref{subsec:discussion_photoevaporation}.

\begin{figure*}
\centering
\includegraphics[width=\hsize]{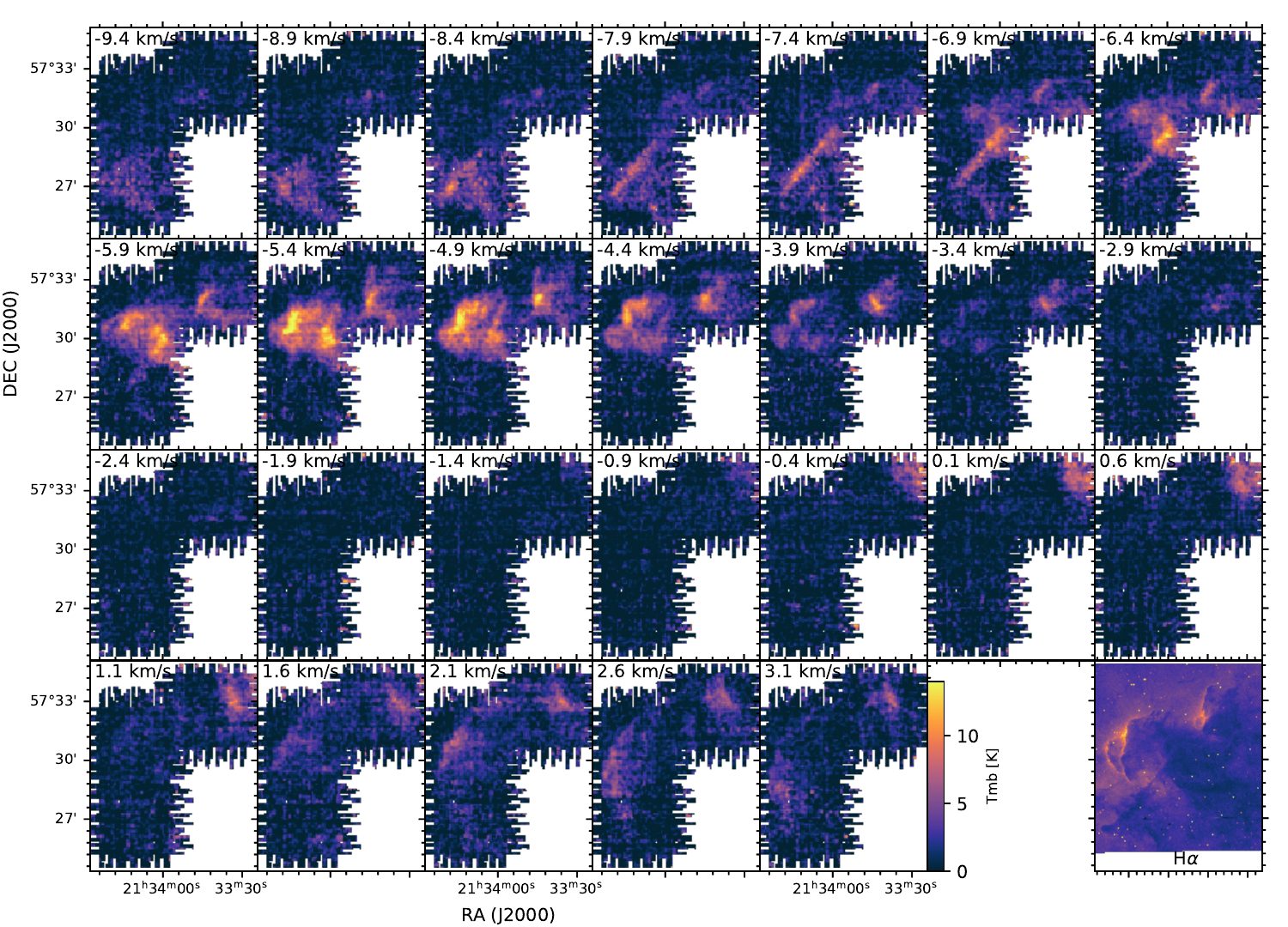}
\caption{Channel maps of \cii\ in IC~1396B (16\arcsec\ resolution). The bottom-right panel shows the H$\alpha$ image from the archive of the INT (2.5 m \textit{Isaac Newton} Telescope)/WFC (Wide Field Camera) Photometric H$\alpha$ Survey of the Northern Galactic Plane \citep[IPHAS;][]{Barentsen2014} for comparison.}
\label{figure:channelmap_IC1396B_CII}
\end{figure*}

We quantitatively analyzed the velocity profile in the same manner as in \citet{Sicilia-Aguilar2019}; we fit the line with multiple Gaussians and determined the peak velocity and the width of the line where the intensity is above 10\% of the peak intensity (Fig.~\ref{figure:width_IC1396A}).  As discussed in \citet{Sicilia-Aguilar2019}, we used the fitted spectra to derive these quantities so that they are not affected by the noisy data, and the derived quantities are not affected by the non-uniqueness of the fit because we do not interpret the individual Gaussian components but only use the sum of the fitted Gaussians.  The peak velocity here represents the absolute peak, independent of the structure of the line wings (i.e., the skewness of the profile) in contrast to the standard moment 1. The width above 10\% intensity better represents weak wings than the width above 50\%.  We used the map with a resolution of 25\arcsec\ and limited the analysis to spectra with S/N greater than 8 in order to avoid an over-interpretation of low S/N spectra. Figure~\ref{figure:width_IC1396A} shows the derived line width and peak velocity for \thco(3-2), \ceio(3-2), and \cii. The peak velocity maps clearly show that \cii\ is overall blueshifted compared to \thco(3-2) and \ceio(3-2). The \cii\ peak velocity map shows that the northwestern and southwestern part of the BRC have bluer peak velocities ($<\!-9.0$~\kms), which is also visible in \thco(3-2). As mentioned above, this is not due to the blue wing, but to the velocity shift of the peak emission, and it is also presented as the curved shapes of the p-v diagram of cut (b) in Fig.~\ref{figure:pvdiagram_IC1396A}. The jump in the peak velocity at the northwest edge of \cii\ is due to a stronger peak in the velocity component at $-6.5$~\kms\ (Fig.~\ref{figure:channelmap_IC1396A_CII}). The maps of \thco(3-2) and \cii\ width have a similar spatial distribution, in the sense that the line width increases toward the southwest and northwest, as shown with \ceio(2-1) by \citet{Sicilia-Aguilar2019}, but the \cii\ shows an overall broader line width (note that the color scales of the \cii\ width are different from those of the other lines). Very wide \cii\ widths at the northern and southern edges of the map are due to a blue wing (see Fig.~\ref{figure:pvdiagram_IC1396A}).  The region where the width of \thco(3-2) and \ceio(3-2) increases west of the densest part of the BRC match with the peak of the \ci\ 492~GHz integrated intensity (Fig.~\ref{figure:integmap_IC1396A_CI492}). 

\subsubsection{IC~1396B}\label{subsec:result_IC1396B}

\begin{figure*}
\centering
\subfigure{\includegraphics[width=0.279\hsize]{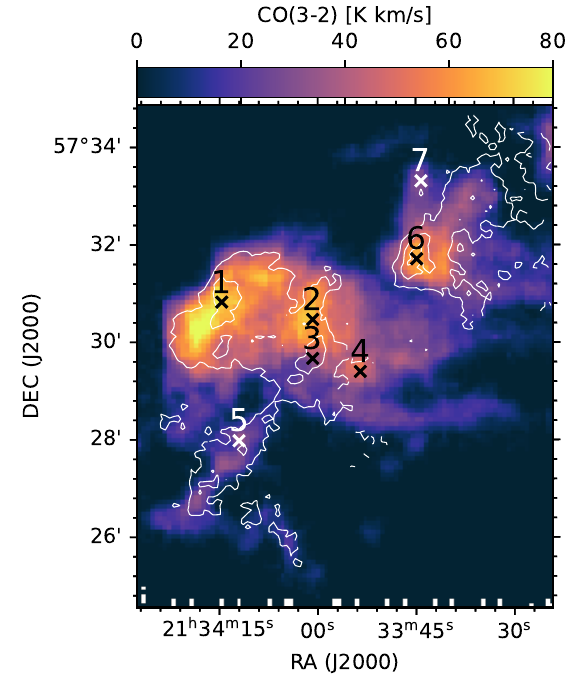}}
\subfigure{\includegraphics[width=0.223\hsize]{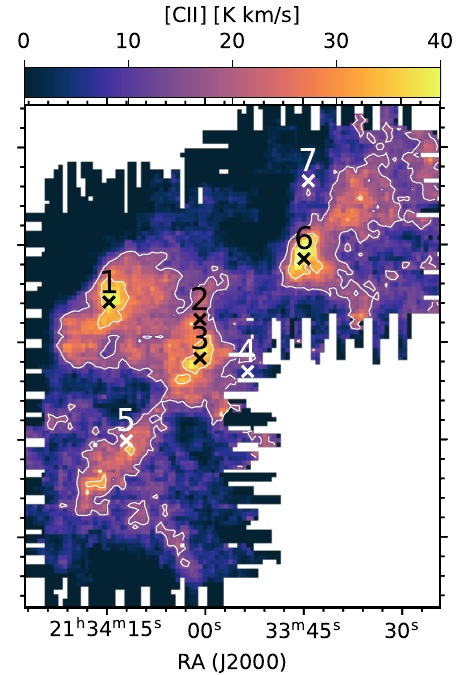}}
\subfigure{\includegraphics[width=0.223\hsize]{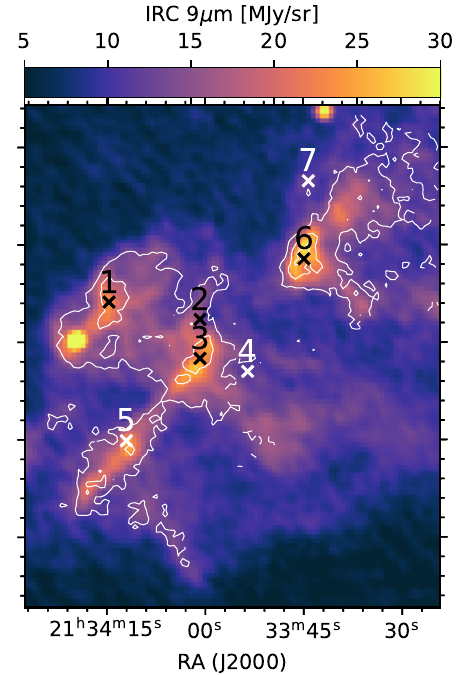}}
\subfigure{\raisebox{0.25cm}{\includegraphics[width=0.25\hsize]{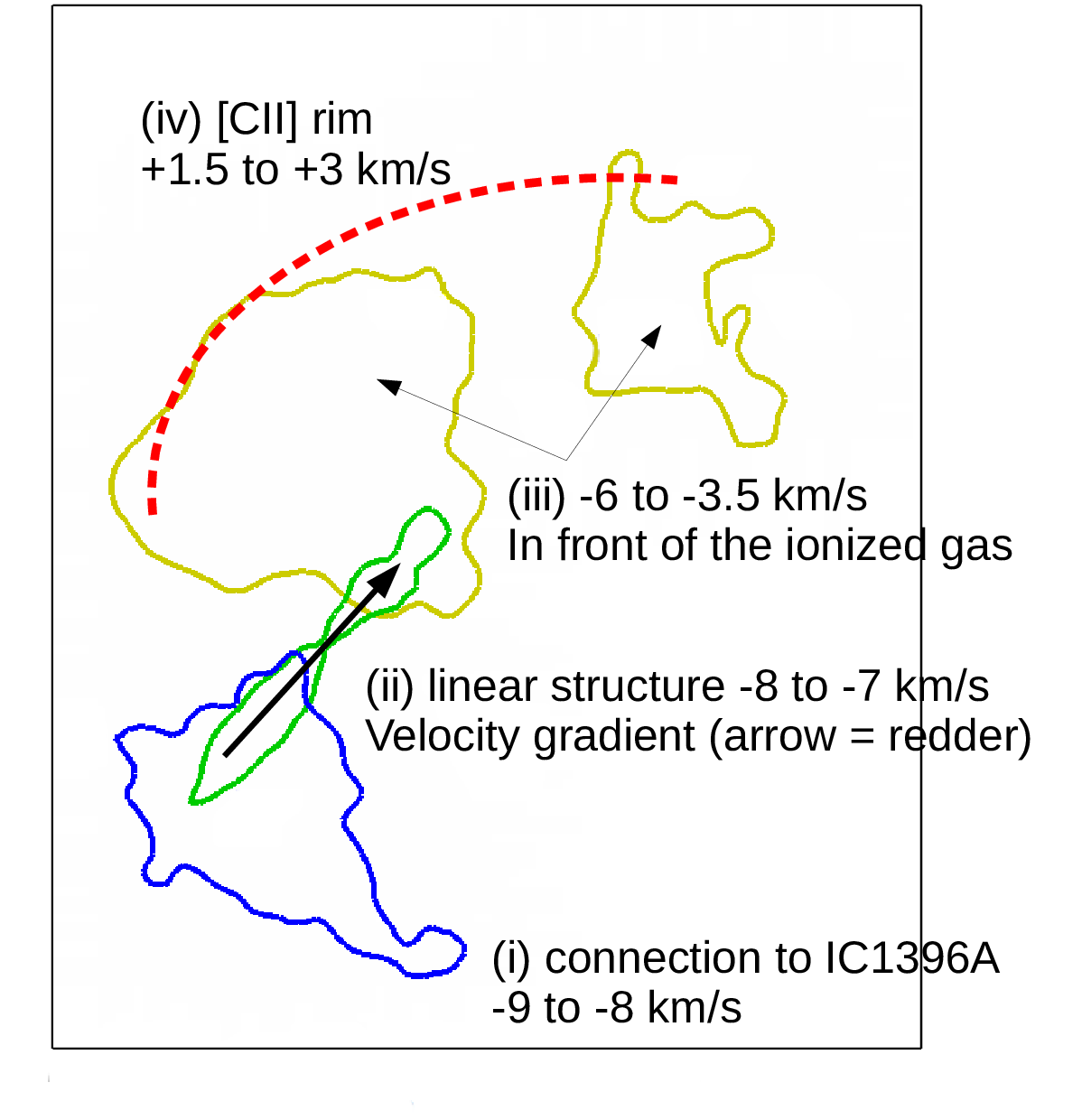}}}
\caption{Integrated intensity maps (colors, 16\arcsec\ resolution) of the different tracers (denoted on the top of each panel) overlaid with contours of \cii\ integrated intensity (solid white lines) in 
IC~1396B. White and black areas have the same meaning as in Fig.~\ref{figure:integmap_IC1396A}. Integrated intensities were calculated over the components detected by the dendrogram analysis (see the main text). Black crosses with numbers mark positions whose spectra are shown in Fig.~\ref{figure:spec_selected_IC1396B}. The right panel shows the schematic decomposition as in Fig.~\ref{figure:integmap_IC1396A}.}
\label{figure:integmap_IC1396B}
\end{figure*}

\begin{figure*}
\centering
\includegraphics[width=\hsize]{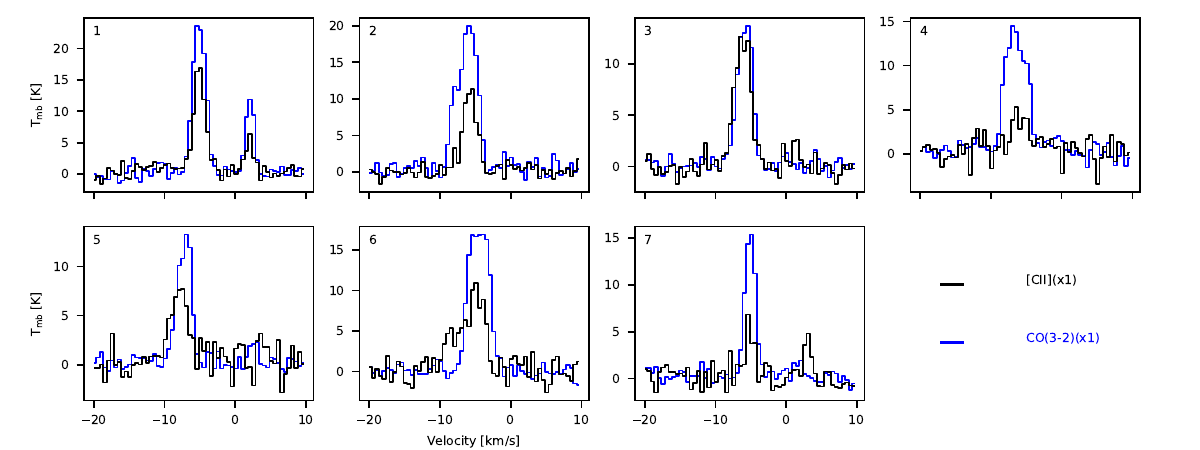}
\caption{Spectra at the selected positions in IC~1396B marked in Fig.~\ref{figure:integmap_IC1396B} (16\arcsec\ resolution).}
\label{figure:spec_selected_IC1396B}
\end{figure*}

Figure~\ref{figure:channelmap_IC1396B_CII} shows the channel maps of \cii, Fig.~\ref{figure:integmap_IC1396B} shows the integrated intensity maps, Fig.~\ref{figure:spec_selected_IC1396B} plots the spectra of selected positions marked in Fig.~\ref{figure:integmap_IC1396B}, and Fig.~\ref{figure:pvdiagram_IC1396B} shows the p-v diagrams for two cuts through IC~1396B.

The spatial distribution of the \cii\ integrated intensity not only follows AKARI/Infrared Camera (IRC) 9\um, but also generally agrees with the CO(3-2) map (Fig.~\ref{figure:integmap_IC1396B}), which is not the case for the other BRCs we study in this paper. The \cii\ channel maps of IC~1396B show a rich velocity structure in this region (Fig.~\ref{figure:channelmap_IC1396B_CII}). At the bottom of Fig.~\ref{figure:integmap_IC1396B}, we sketch the structures identified in IC~1396B.

(i) The southern region appears prominent at a velocity of $-9$ to $-8$\,\kms, similar to the rim in IC~1396A. In the MIR and optical images (Fig.~\ref{figure:IC1396_overview}), a bridging structure between IC~1396A and B is visible, and a similar velocity between the southeastern part of IC~1396B and IC~1396A indicates that they are indeed connected.

(ii) At the velocity of $-8$ to $-7$\,\kms, there is a strong linear structure, where the southeast is more blueshifted, and it becomes more redshifted toward the northwest, connecting to the strongest cloud components at the velocity of $<-7$\,\kms. Figure~\ref{figure:pvdiagram_IC1396B} shows the p-v diagrams along this structure (cut (b)). This linear structure is visible in both \cii\ and CO(3-2) (Fig.~\ref{figure:integmap_IC1396B} and positions 5 in Fig.~\ref{figure:spec_selected_IC1396B}).  There are no known YSOs along this structure \citep{Barentsen2011,Nakano2012}.  Similar straight features have been observed in other regions, for example the "Stick" in Orion \citep{Kong2021} or a linear feature in NGC~7538 \citep{Beuther2022}.  Unlike the Stick, which is modeled by collision-induced magnetic reconnection \citep{Kong2021}, the linear feature here in IC~1396B has no obvious ring or double line structure.  It is also unlikely that the direction of the magnetic field is parallel to this straight feature, because three out of the four BRCs measured by \citet{Soam2018} have the magnetic field parallel to the Galactic plane, which is rather perpendicular to this line.  This may just be the UV-illuminated edge of a part of the structure connecting IC~1396B and A, also visible as a dark lane in the optical \citep[see the H$\alpha$ panel of our Fig.~\ref{figure:channelmap_IC1396B_CII} as well as Fig. 2 of ][]{Barentsen2011}, which happens to look straight.  The \cii\ spectrum along this straight feature (position 5 in Fig.~\ref{figure:spec_selected_IC1396B}) shows a slightly blueshifted profile and a blue wing compared to CO(3-2).

(iii) At a velocity of $-6$ to $-3.5$~\kms, the channel maps show two bright \cii\ clouds (Fig.~\ref{figure:channelmap_IC1396B_CII}).  The H$\alpha$ map \citep[the bottom right panel of Fig.~\ref{figure:channelmap_IC1396B_CII}, ][]{Barentsen2011} outlines these clouds as a silhouette with bright rims, strongly suggesting that these clouds are in front of the ionized gas along the line of sight.

(iv) There is a clearly separated velocity component in both \cii\ and CO(3-2) at $+1.5$ to $+3$\,\kms\ (e.g., position 1 in Fig.~\ref{figure:spec_selected_IC1396B}). The p-v diagrams in Fig.~\ref{figure:pvdiagram_IC1396B} do not indicate any connection between the positive and negative velocity components. The positive velocity components form a large rim structure, facing east-northeast, although the direction of the exciting star is more east-southeast. The p-v diagram of cut (b) in Fig.~\ref{figure:pvdiagram_IC1396B} does not show a clear velocity gradient of the positive velocity component along the rim. There is no corresponding structure visible in the H$\alpha$ map. This may be a separate cloud behind the ionized gas, illuminated from the front by the exciting star.  The northern part of the velocity component $>0$~\kms\ is dominated by the \cii\ emission and CO(3-2) is very faint (west side of the cut (b) in Fig.~\ref{figure:pvdiagram_IC1396B} and position 7 in Fig.~\ref{figure:spec_selected_IC1396B}).

Except for a blue wing in the \cii\ emission around position 5 and 6 (Fig.~\ref{figure:spec_selected_IC1396B}), the \cii\ profile shows no significant broadening compared to CO(3-2).

\begin{figure*}
\centering
\includegraphics[width=\hsize]{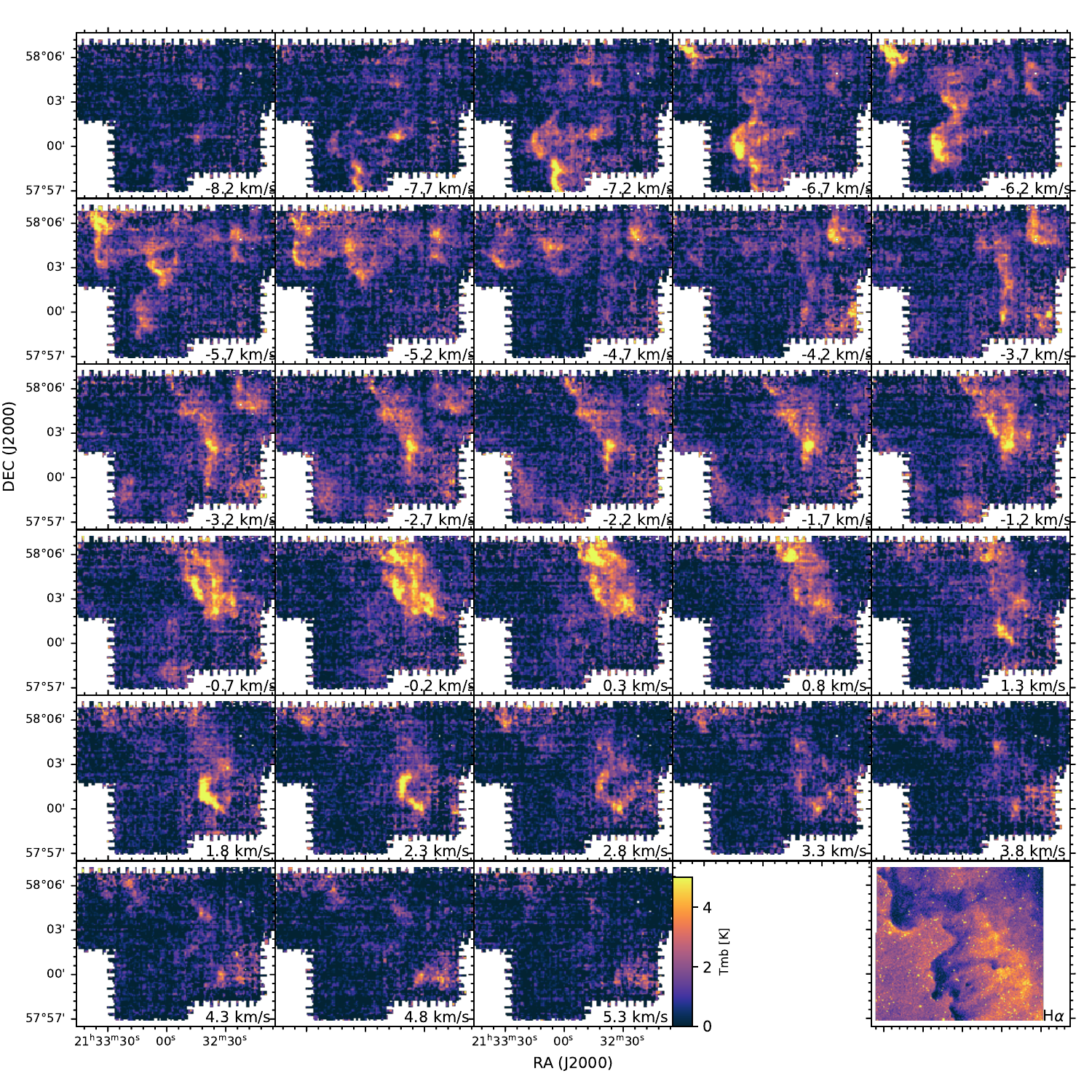}
\caption{Same as Fig.~\ref{figure:channelmap_IC1396B_CII} (16\arcsec\ resolution) but for the \cii\ in IC~1396D.}
\label{figure:channelmap_IC1396D_CII}
\end{figure*}

\subsubsection{IC~1396D} \label{subsec:result_IC1396D}

\begin{figure*}
\centering
\includegraphics[width=0.525\hsize]{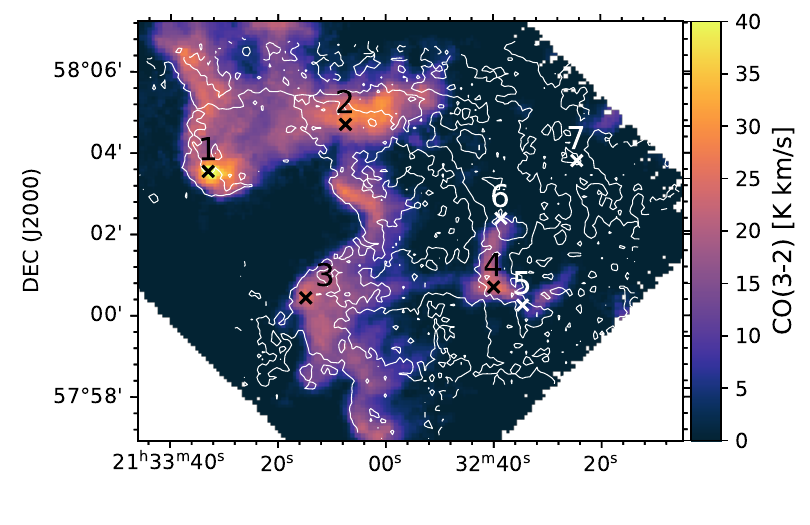}
\vspace{-0.5cm}
\includegraphics[width=0.47\hsize]{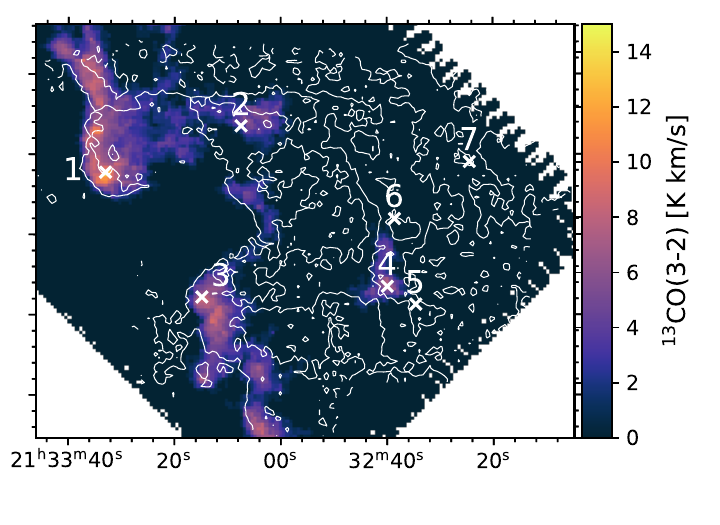}
\includegraphics[width=0.525\hsize]{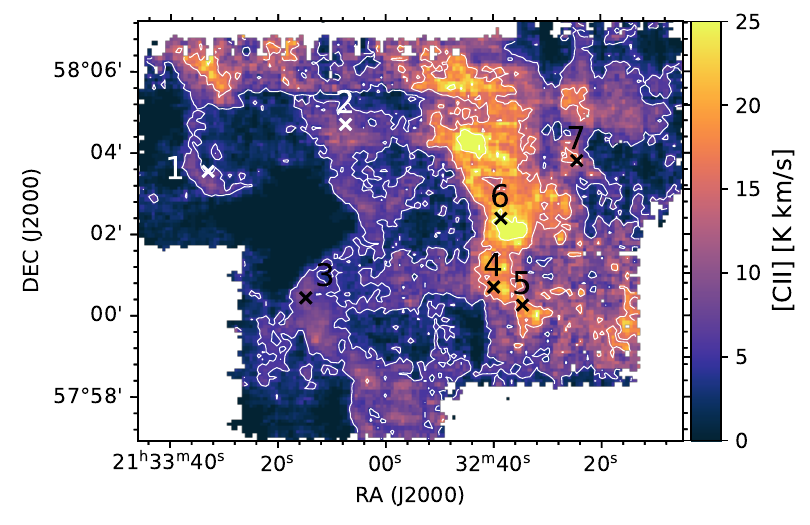}
\vspace{0.cm}
\includegraphics[width=0.47\hsize]{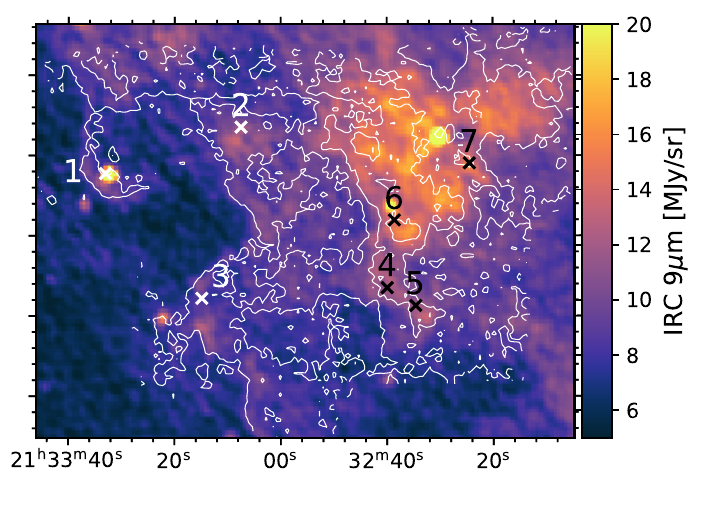}
\includegraphics[width=0.45\hsize]{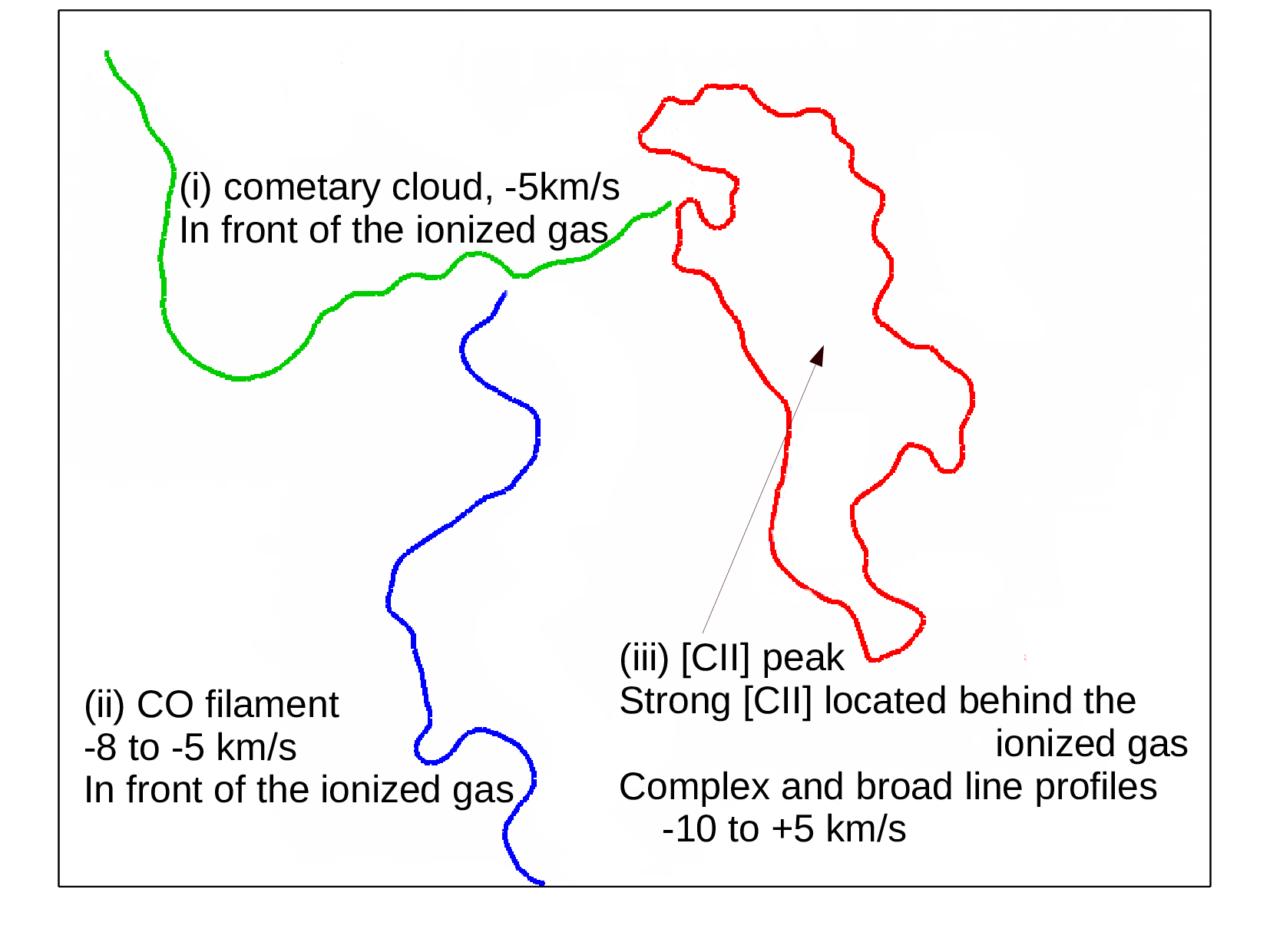}
\caption{Integrated intensity maps (colors, 16\arcsec\ resolution) of the different tracers (denoted on the right of each panel) overlaid with contours of \cii\ integrated intensity (white lines) in IC~1396D. White and black areas have the same meaning as in Fig.~\ref{figure:integmap_IC1396A}. Integrated intensities were calculated over the components detected by the dendrogram analysis (see the main text). Black crosses with numbers mark positions whose spectra are shown in Fig.~\ref{figure:spec_selected_IC1396D}. The bottom panel shows the schematic decomposition as in Fig.~\ref{figure:integmap_IC1396A}.}
\label{figure:integmap_IC1396D}
\end{figure*}

\begin{figure*}
\centering
\includegraphics[width=\hsize]{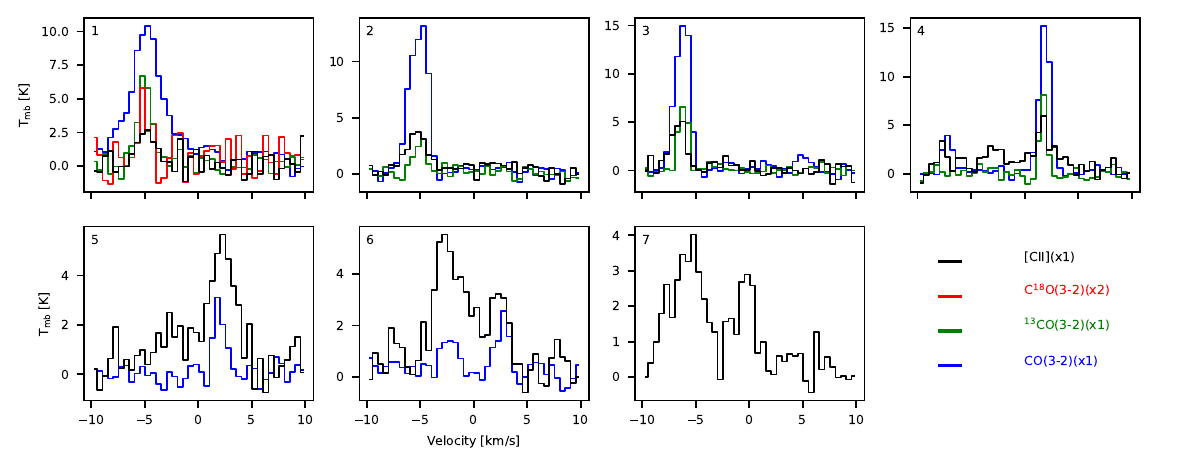}
\caption{Spectra at the selected positions in IC~1396D marked in Fig.~\ref{figure:integmap_IC1396D} (16\arcsec\ resolution)}
\label{figure:spec_selected_IC1396D}
\end{figure*}


IC~1396D shows a complex structure with multiple velocity components as shown in the channel maps (Fig.~\ref{figure:channelmap_IC1396D_CII}). The CO and \cii\ integrated maps (Fig.~\ref{figure:integmap_IC1396D}) look very different, but individual velocity components are often common in the spectra (Fig.~\ref{figure:spec_selected_IC1396D}).  At the bottom of Fig.~\ref{figure:integmap_IC1396D}, we sketch the structures identified in IC~1396D.

(i) A cometary-shaped cloud in the north is a BRC with an IRAS source identified by \citet{Sugitani1991} (SFO34). \cii\ at the rim is faint but detected (Fig.~\ref{figure:channelmap_IC1396D_CII} around $-5$\kms). The rim is also visible in the IRAC 8\um\ image of \citet{Rebull2013} and AKARI 9\um\ (Fig.~\ref{figure:integmap_IC1396D}). The H$\alpha$ image (Fig.~\ref{figure:channelmap_IC1396D_CII} bottom right panel) also shows a bright rim just outside the dark cometary cloud, indicating that they are located in front of the ionized gas along the line of sight.

(ii) CO filaments extend to the south, and the same structure is visible in \cii\ ($-8$ to $-5$~\kms). This filamentary structure appears as dark clouds in H$\alpha$ (Fig.~\ref{figure:channelmap_IC1396D_CII} bottom right panel), with a sharp boundary on the eastern side toward the exciting source, indicating that they are also located in front of the ionized gas.

(iii) The integrated intensity peak of \cii\ is located west of the CO filament, where the 9\um\ emission is also bright, and further away from the star in projection than the CO filament. The velocity structure is complex (Fig.~\ref{figure:pvdiagram_IC1396D}) and some components have a very broad line profile (positions 4--7 in Fig.~\ref{figure:spec_selected_IC1396D}). This indicates that the region with strong \cii\ receives the UV radiation from the exciting source without being shielded by the BRC and the CO filament in the north-south direction, but they are displaced along the line of sight.  This is also supported by the gradual increase in H$\alpha$ emission from the CO filaments toward the west, and the fact that the boundary to the next cloud (at the \cii\ peak) does not appear as sharp as the blueshifted BRC and the CO filament. The presence of multiple velocity components in \cii\ also suggests a 3D distribution of C$^+$. An ionized gas origin of the \cii\ emission is unlikely, or at least not significant, because of the associated weak CO emission ($-5$ to 0~\kms\ seen at position 6) and a good correlation with the 9\um\ emission, which traces the polycyclic aromatic hydrocarbon (PAH) emission. 

In a simple picture of an expanding shell structure created by the evolving \hii\ region, the blueshifted clouds are in front of the \hii\ region along the line of sight, moving toward us, and the redshifted clouds are behind the \hii\ region.  This picture is consistent with the structure observed in IC~1396D: the blueshifted BRC and the associated north-south CO filamentary structure are located closer to the observer, while the western cloud with strong \cii\ is located behind the ionized gas.

When CO and \cii\ velocity components are associated with each other (e.g., position 1 in Fig.~\ref{figure:spec_selected_IC1396D}), in most cases no significant broadening of the \cii\ emission is visible. At position 1, this means that in the BRC with a cometary shape, there is no indication of a velocity shift in any line emission, and thus no sign of ablated surface or photoevaporation flows.

\subsubsection{IC~1396E}\label{subsec:result_IC1396E}

\begin{figure*}
\centering
\includegraphics[width=\hsize]{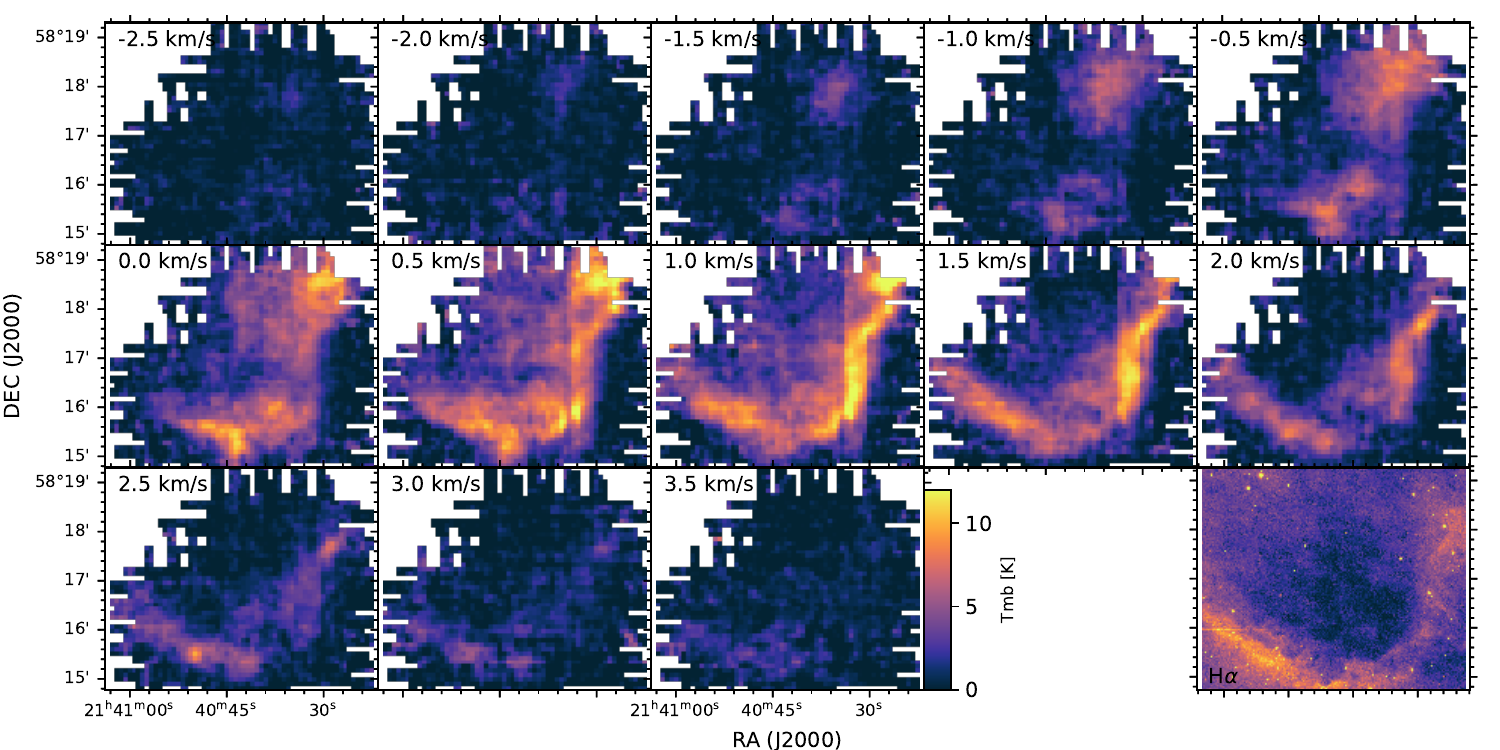}
\caption{Same as Fig.~\ref{figure:channelmap_IC1396B_CII} (16\arcsec\ resolution) but for the \cii\ in IC~1396E.}
\label{figure:channelmap_IC1396E_CII}
\end{figure*}

\begin{figure*}
\centering
\includegraphics[width=0.35\hsize]{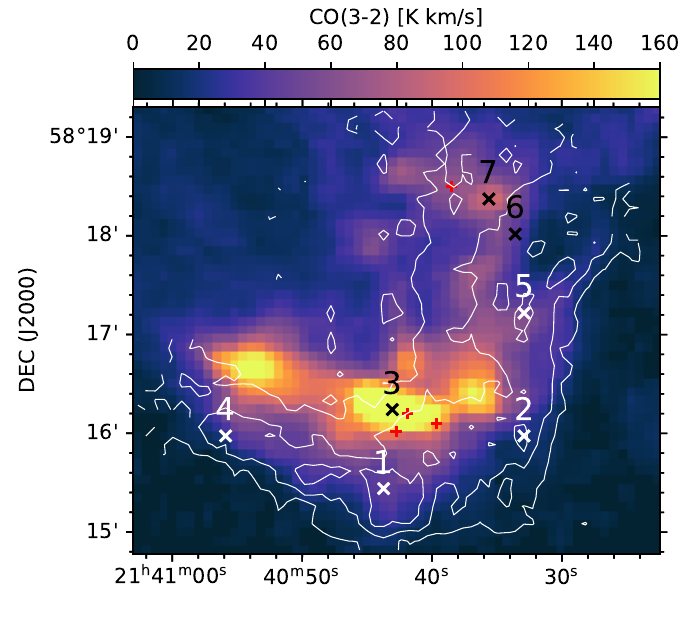}
\includegraphics[width=0.30\hsize]{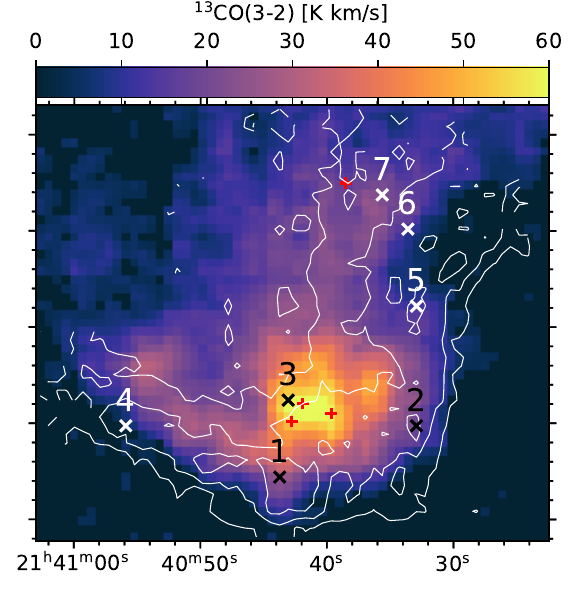}
\includegraphics[width=0.30\hsize]{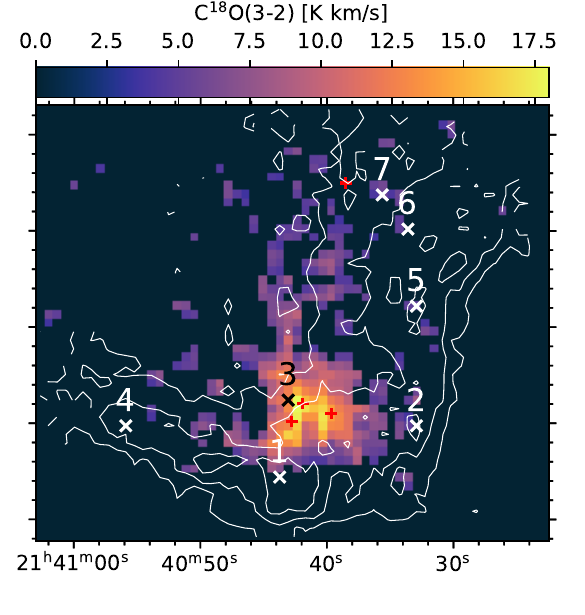}
\includegraphics[width=0.35\hsize]{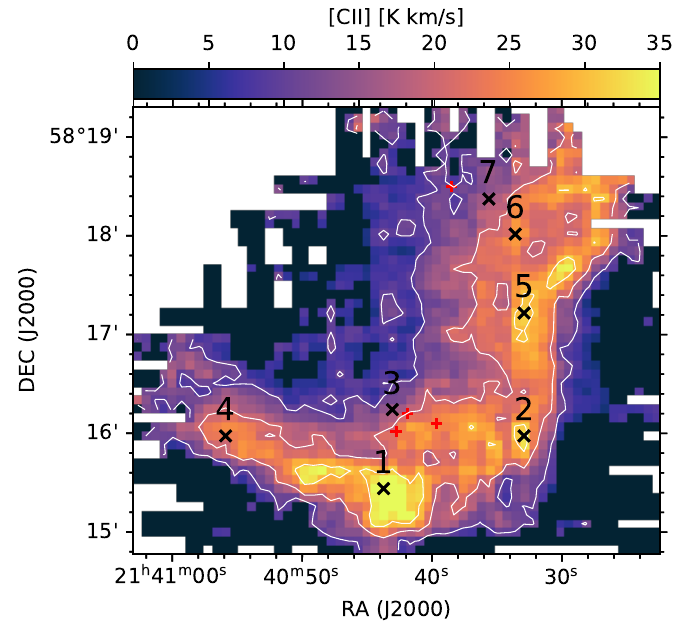}
\includegraphics[width=0.30\hsize]{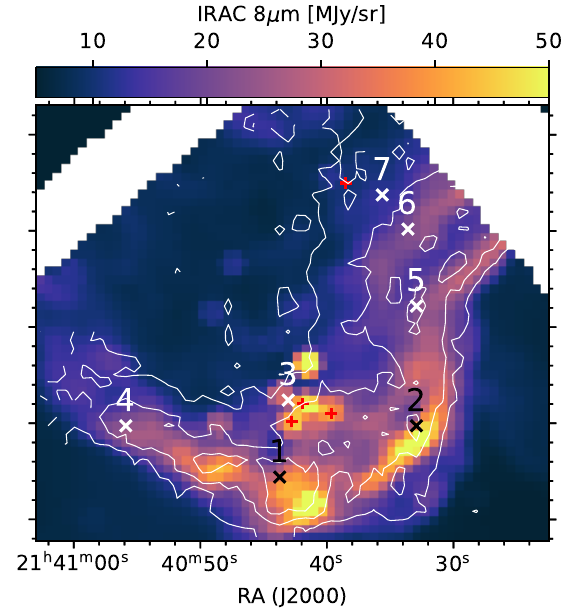}
\includegraphics[width=0.33\hsize]{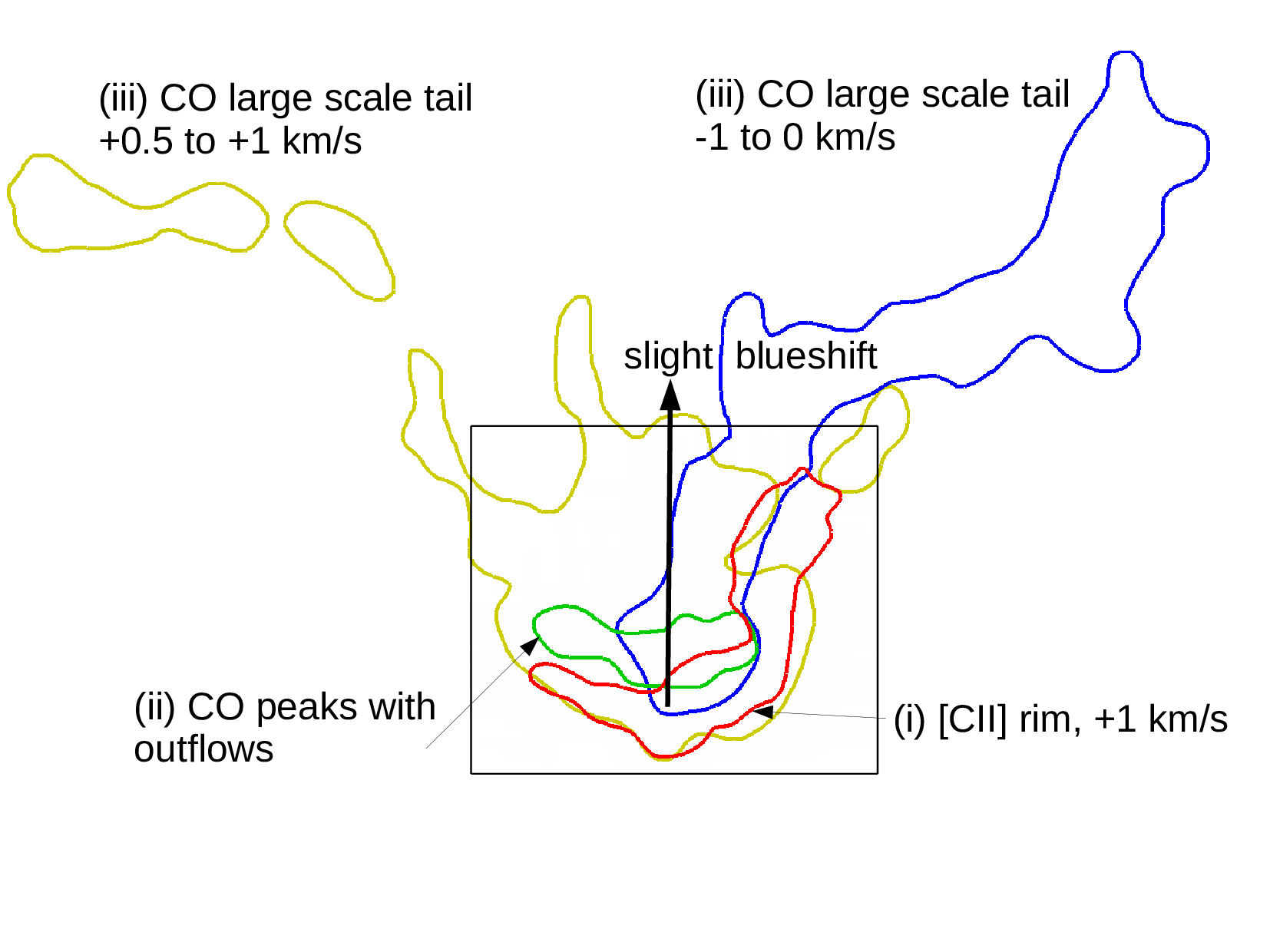}
\caption{Integrated intensity maps (colors, 16\arcsec\ resolution) of the different tracers (denoted on the top of each panel) overlaid with contours of \cii\ integrated intensity (white lines) in IC~1396E. White and black areas have the same meaning as in Fig.~\ref{figure:integmap_IC1396A}. The bottom right panel shows the schematic decomposition as in Fig.~\ref{figure:integmap_IC1396A}. Integrated intensities were calculated by direct integration over the velocity range of ($-5$, $8$)~\kms, except for CO(3-2) with the range of ($-10$, $10$)~\kms. Red pluses are the positions of outflow-driving sources: BIMA1--3 \citep[southern ones,][]{Beltran2002} and the source C \citep[northern one,][]{Codella2001}. Black crosses with numbers mark positions whose spectra are shown in Fig.~\ref{figure:spec_selected_IC1396E}.}
\label{figure:integmap_IC1396E}
\end{figure*}

\begin{figure*}
\centering
\includegraphics[width=0.9\hsize]{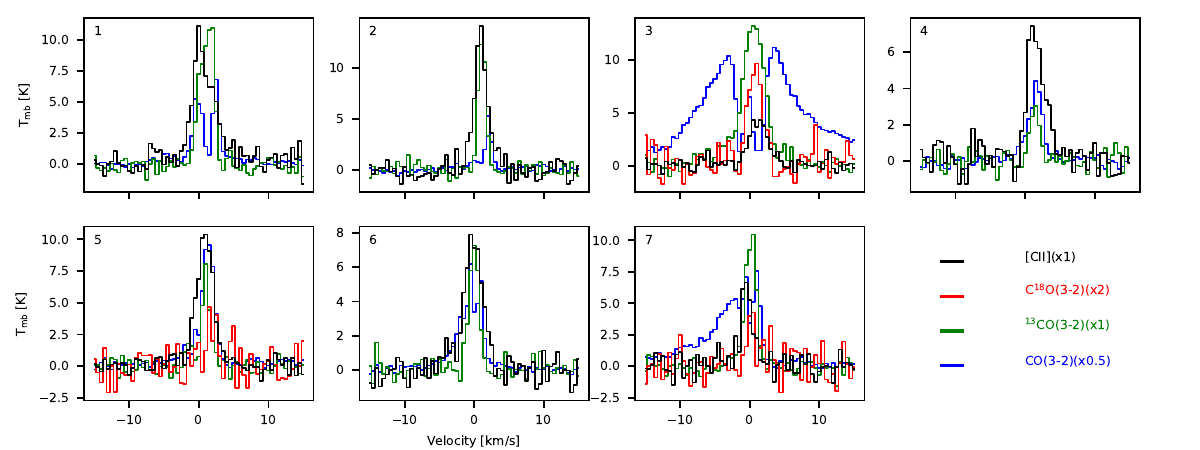}
\caption{Spectra at the selected positions in IC~1396E marked in Fig.~\ref{figure:integmap_IC1396E} (16\arcsec\ resolution).}
\label{figure:spec_selected_IC1396E}
\end{figure*}

\begin{figure*}
\centering
\includegraphics[width=0.28\hsize]{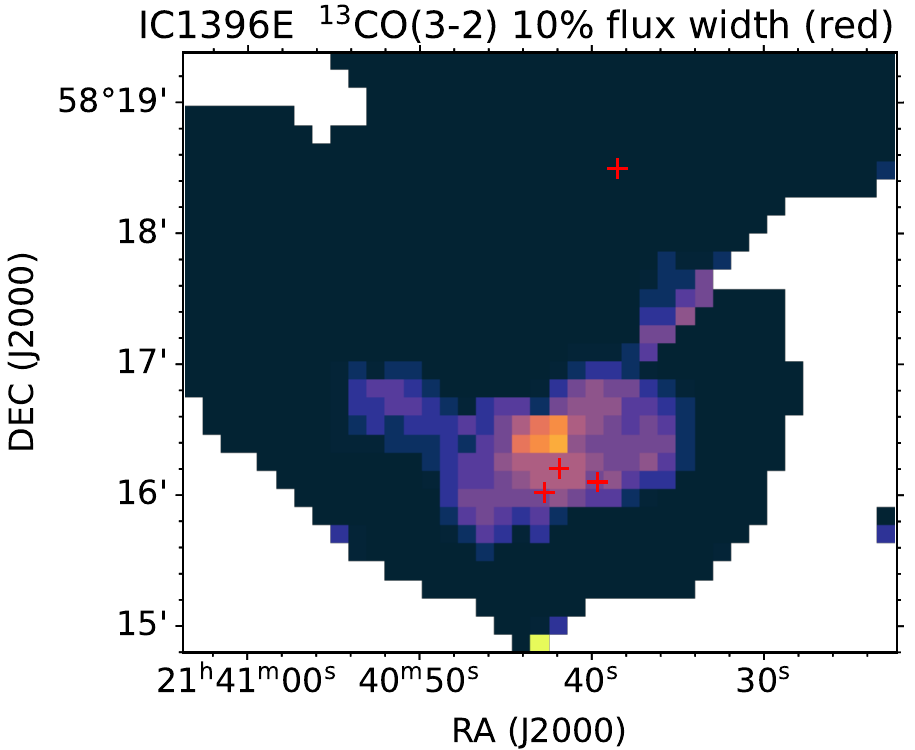}
\includegraphics[width=0.22\hsize]{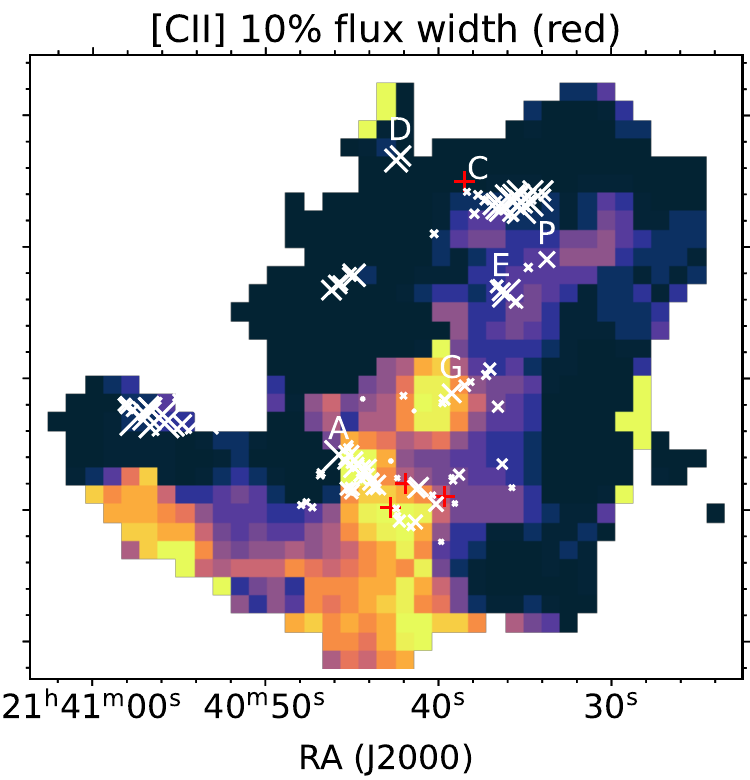}
\includegraphics[width=0.22\hsize]{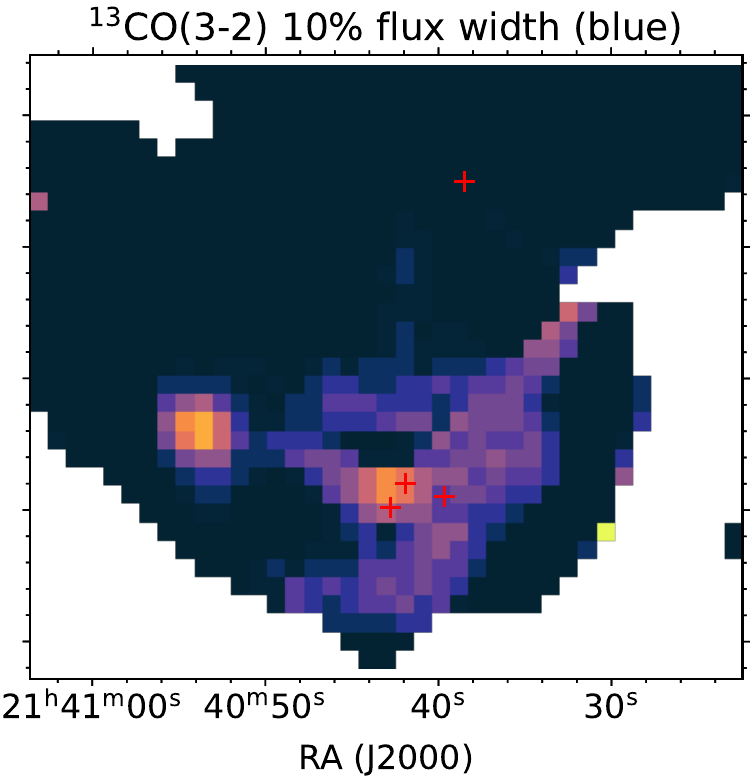}
\includegraphics[width=0.265\hsize]{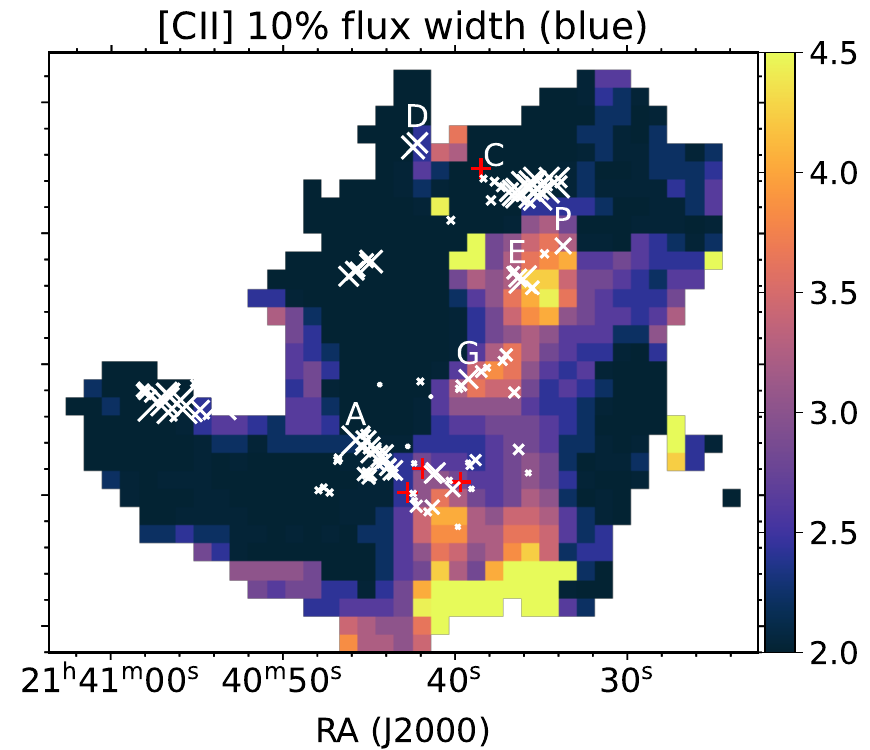}
\caption{Red wing (left two panels) and blue wing (right two panels) line width at 10\% of the peak intensity for \thco(3-2) and \cii\ in IC~1396E at 25\arcsec\ resolution. Red pluses are the positions of the outflow-driving sources: BIMA1--3 \citep[southern ones,][]{Beltran2002} and the source C \citep[northern one,][]{Codella2001} as in Fig.~\ref{figure:integmap_IC1396E}. The crosses on the \cii\ panels are the positions of the H$_2$ knots from \citet{Beltran2009}, with the size of the crosses corresponding to the H$_2$ flux. Some of the H$_2$ knots clusters are labeled according to \citet{Beltran2009}.}
\label{figure:width_IC1396E}
\end{figure*}

Figure~\ref{figure:channelmap_IC1396E_CII} shows the channel maps of \cii, Fig.~\ref{figure:integmap_IC1396E} shows the integrated intensity maps, Fig.~\ref{figure:spec_selected_IC1396E} plots the spectra of selected positions marked in Fig.~\ref{figure:integmap_IC1396E}. At the bottom of Fig.~\ref{figure:integmap_IC1396E}, we again sketch the structures identified in IC~1396E.

(i) The channel maps of \cii\ (Fig.~\ref{figure:channelmap_IC1396E_CII}) show that the rim is clearly outlined at a velocity of about 1~\kms, while the inside of the BRC or the northern part of the map has somewhat blueshifted emission ($-1$ to $0$~\kms). The \cii\ integrated intensity traces the edge of the bright rim, illuminated by the exciting sources from the south (Fig.~\ref{figure:integmap_IC1396E}). The rim is also visible in \thco\ at 1--2~\kms. We discuss the implication of the velocity gradient toward the inside of the BRC in Sect.~\ref{sec:discussion}.

(ii) In contrast to the \cii\ emission, CO stems mainly from the inside of the BRC. As shown in previous works (see Sect.~\ref{sec:intro_IC1396}), the \twco\ lines show a very prominent outflow signature in their spectra, together with a strong self-absorption feature in IC~1396E (Fig.~\ref{figure:spec_selected_IC1396E}). BIMA2, an intermediate-mass protocluster and the driving source of the central outflow, is located west of position 3 in Fig.~\ref{figure:integmap_IC1396E}. Source C, the driving source of the northern outflow \citep{Codella2001}, is located east of position 7 (between the two CO peaks). 

(iii) The channel maps of \thco\ at a larger scale (Fig.~\ref{figure:channelmap_large_IC1396E}) than the area mapped in \cii\ show the two tails with different velocities as already discussed by \citet{Patel1995} (Sect.~\ref{sec:intro}). The eastern tail connects to the rim (i) at its most redshifted velocity.

The effect of outflows on \cii\ is not obvious in the individual spectra (Fig.~\ref{figure:spec_selected_IC1396E}), but the analysis of the velocity widths (Fig.~\ref{figure:width_IC1396E}) shows its influence in the \cii\ spectra. We performed the same analysis as in IC~1396A, fitting the spectra with multiple Gaussians. To account for the skewed shape of the spectra, we derived the red and blue velocity widths separately, namely as the width from the peak velocity to the velocity where the intensity drops to 10\% of the peak intensity on the respective side \citep{Sicilia-Aguilar2019}. Figure~\ref{figure:width_IC1396E} shows the results for \thco(3-2) and \cii.  The \thco(3-2) width shows significant enhancements at the clumps produced by the central outflow. In the \cii\ width maps, there are two local peaks in the middle of the BRC, associated with the H$_2$ knots identified by \citet{Beltran2009}; one peak in the red velocity with the chain of \hh\ knots G, and the other in the blue velocity with the \hh\ knots E and P.  These results suggest that the \cii\ emission traces the ambient gas interacting with the outflows.  

The width of the \cii\ red velocity (upper-right panel of Fig.~\ref{figure:width_IC1396E}) shows that the eastern side of the rim has a broader line in \cii. It is visible as a wing in the \cii\ spectrum at position 4 (Fig.~\ref{figure:spec_selected_IC1396E}). The CO(3-2) peak and the H$_2$ knots are located about 1\arcmin\ (0.27~pc) to the north from the rim. At position 4, the peak velocity of \cii\ and \thco(3-2) match, and this red wing in \cii\ may be a sign of photoevaporation flow (see Sect.~\ref{sec:discussion}). On the other hand, at the intensity peak of the rim (position 1), the \thco(3-2) peak velocity and the absorption peak of CO(3-2) correspond to the dip in the \cii\ line profile; thus, the \cii\ profile is likely affected by self-absorption.

\subsection{\thcii\ in IC~1396A} \label{subsec:13CII_IC1396A}

\begin{figure*}
\centering
\includegraphics[width=0.9\hsize]{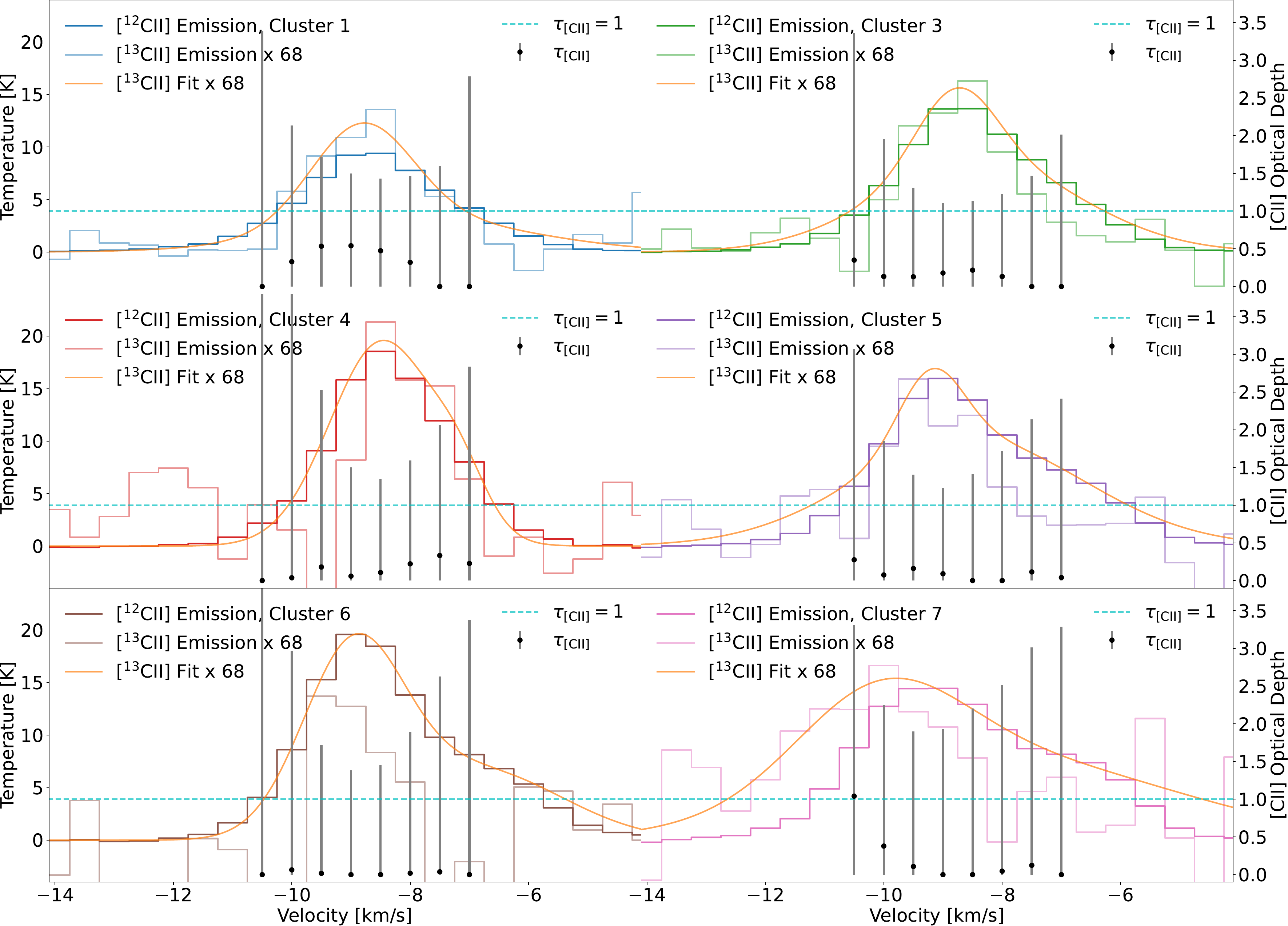}
\caption{Averaged spectra in each dendrogram cluster (Fig.~\ref{figure:2d_dendro_IC1396A}) in IC~1396A except cluster 2, which does not have a high enough S/N for the \thcii\ analysis. Colored stepped data are the \cii\ spectra (thick) and the combined and scaled \thcii\ spectra (thin). The orange curve is the fit to the \thcii\ spectra using two Gaussian profiles. Derived optical depths of \cii\ are shown as black data points with error bars. See the main text for details.}
\label{figure:dendro_spec_tau_IC1396A}
\end{figure*}

We analyzed the \thcii\ hyperfine emission lines in IC~1396A, where the S/N of the \cii\ observations is the highest. The \thcii\ $^2$P$_{2/3}$-$^2$P$_{1/2}$ transition splits into three hyperfine components with relative intensities of 0.625, 0.25 and 0.125 \citep{Ossenkopf2013}. The strongest component ($F=2-1$) is $11.2$~\kms\ away from the \cii\ line and often blended with its wing, but it can be recovered using a polynomial fit to the wing of the \cii\ line. The S/N of individual positions in the map is not sufficient for a \thcii\ analysis; we thus averaged the spectra over larger areas. To determine which area to average, we performed a 2D dendrogram analysis on the \cii\ integrated intensity using the same parameters as in the 3D dendrogram analysis (Sect.~\ref{subsec:integrated_intensities}) except for the number of pixels for a leaf. A 3D dendrogram is not appropriate for this purpose because the area to average should be common at all velocities, so we change the number of pixels for a leaf to $3^2$, suitable for a 2D dendrogram. Figure~\ref{figure:2d_dendro_IC1396A} shows the seven clusters identified. We averaged the spectra within each cluster to disentangle the \thcii\ emission from the noise floor. We fit the baseline around each hyperfine line by a second-order polynomial and subtracted it (see Appendix~\ref{app:13cii_baseline} for the discussion on the impact on the order of the baseline). We have a clear detection of the \thcii\ $F=2-1$ line in most of the clusters. In a second step we average the three hyperfine lines weighted by their relative intensities as described in \citet{Guevara2020}. The averaged \thcii\ emission scaled up by the local ${\rm ^{12}C/^{13}C}$ abundance ratio $\alpha = 68 \pm 15$ \citep{Milam2005} is shown in Fig.~\ref{figure:dendro_spec_tau_IC1396A}. We find that the scaled up \thcii\ line only slightly overshoots the \cii\ line, indicating weak optical depth effects. 

We then quantified the velocity-resolved optical depth. Since the \thcii\ line is weak, it is easily affected by baseline or noise fluctuations. 
Therefore, we performed a fit to the averaged \thcii\ line using a superposition of two Gaussian profiles. In some velocity bins the scaled \thcii\ are weaker than the \cii\ intensity, which suggests fully optically thin \cii\ within errors. To guide the fit into a physically meaningful solution and to avoid numerical problems when calculating the optical depth, we replaced the scaled \thcii\ with the \cii\ intensity for those data points where the former is smaller than the latter during the fit. This is equivalent to assigning an optical depth of zero. This adjustment was necessary due to significant baseline fluctuations, mostly at the wings of the \thcii\ line where the S/N is lowest.
The fitted \thcii\ spectra are shown by the orange curve in each panel in Fig.~\ref{figure:dendro_spec_tau_IC1396A}. We use the fitted \thcii\ for the following analysis. We assume that the \cii\ emission originates from a layer that can be described by a single excitation temperature. The lack of absorption dips in the \cii\ spectra that would result from additional colder material along the line of sight supports our approximation. In this simple scenario, we can determine the optical depth by comparing the intensities of the two isotopes \citep{Ossenkopf2013}:
        
        \begin{equation}
                \frac{T_{\mathrm{mb, ^{12}C}}(v)}{T_{\mathrm{mb, ^{13}C}}(v)}= \frac{1-e^{-\tau(v)}}{\tau(v)} \alpha~.
                \label{eq:optical_depth}
        \end{equation}
        
\noindent

 The corresponding error is derived following the Eq. (2) in \citet{Kabanovic2022}. It propagates the uncertainties of \cii\ and \thcii\ intensities, and the ${\rm ^{12}C/^{13}C}$ abundance ratio $\alpha$, using the first-order Taylor expansion of each component. The dominant term in the optical depth error comes from the error of \thcii\ ($>=75$\%).
        The resulting velocity-resolved optical depth is shown by the black dots with error bars in Fig.~\ref{figure:dendro_spec_tau_IC1396A}. As mentioned above, we forced the optical depth to zero when the \cii\ intensity is equal or exceeds the $\alpha$-scaled \thcii\ intensity. Consequently only positive error bars indicate physical meanings, and thus shown. The derived $\tau$ values are mostly smaller than 1 as indicated by the dashed cyan line in each panel, and it is compatible with the optically thin \cii\ emission within the error bars. While it cannot be completely excluded that the \cii\ emission is optically thick in a small area of locally high column density, we conclude that the optical depth does not significantly affect the observed \cii\ emission.

\subsection{Heating efficiency in IC~1396A}\label{subsec:heating_eff}

\begin{figure}
\centering
\includegraphics[width=0.9\hsize]{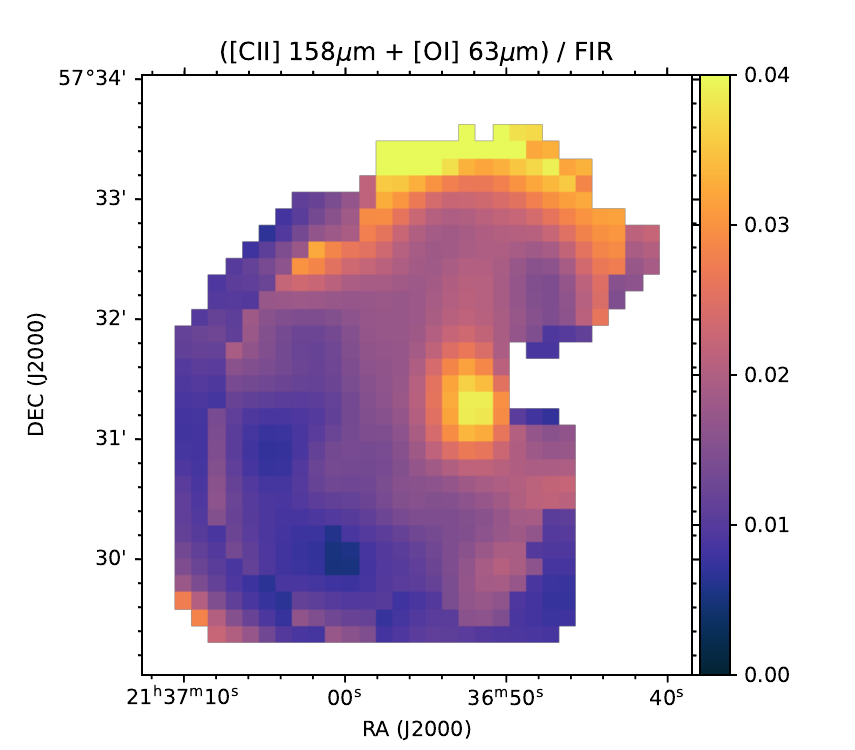}
\caption{(\cii$+$\oi\ 63\um)/FIR integrated intensity ratio in IC~1396A at 25\arcsec\ resolution. A jump in values along the shape of the rim is reflecting the \oi\ 63\um\ detection limit (see Fig.~\ref{figure:integmap_IC1396A} for the area where \cii\ and \oi\ 63\um\ are detected.}
\label{figure:IC1396A_CIIOIdivFIR}
\end{figure}

We investigated the heating efficiency by calculating the (\cii\ 158\um\ $+$ \oi\ 63\um)/FIR in IC~1396A. We fit a gray body with the frequency dependence of the dust absorption coefficient $\beta = 2$ to the PACS 70\um, 160\um, and SCUBA-2 850\um\ and derived the total FIR as the integral of the fitted function. In contrast to the fitted temperature and dust column density, the integrated FIR is insensitive to the assumed $\beta$. Figure~\ref{figure:IC1396A_CIIOIdivFIR} shows the spatial distribution of (\cii\ 158\um\ $+$ \oi\ 63\um)/FIR. It has a value of 3--4\% in the cavity around LkH$\alpha$ 349c and the northern part of the BRC. When interpreting this quantity as heating efficiency and comparing it with the values in previous works, two things have to be considered. One is that we refer to FIR, the total FIR flux only, while some literature reference is made to the ``TIR.'' The TIR is defined as the total infrared flux at 3--1100\um\ \citep{DaleHelou2002} and better traces the total input energy, because it includes the emission from all dust grains, including PAHs and very small grains (VSGs). The ratio of FIR to TIR varies depending on the environment but is typically about 50\% \citep{Pabst2021,Okada2013}. Here we use FIR due to lack of sufficient MIR data. Another difference is that the cooling line flux does include the \oi\ emission, which is often ignored due to its rare availability in observations compared to \cii. In IC~1396A the ratio of \oi\ 63\um/\cii\ ranges from 0.3--1.2. Taking these two differences into account, the heating efficiency in IC~1396A is similar to the upper bounds of the data shown, for example, in \citet{Pabst2021}.

The spatial distribution of the heating efficiency in Fig.~\ref{figure:IC1396A_CIIOIdivFIR} indicates that the variation is not due to a decrease in the photoelectric heating efficiency on PAHs. The photoelectric heating efficiency is theoretically expressed as a function of the charging parameter, $\gamma = G_0 T^{0.5}/n_\textrm{e}$, where $G_0$ is the radiation field strength, $T$ is the gas temperature, and $n_\textrm{e}$ is the electron density \citep{Bakes1994}. As discussed in Sect.~\ref{subsec:result_IC1396A}, the UV radiation in the southern part of the BRC is likely shielded by the dense molecular gas west of the rim. A low UV field, cold and dense gas results in a low value of $\gamma$, which should increase the abundance of neutral PAHs and thus increase the heating efficiency. However, Fig.~\ref{figure:IC1396A_CIIOIdivFIR} presents a contradictory spatial distribution; the heating efficiency is clearly lower in the southern region than in the northern region. Furthermore, the estimated\footnote{\citet{Okada2012} derived the UV field strength and gas density at five positions in IC~1396A using a simple single clump PDR model. Assuming that most of the electrons are provided by carbon ionization and the carbon abundance is $1.6\times 10^{-4}$ \citep{Sofia2004}, we can derive $G_0/n_\textrm{e}$, which is about 10 or less. If the gas temperature is 100~K \citep[c.f. ][]{RoelligOssenkopf2022}, this gives $\gamma$ in the order of 100.} $\gamma$ in IC~1396A is on the order of 100, a value for which  the heating efficiency is high and not sensitive to the exact value of $\gamma$ \citep{Bakes1994,WD2001pe}. This also supports the view that the spatial difference of the measured heating efficiency cannot be attributed to the charging status of PAHs.

One possible reason for different heating efficiencies in the southern and northern regions of the BRC is a difference in the dust properties. \citet{WD2001pe} modeled the photoelectric heating efficiency with different size distributions of dust grains listed in \citet{WD2001}. Models with strong enhancement at the size of $<10$\AA\ (the C abundance in very small grains is parameterized as $b_\textrm{c}$) show higher efficiencies, and models with $R_V = 5.5$ ($R_V$ is the ratio of visual extinction to reddening), representing dense environments, have lower efficiencies compared to models with $R_V = 3.1$ \citep[see also Fig. 6 in ][]{Okada2013}. If the dust in the southern part is composed of fewer small grains relative to the large grains (i.e., larger $R_V$ and smaller $b_\textrm{c}$), this is consistent with a lower heating efficiency in the southern region compared to the northern region.

\subsection{\cii--MIR correlation}\label{subsec:CII-MIR_correlation}

\begin{figure}
\centering
\includegraphics[width=0.9\hsize]{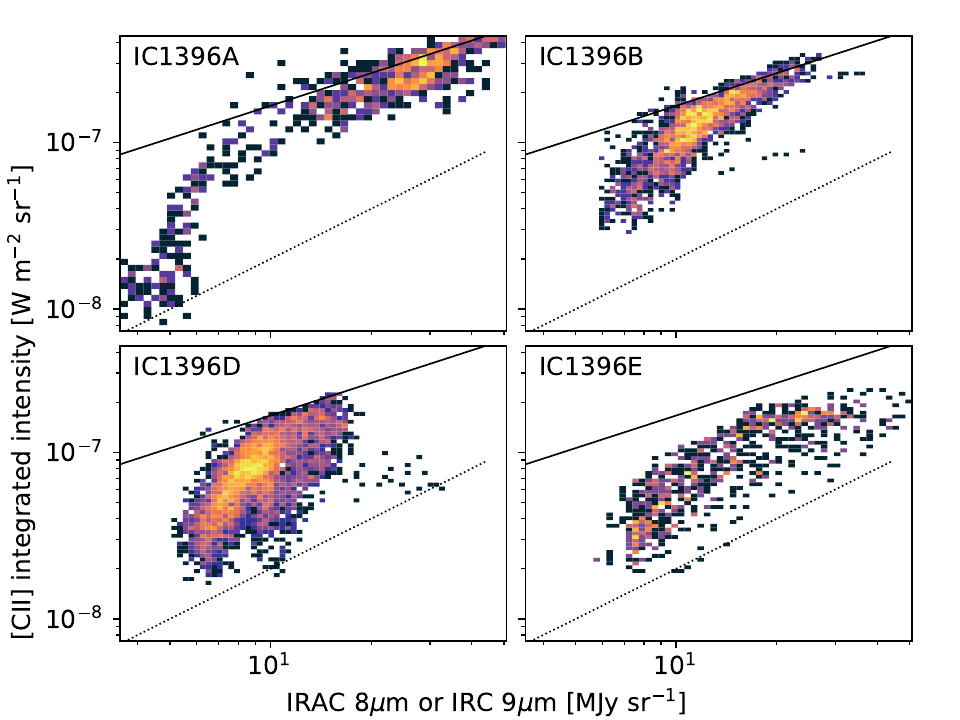}
\caption{Density plot of the correlation between \cii\ integrated intensities and IRAC 8\um\ (IC~1396A and E) or IRC 9\um\ (IC~1396B and D) at 25\arcsec\ resolution. The solid line is the relation derived in the Orion A region by \citet{Pabst2021}, adapted to this study (see the main text). The dotted line indicates the slope of a linear relation (slope of unity in the log scale) with an arbitrary scaling (offset in the log scale).}
\label{figure:CII_MIR_correlation}
\end{figure}

In all subregions a clear correlation between the \cii\ integrated intensities and the MIR photometric observations (IRAC 8\um\ or IRC 9\um) is observed (Fig.~\ref{figure:CII_MIR_correlation}). We compare it with the results in the Orion A region \citep{Pabst2021}, shown as a solid line in Fig.~\ref{figure:CII_MIR_correlation}. We note that their relation is adapted for Fig.~\ref{figure:CII_MIR_correlation} because they use the extended calibration factor (0.74) and filter width (2.9\um) for the IRAC 8\um\ flux (Pabst, priv. communication), which is not considered in this study. The correlation in IC~1396 agrees with that in Orion A for the flux range of \cii\ ($\geq 10^{-7}$\,W\,m$^{-2}$\,sr$^{-1}$ = $10^{-4}$\,erg\,s$^{-1}$\,cm$^{-2}$\,sr$^{-1}$) in IC~1396A and B, while we clearly see a turnover in fainter areas. This turnover is probably due to emission along the line of sight, because the \cii\ observations are velocity-resolved and only the velocity components associated with the individual BRCs are extracted, while the IRAC 8\um\ and IRC 9\um\ represent the total emission along the line of sight. If we subtract about $5$~MJy sr$^{-1}$ from the IRAC 8\um\ or IRC 9\um\ data, the full range of the data in Fig.~\ref{figure:CII_MIR_correlation} follow the correlation of the black line in IC~1396A, B, and D.

In IC~1396E the \cii\ emission is overall about a factor of 2 weaker than in other regions in the relation to the 8\um\ or 9\um. This could be either due to the contribution of very hot thermal dust to 8\um\ in IC~1396E with active YSOs and jets inside the BRCs, or due to a larger \oi\ contribution to the cooling in a higher density environment in IC~1396E, although the latter remains speculative without available \oi\ data.

\section{Discussion} \label{sec:discussion}

\subsection{Searching for the signature of photoevaporation} \label{subsec:discussion_photoevaporation}

The interpretation of the velocity structure depends on whether the BRCs are located in front of or behind the ionized gas \citep{Okada2015proc}. As described in Sect.~\ref{sec:intro_IC1396}, the optical and infrared images suggest that IC~1396A is located in front of the ionized gas and illuminated from the backside along the line of sight.  Blueshifted lines (shifted relative to the other spectral lines or other parts of the BRC) trace gas moving faster toward the observer.  As shown in Sect.~\ref{subsec:result_IC1396A}, the peak velocity of \cii\ in IC~1396A is overall blueshifted compared to \thco(3-2). This \cii\ velocity structure is inconsistent with a photoevaporating flow, which would be a flow toward the ionized gas and thus redshifted. We also find no other evidence of photoevaporating gas in \cii\ in IC~1396A. The redshifted ring in the channel maps of IC~1396A (at $-6$\,\kms) is not consistent with photoevaporating gas because part of this ring is also visible in \thco(3-2) (position 4 in Fig.~\ref{figure:spec_selected_IC1396A}).

For IC~1396E, comparison with the mean velocity of the ionized gas from \citet{Pedlar1980} ($-1.9$~\kms) would indicate that the BRC is behind the expanding ionized gas \citep{Patel1995}. However, the central velocity derived from the H166$\alpha$ observations in \citet{Pedlar1980} actually ranges from $-4.5$\,\kms\ to $+0.6$\,\kms\ at different positions in IC~1396, with an error of $\sim 2$\kms, and thus the velocity alone does not give a clear picture of the geometry. The sharp boundary in the H$\alpha$ image (Fig.~\ref{figure:channelmap_IC1396E_CII} bottom right panel) rather suggests that IC~1396E is in front of the ionized gas. On the other hand, the contrast between the inside and outside of the BRC in H$\alpha$ is not as strong as in IC~1396A \citep{Barentsen2011}. If the BRC is completely in front of the ionized gas, H$\alpha$ should be almost completely extinct, considering that the extinction is estimated to be $A_v\sim 20$ mag \citep{Nisini2001}. IC~1396E could be located in the middle of the ionized gas along the line of sight. With this uncertainty in the geometry, it is difficult to conclude whether the redshifted component in the eastern side of the \cii\ rim in IC~1396E is a sign of a photoevaporating flow. It potentially traces the photoevaporating gas leaving the BRC on the backside of the rim along the line of sight toward the ionized gas, but it is not excluded that this is an additional local distortion by the outflows.

In both IC~1396A and IC~1396E, there is a gradient of the velocity in \cii\ from the rim (most redshifted) to the rear side of the BRC (the side opposite to the exciting sources), and the blue wings at northwest and southwest of IC~1396A are prominent. Our interpretation is that the \cii\ gas comes dominantly from the surface of the BRC, and is blown away from the BRC toward the observer, and from the rim to the rear side of the BRC. The curved shape of the p-v diagram in IC~1396A (Fig.~\ref{figure:pvdiagram_IC1396A}) supports this interpretation of ablation flows.

\subsection{3D kinematic structures and comparison of different BRCs}\label{subsec:discussion_gaia}

\citet{Pelayo-Baldarrago2023} identified seven subclusters in IC~1396 based on parallax and proper motion measurements from \textit{Gaia}. Subclusters A, B, E, and F have a large number of members to be statistically significant and are relatively free from contamination by sources not associated with IC~1396. As mentioned in Sect.~\ref{sec:intro}, they found that these subclusters are clearly distinct in proper motion and partly in age. Inside the field of our observations, IC~1396A has a clear association with their subcluster A (six \textit{Gaia} sources in the field of our \cii\ map) and IC~1396E with their subcluster B (nine \textit{Gaia} sources in the field of our \cii\ map). IC~1396D has a dispersed population (with each of their subclusters A, E, and F having a single \textit{Gaia} source within the field of our \cii\ map), and IC~1396B has only one \textit{Gaia} source belonging to subcluster A within the field of our \cii\ map.

The distance to the subclusters A and B determined by \citet{Pelayo-Baldarrago2023} is $908 \pm 73$~pc and $911\pm 75$~pc, respectively. The projected diameter of IC~1396 on the sky (basically the width of Fig.~\ref{figure:IC1396_overview}) is approximately 45~pc at a distance of 925~pc. Given the considerable uncertainty in the derived distance, it remains unclear whether IC~1396E is located in front or behind the ionized gas. A comparison of the distances of the \textit{Gaia} sources that overlap with the \cii\ map in IC~1396A and IC~1396E does not provide a conclusive result either.

We can, however, combine the mean proper motion of the subclusters A and B \citep{Pelayo-Baldarrago2023} and the \cii\ line-of-sight velocity to derive the relative 3D motion between IC~1396A and E. Table~\ref{table:gaia} summarizes the velocity calculated by the proper motion and the \cii\ line-of-sight velocity. It is evident that IC~1396A and IC~1396E are moving away from each other in 3D (both in the plane of the sky and along the line of sight), which is consistent with an expansion of the entirety of IC~1396  \citep{Patel1995,Pelayo-Baldarrago2023}.

\begin{table}
\caption{Proper motion and line-of-sight velocity in IC~1396A and E.}
\label{table:gaia}
\centering
\begin{tabular}{ccrrr}
\hline\hline
Region & subcluster$^\mathrm{a}$ & $v_\alpha^\mathrm{a,b}$ & $v_\delta^\mathrm{a,b}$ & $v_\mathrm{LOS}^\mathrm{c}$\\
&& [\kms] & [\kms] & [\kms]\\
\hline
IC~1396A & A (total) & 0.00 & 0.00 & --\\
& A (\cii\ map$^\mathrm{d}$) & $-0.87$ & $-1.52$ & $-9.0$ \\
IC~1396E & B (total) & $0.79$ & $6.96$ & -- \\
& B (\cii\ map$^\mathrm{d}$) & $1.63$ & $6.66$ & $1.0$\\
\hline\hline
\end{tabular}
\begin{list}{}{\setlength{\itemsep}{0ex}}
\item[$^\mathrm{a}$] \citet{Pelayo-Baldarrago2023}
\item[$^\mathrm{b}$] Averaged projected velocity along RA and Dec. relative to the average of subcluster A.
\item[$^\mathrm{c}$] This study.
\item[$^\mathrm{d}$] Averaged velocity is calculated using only the sources within the field of the \cii\ map in this study.
\end{list}
\end{table}

IC~1396A and E have more, young sources listed by \citet{Pelayo-Baldarrago2023} and embedded YSOs identified in the literature, as mentioned in Sect.~\ref{sec:intro}, while those in IC~1396B and D are less prominent. This difference is not attributable to the current projected distance of each BRC from the exciting sources. It rather must be associated with the presence of high density clumps in IC~1396A and E \citep{Sicilia-Aguilar2014,Serabyn1993}. The question whether these BRCs have a higher gas density in general will be addressed in a follow-up paper by deriving physical properties using PDR models.

\section{Summary}\label{sec:summary}

We investigated the dynamical and physical structures of four BRCs in IC~1396 using velocity-resolved observations of \cii\ 158\um, \oi\ 63\um,\ and 145\um\ with (up)GREAT on board SOFIA, combined with low-$J$ (3-2) transitions of CO, \thco, and \ceio\ from the JCMT archive. The individual results are as follows:

\begin{itemize}
\item In IC~1396A, the peak velocity of \cii\ is blueshifted compared to \thco(3-2) and \ceio(3-2), especially on the back side (the side farthest from the exciting sources). Because IC~1396A is likely in front of the ionized gas along the line of sight, the blueshifted \cii\ cannot be explained by photoevaporation.\ It instead indicates that gas was blown off from the rim to the back side of the BRC.
\item The spatial distribution and the self-absorption signature of the emission lines suggest that the northern part of IC~1396A is clumpier and the UV radiation reaches the middle of the BRC, while the southern part is strongly shielded from the UV radiation by a high column of hot and dense gas.
\item Our analysis of \thcii\  in IC~1396A did not reveal any evidence that the \cii\ emission has a significant optical depth.
\item The heating efficiency measured by the (\cii\ + \oi\ 63\um)/FIR is 3--4\% in IC~1396A. This is similar to the upper limits from previous studies of other regions, taking the different definitions of efficiency in different studies into account. The southern part of IC~1396A has a lower efficiency than the northern part, which may be due to fewer small grains and a high density.
\item In IC~1396B and IC~1396D, distinct velocity components are observed in both the \cii\ and CO emission, with no indication that they are physically connected. They are probably independent clouds behind and in front of the ionized gas along the line of sight.
\item The width of the \cii\ emission in IC~1396E indicates an interaction between outflows and the ambient gas.
\item Except for IC~1396E, the \cii\ intensity shows a good correlation with the 8\um\ or 9\um\ emission in an intensity range that is significantly higher than that of the background emission, which is quantitatively consistent with previous studies.
\item By combining the line-of-sight \cii\ velocity with the proper motion measured by \textit{Gaia}, we confirm that IC~1396A and IC~1396E are moving away from each other in 3D, which supports the scenario that the entire region is expanding.
\end{itemize}
The interpretation of the \cii\ and \oi\ 63\um\ spectral shape and intensity strongly depends on their optical depth. This study demonstrates that the combination of \thcii\ with \cii, and \oi\ 145\um\ with \oi\ 63\um\ leads to conclusive results because we can disentangle the effect of optical depth and the intrinsic dynamics in the velocity profile.

\begin{acknowledgements}
This work is based in part on observations made with the NASA/DLR Stratospheric Observatory for Infrared Astronomy (SOFIA). SOFIA is jointly operated by the Universities Space Research Association, Inc. (USRA), under NASA contract NAS2-97001, and the Deutsches SOFIA Institut (DSI) under DLR contract 50 OK 0901 to the University of Stuttgart. 
This work is carried out within the Collaborative Research Centre 956, sub-project A4 and C1, funded by the Deutsche Forschungsgemeinschaft (DFG) – project ID 184018867. 
This research used the facilities of the Canadian Astronomy Data Centre operated by the National Research Council of Canada with the support of the Canadian Space Agency. This work is partly based on observations with AKARI, a JAXA project with the participation of ESA. This research made use of astrodendro, a Python package to compute dendrograms of Astronomical data (http://www.dendrograms.org/). The authors wish to recognize and acknowledge the very significant cultural role and reverence that the summit of Maunakea has always had within the indigenous Hawaiian community.  We are most fortunate to have the opportunity to conduct observations from this mountain.
N.S. acknowleges supprot by the BMWI via DLR, project number 50 OR 2217 (FEEDBACK-plus).
S.K. acknowledges support  by the BMWI via DLR, project number 50 OR 2311 (Orion-Legacy).
Y.O. et al. gratefully acknowledges the Collaborative Research Center 1601 (SFB 1601 sub-project A6) funded by the Deutsche Forschungsgemeinschaft (DFG, German Research Foundation) – 500700252.

\end{acknowledgements}

\bibliographystyle{aa}
\bibliography{mydatabase}
\begin{appendix}

\FloatBarrier

\section{Baseline correction using PCA} \label{app:pca}

Principal component analysis was used to characterize and correct for the instrument-origin baseline structures in the spectra. Its application to different types of observations is described in detail in Buchbender et al. (in prep.).  Here, we summarize the procedure applied to the chopped OTF data presented in this paper and present the improvement in the data quality.

\begin{figure}
\centering
\includegraphics[width=0.85\hsize]{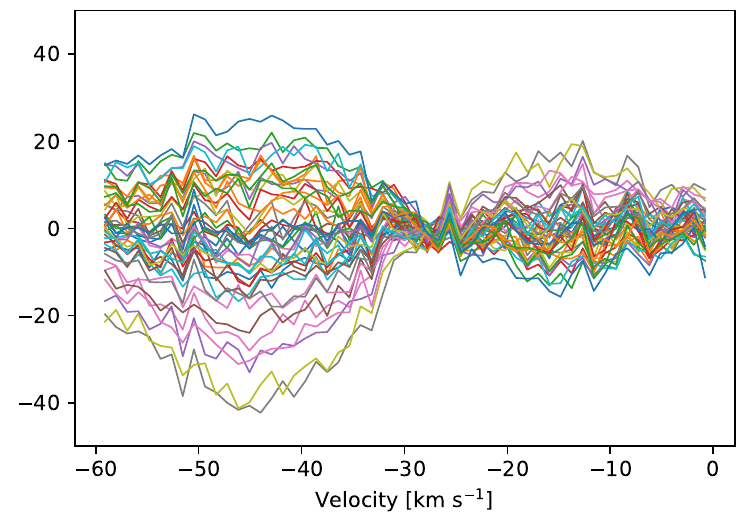}
\caption{Difference of adjacent OFF measurements in one of the chopped OTF lines in the flight on March 6, 2020, for HFAV pixel 0. Each spectrum is smoothed to the velocity resolution of about 0.5 \kms. The vertical axis is in Kelvin.}
\label{figure:skydiff_34800}
\end{figure}

The basic idea is to extract principal components from the OFF (sky) measurements, and subtract these components from the ON spectra. This ensures that the corrected baseline components are not affected by the line emission in on-source spectra. The whole procedure is written in python with the pygildas interface, so that the program can be called from GILDAS (Buchbender et al. in prep.). Figure~\ref{figure:skydiff_34800} shows as an example the difference of adjacent OFF measurements (OFF-OFF) in a chopped OTF line for the HFAV pixel 0, clearly showing a common broad baseline pattern. This is inferred to be caused by the gain drift of the instrument, and it also appears in the ON spectra. While the patterns are different for different pixels, the pattern of the same pixel is stable in each observing leg. Therefore, we construct the principal components from the set of OFF-OFF spectra for each pixel in each flight. We used only the velocity range of $-57$ to 0\,\kms\ for \oi\ 63\um\ (HFA) and $-57$ to $30$\,\kms\ for \cii\ (LFA), which covers the expected range of the emission line well and excludes the velocity of the atmospheric line. We resampled each OFF-OFF spectrum in velocity to the goal spectral resolution (0.5\,\kms), subtracted the mean over this velocity range as an offset from each OFF-OFF spectrum, and normalized them with respect to the standard deviation.  For all six flights, the number of OFF-OFF measurements per pixel per flight is more than 1600, which is statistically sufficient to extract the principal components.


\begin{figure}
\centering
\includegraphics[width=0.9\hsize]{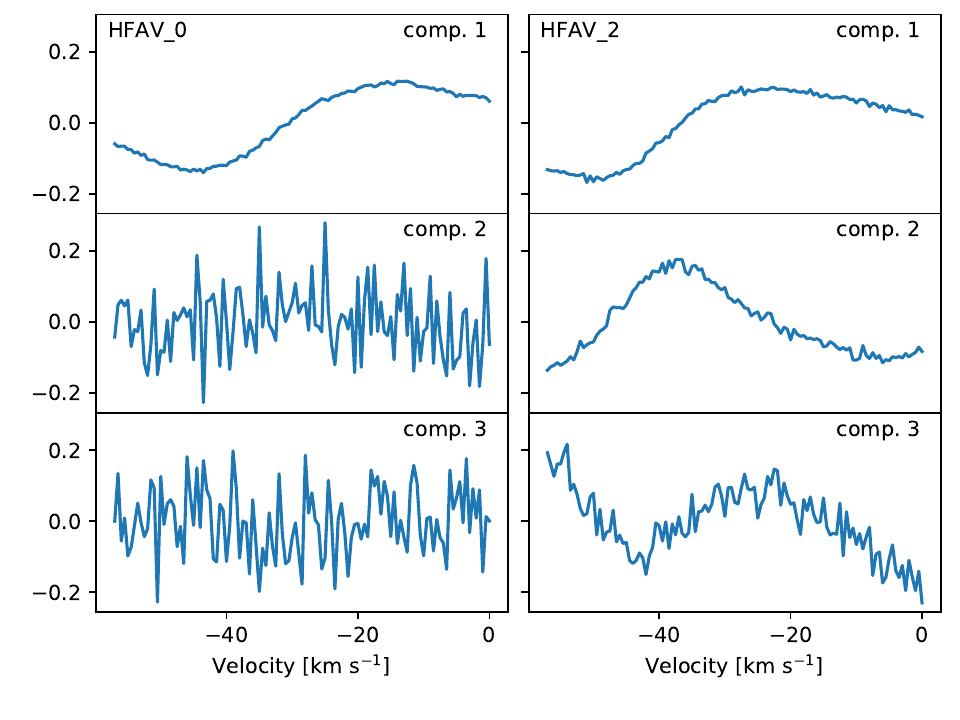}
\caption{Examples of derived principal components (HFAV pixels 0 and 2 for the flight on March 6, 2020).}
\label{figure:pca_components}
\end{figure}

\begin{figure}
\centering
\includegraphics[width=0.9\hsize]{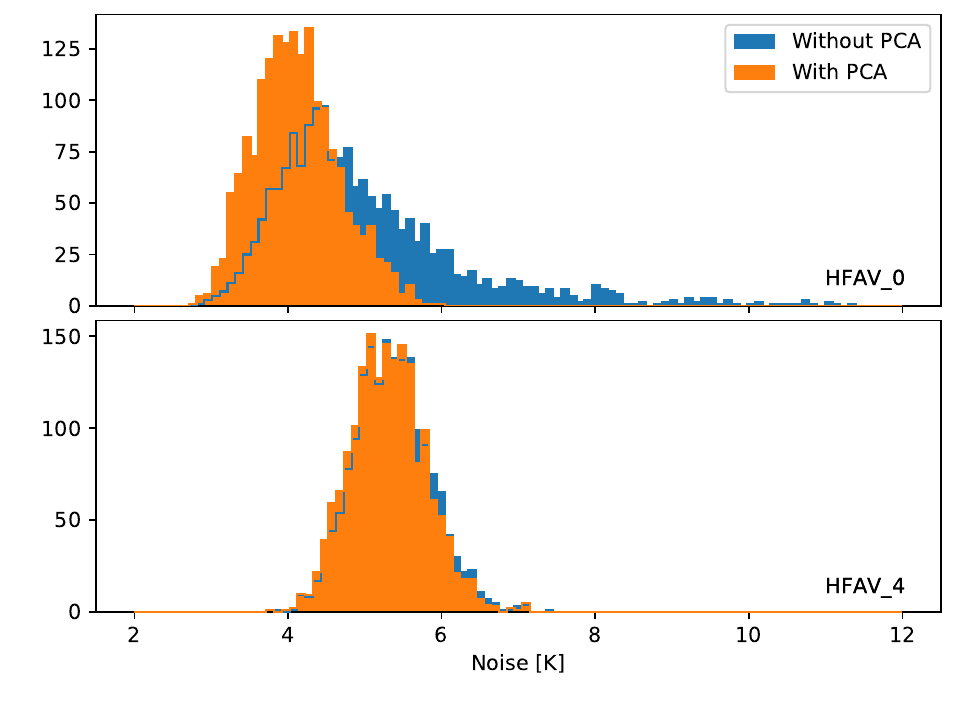}
\caption{Examples of the baseline rms with and without the PCA (HFAV pixels 0 and 4 for the flight on March 6, 2020).}
\label{figure:rms_pca}
\end{figure}

Figure~\ref{figure:pca_components} shows examples of the first three principal components (HFAV pixels 0 and 2, flight on March 6, 2020). Mathematically, one can derive as many components as the number of input spectra. We used only those components that show a broad baseline structure by comparing the standard deviation of a narrow and broad velocity range. We measured a running mean and standard deviation of 5 channels in the 0.5 \kms\ bin, and we used the component if the median of the running standard deviation is less than 0.7 of the standard deviation of the entire velocity range used. In the examples in Fig.~\ref{figure:pca_components}, only component 1 is used for HFAV\_0, while three components are used for HFAV\_2. An alternative criterion could be the percentage of variance explained by each of the selected components. This works well when there is a clear dominant component, but it is difficult to decide on a consistent number as a criterion for differently behaving pixels and receivers.

After selecting the components as above, the contributions of these components to the individual ON spectra are calculated and then subtracted. The ON spectra are also resampled in velocity to the goal spectral resolution before the principal components are subtracted.

Figure~\ref{figure:rms_pca} shows typical histograms of the baseline rms over the entire velocity range used. The plot of HFAV\_0 represents cases where the PCA clearly improves the baseline quality, where the rms distribution becomes narrower and generally shifts to lower noise after the PCA is applied. The plot of HFAV\_4 represents cases where the original data are already good enough that the PCA has no significant effect. This is the case for all LFA data. 

\FloatBarrier

\section{Comparison of different methods for generating integrated intensity maps}\label{app:integrated_intensity}

\begin{figure*}
\centering
\includegraphics[width=0.355\hsize]{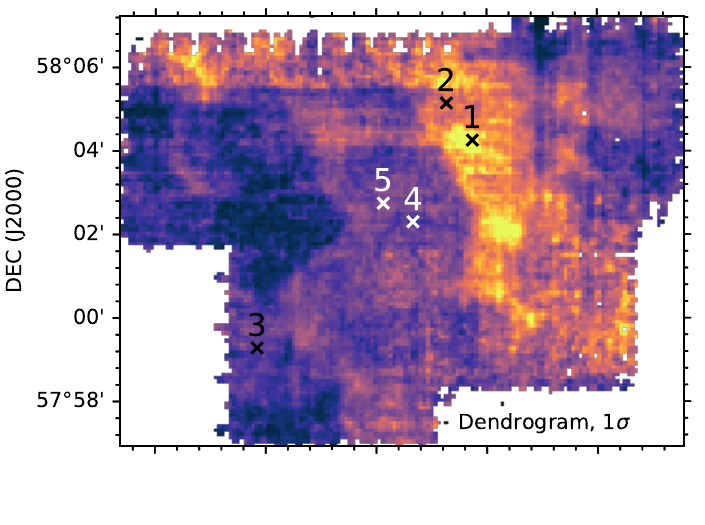}
\includegraphics[width=0.315\hsize]{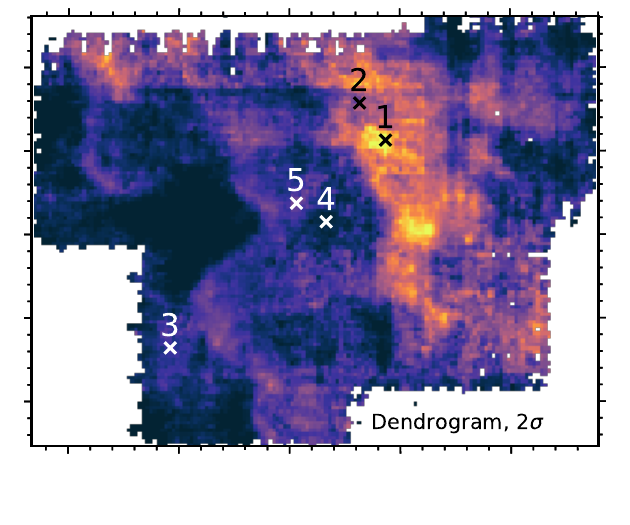}
\vspace{-0.5cm}
\includegraphics[width=0.315\hsize]{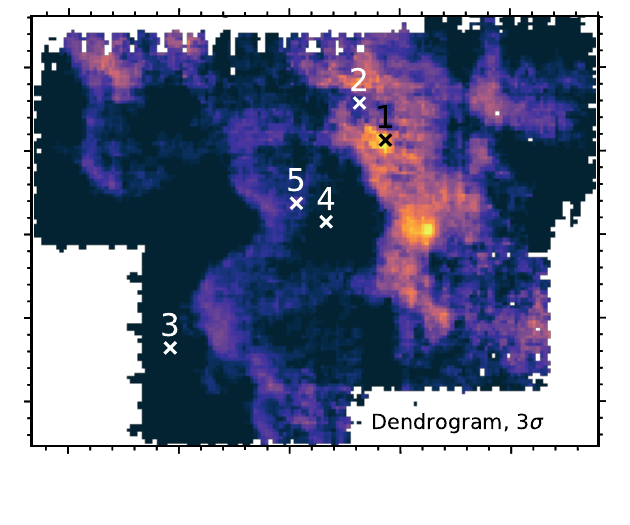}
\vspace{-0.5cm}
\includegraphics[width=0.355\hsize]{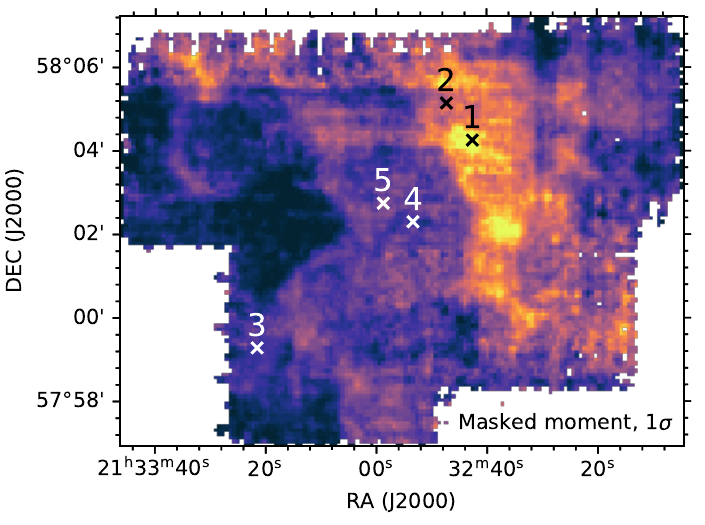}
\includegraphics[width=0.315\hsize]{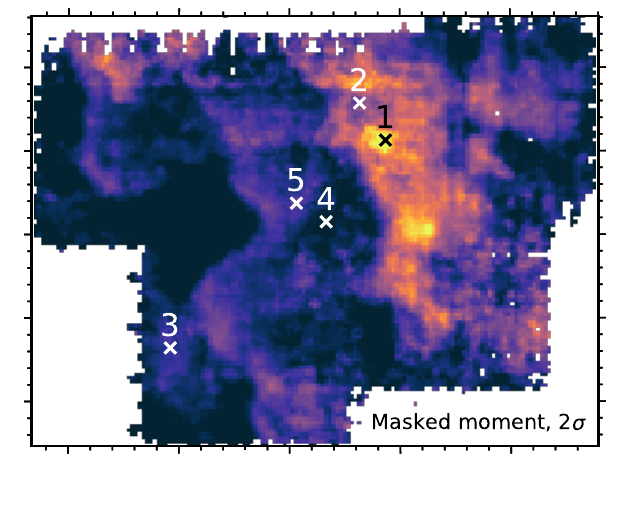}
\includegraphics[width=0.315\hsize]{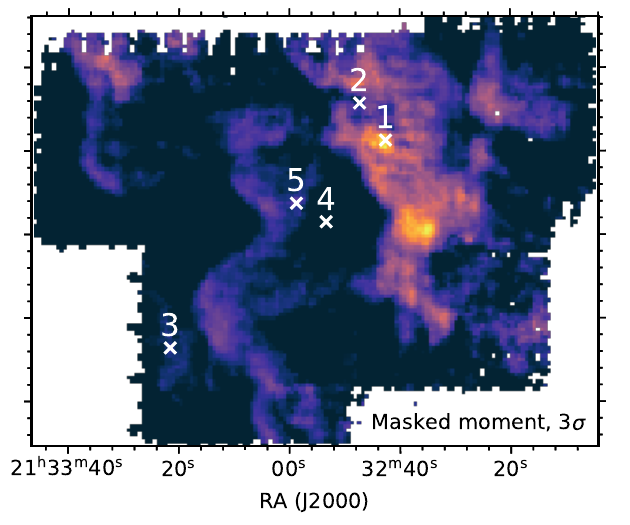}
\hspace{1cm}
\includegraphics[width=0.4\hsize]{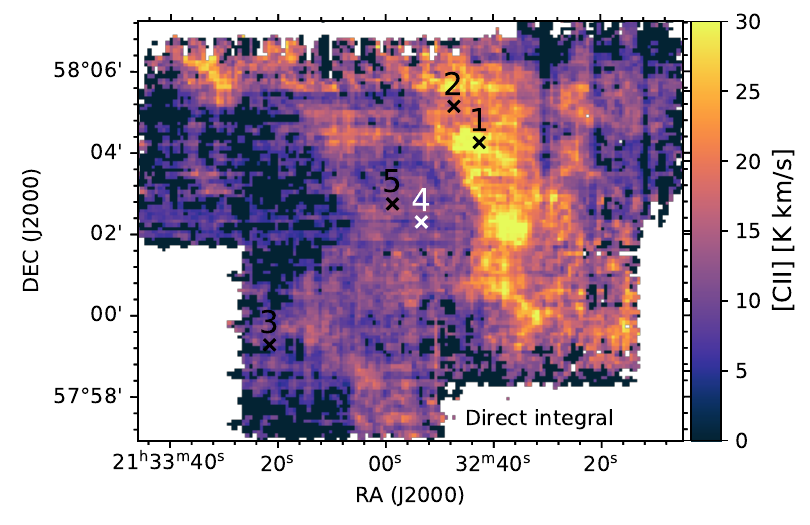}
\caption{Integrated intensity maps of IC~1396D derived using different methods and thresholds at 16\arcsec\ resolution. The upper three panels are from the 3D dendrogram method, with thresholds of $1\sigma$, $2\sigma$, and $3\sigma$ from left to right. The middle three panels are from the masked moment method, also with the threshold of $1\sigma$, $2\sigma$, and $3\sigma$ from left to right. The bottom panel is the direct integration of the spectra with the velocity range from $-12$~\kms\ to $+10$~\kms. See the main text for details of the different methods. Black crosses with numbers mark positions where the masks along the spectral axis are shown in Fig.~\ref{figure:compare_masks_integrated_intensities}.}
\label{figure:integmap_IC1396D_CII_compare}
\end{figure*}

\begin{figure*}
\centering
\includegraphics[width=0.36\hsize]{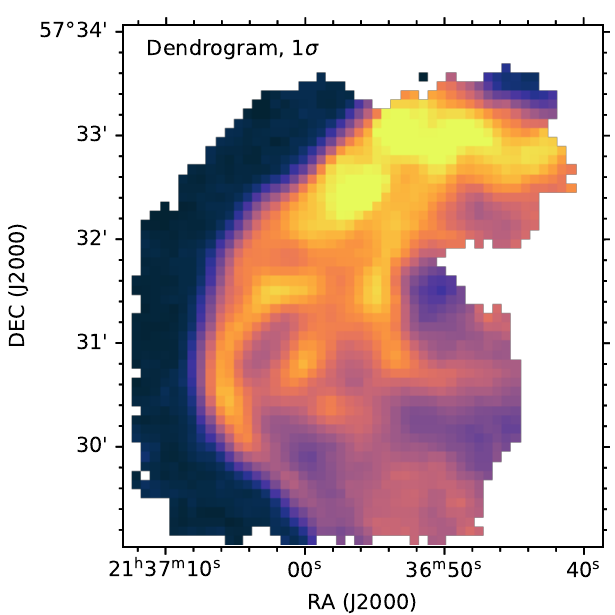}
\includegraphics[width=0.303\hsize]{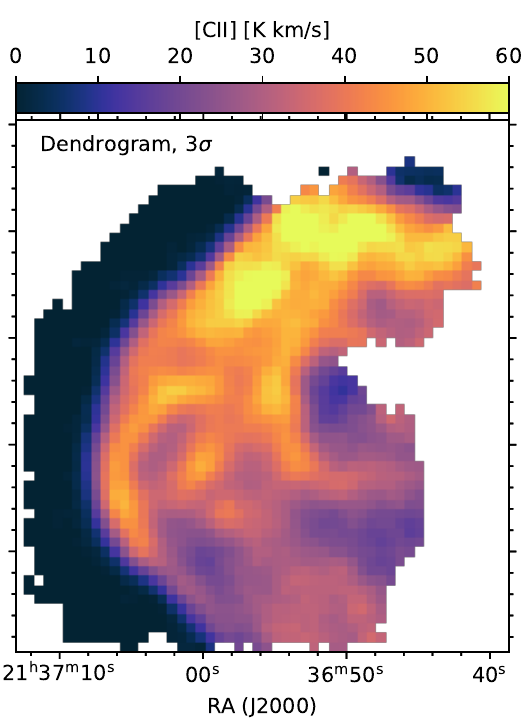}
\includegraphics[width=0.295\hsize]{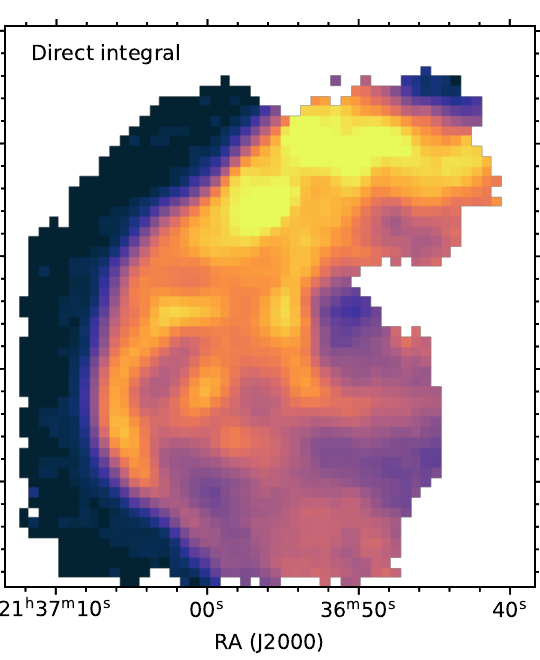}
\caption{Integrated intensity maps of IC~1396A derived using the 3D dendrogram method, with thresholds of $1\sigma$ (left) and $3\sigma$ (middle), and the direct integration of the spectra in the velocity range $-13$~\kms\ to $-4$~\kms. The color scale is the same for all panels.}
\label{figure:integmap_IC1396A_CII_compare}
\end{figure*}

In this appendix we compare the integrated intensity maps obtained using different methods: direct integration of the spectra with a fixed velocity range, 3D dendrograms, and the masked moment method. It should be noted that the purpose of the integrated intensity maps presented in this paper is to identify spatial structures and to compare the spatial distribution of different emission lines. The precise selection of methods and thresholds does not significantly impact the outcome of the analysis in this paper.

When using the 3D dendrogram analysis to create a mask for integrated intensity maps, the parameter of a minimum significance of a leaf is less important because we do not distinguish the individual detected clusters, while the minimum intensity limit directly affects the results because its high value cuts off weak emission. The top three panels in Fig.~\ref{figure:integmap_IC1396D_CII_compare} compare the derived integrated intensity maps using $1\sigma$, $2\sigma$, and $3\sigma$ as the minimum intensity threshold, where $\sigma=0.6$~K is the median rms noise in IC~1396D, and the bottom panel shows the direct integrated intensity using the velocity range from $-12$~\kms\ to $+10$~\kms. The $3\sigma$ threshold not only loses extended faint emission, but also suppresses the intensities at bright regions.

On the other hand in Fig.~\ref{figure:integmap_IC1396A_CII_compare} we show the same comparison for IC~1396A, which has an overall higher S/N. This demonstrates that the results are essentially indistinguishable for high S/N data.

\begin{figure}
\centering
\includegraphics[width=\hsize]{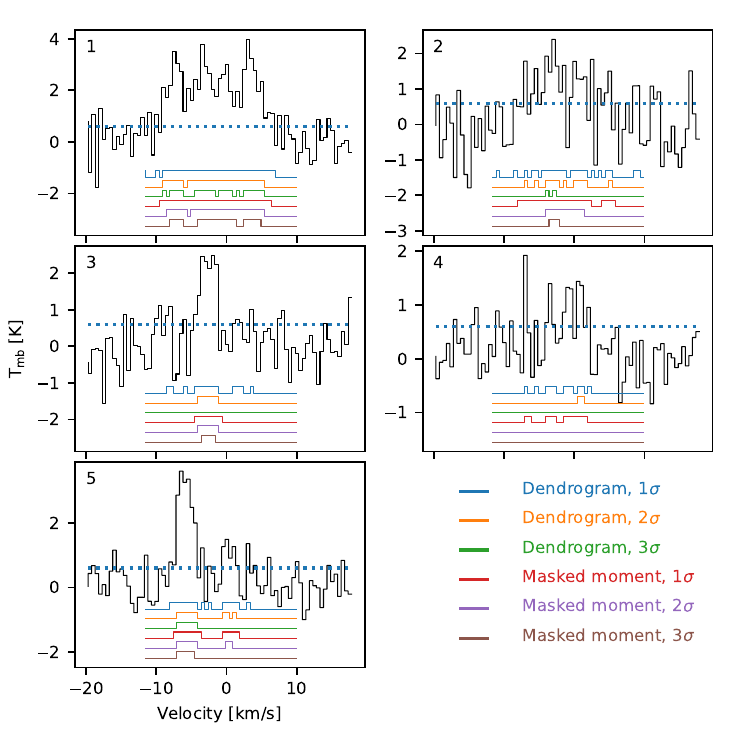}
\caption{Masks along the spectral axis at the positions marked in Fig.~\ref{figure:integmap_IC1396D_CII_compare} derived using different methods for calculating the integrated intensity. The blue dotted line shows the $1\sigma$ level. Note that the mask is calculated by averaging the $3\times 3$ spatial pixels, but the spectrum shown in each panel is the one at the central pixel.}
\label{figure:compare_masks_integrated_intensities}
\end{figure}

We also compare the integrated intensity maps using the dendrogram-based masks with the masked moment method \citep{Adler1992}. This method also creates a 3D mask for integration based on the 3D average of a small volume around each pixel. We used 3 pixels here to be comparable to the minimum cluster size of the dendrogram analysis. For each spatial and spectral bin, we calculated an average over $3\times3\times3$ (spatial and spectral) pixels, and if the average was greater than a certain threshold, we assigned a value of 1 to the 3D mask of the central pixel. We then derived the integrated intensity map by integrating the 3D spectral cube over this mask, as with the dendrogram-based masks. We also used $1\sigma$, $2\sigma$, and $3\sigma$ as threshold values (Fig.~\ref{figure:integmap_IC1396D_CII_compare} middle row). Overall, the derived integrated intensity maps look very similar between the dendrogram-based mask method and the masked moment method when using the same threshold.

Figure~\ref{figure:compare_masks_integrated_intensities} shows examples of the mask at the selected positions indicated in Fig.~\ref{figure:integmap_IC1396D_CII_compare}. The masks over a spectral axis at these positions are shown as 0-1 step plots. The example of position 1 clearly shows that the $3\sigma$ threshold masks too much structure, indicating that a $2\sigma$ or $1\sigma$ threshold should be used instead. This is also consistent with experience with the masked moment method \citep[e.g.,][]{Simon2001}. An advantage of the dendrogram-based method is that a strong outlier does not affect neighboring pixels. Sometimes the masked spectral bins are shifted by one bin compared to the dendrogram-based masks (Fig.~\ref{figure:compare_masks_integrated_intensities}). On the other hand, the masked moment method is more robust against small fluctuations around the threshold (see position 2 in Fig.~\ref{figure:compare_masks_integrated_intensities}).

\FloatBarrier

\section{Impact of different baseline subtractions on the \cii\ optical depth estimates} \label{app:13cii_baseline}

\begin{figure}
    \centering
    \includegraphics[width=\hsize]{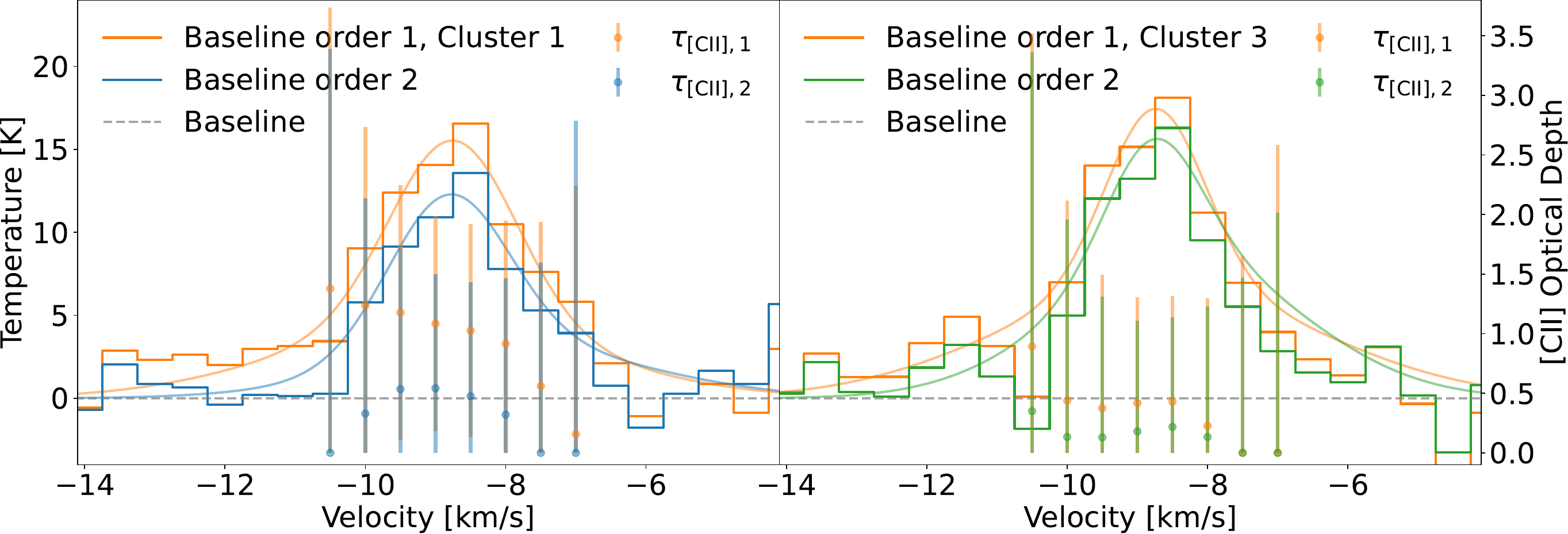}
    \caption{Comparison of different orders of the baseline used to extract \thcii\ hyperfine components for cluster 1 and cluster 3. Colored stepped data are the combined and scaled \thcii\ spectra subtracted by the baseline with the first and second-order polynomial. Curves with corresponding colors are the fit to the \thcii\ spectra using two Gaussian profiles, and derived optical depths of \cii\ are shown as data points with error bars.}
    \label{figure:13cii_compare_baseline}
\end{figure}

As mentioned in Appendix~\ref{app:pca}, the overall baseline quality for the \cii\ spectra in IC~1396A is good enough so that the PCA has no significant effect. However, for the \thcii\ hyperfine components, especially F$=2-1$ is affected by a wing of the \cii\ profile, and an additional baseline subtraction around each hyperfine component is required. Here we compare the derived \thcii\ spectra and \cii\ optical depths when using the second-order polynomial (Fig.~\ref{figure:dendro_spec_tau_IC1396A}) and first-order polynomial (linear; Fig.~\ref{figure:13cii_compare_baseline}). Except for clusters 1 and 3, the best fit second polynomial is very close to linear, so there is no difference in both cases (therefore not shown in Fig.~\ref{figure:13cii_compare_baseline}). Cluster 1 (left panel of Fig.~\ref{figure:13cii_compare_baseline}) has the largest difference depending on the order of the baseline among all clusters. The \thcii\ spectrum at a velocity of between $-14$~\kms\ and $-10.5$~\kms\ indicates that the second-order polynomial (blue data in the figure) better characterizes the baseline. If we use a linear baseline, the derived optical depths are overall higher (around unity) than the calculation with the second-order polynomial baseline. However, the velocity dependence of the optical depth is more reasonable with the second-order baseline; the optical depth has a peak at a velocity close to the \thcii\ peak velocity, instead of having the highest optical depth at the edge of the line profile as shown with orange data points in Fig.~\ref{figure:13cii_compare_baseline} left panel. Therefore, we subtracted the second-order polynomial as the baseline.

\FloatBarrier

\section{Atmospheric model fit in the \oi\ 145\um\ observations} \label{app:atm_oi145}

\begin{figure}[h]
\centering
\includegraphics[width=\hsize]{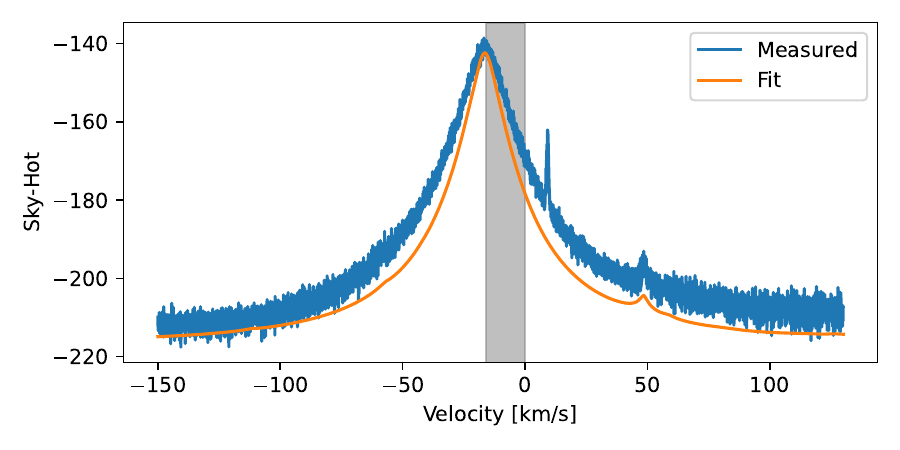}
\caption{Example of the "sky minus hot" spectra (the hot-load measurement subtracted from the OFF measurement and the atmospheric model fit in the \oi\ 145\um\ observations toward IC~1396A (March 6, 2020; LFAV pixel 1). The shaded velocity range is the plotted range in Fig.~\ref{figure:spec_selected_IC1396A}.}
\label{figure:IC1396A_OI145_S-H}
\end{figure}

The \oi\ 145\um\ is located on a shoulder of a significant atmospheric ozone feature in the signal sideband. Figure~\ref{figure:IC1396A_OI145_S-H} shows an example of the atmospheric model fit in the "sky minus hot" spectrum, which is the difference between the OFF and hot-load measurement. Because the precise profile of the ozone feature depends on the pressure-temperature profile of the atmospheric layers, which is not precisely determined during the calibration process, the fit is often suboptimal as illustrated in this figure. This discrepancy results in a significant uncertainty in the absolute intensity calibration of the \oi\ 145\um. However, we focus in this paper on the line profile of \oi\ 145\um\ within the velocity range that is gray shaded in Fig.~\ref{figure:IC1396A_OI145_S-H}. The atmospheric feature in this range has a smooth and monotonic profile, and thus the derived \oi\ 145\um\ line profile is reliable despite an imperfect fit of the atmospheric model.

\section{Supplementary figures}

\begin{figure}
\centering
\includegraphics[width=\hsize]{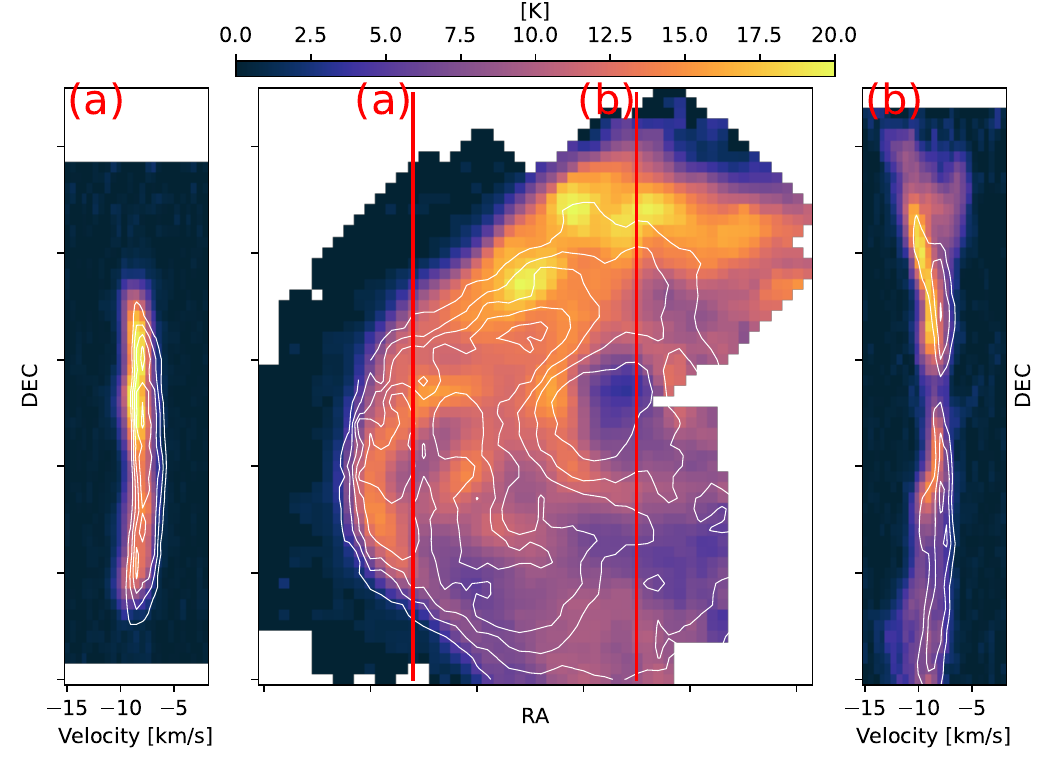}
\caption{Overlay (center) of the integrated intensity of the \cii\ (color) and \thco(3-2) (white contours) of IC~1396A; the two cuts, (a) and (b) (red lines), mark where the p-v diagrams shown on the left and right were made. In the p-v diagrams the color map is \cii\ (scale shown in the color bar) and the contours are \thco(3-2) (with a spacing of 2.5~K). The spatial positions are vertically aligned between the p-v diagrams and the map. All images have a resolution of 16\arcsec.}
\label{figure:pvdiagram_IC1396A}
\end{figure}

\begin{figure}
\centering
\includegraphics[width=0.9\hsize]{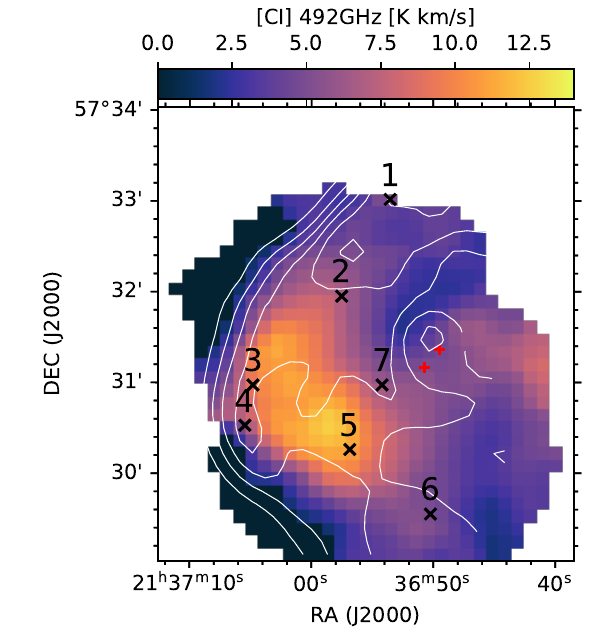}
\caption{Integrated intensity map of the \ci\ 492~GHz (colors) in IC~1396A as presented in \citet{Okada2012}, overlaid with contours of \cii\ integrated intensity, both at 25\arcsec\ resolution. Marks have the same meaning as in Fig.~\ref{figure:integmap_IC1396A}.}
\label{figure:integmap_IC1396A_CI492}
\end{figure}

\begin{figure}
\centering
\includegraphics[width=0.9\hsize]{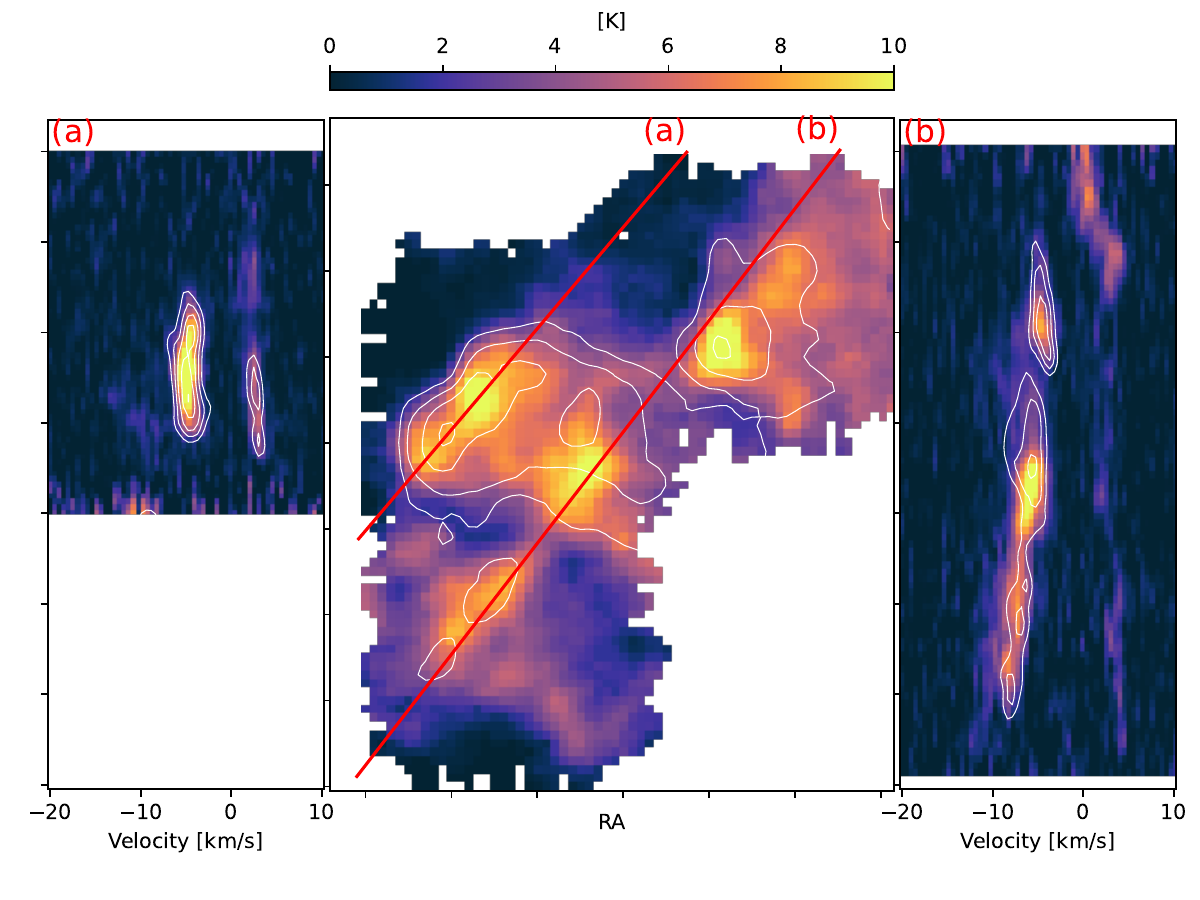}
\caption{Same as Fig.~\ref{figure:pvdiagram_IC1396A} but for IC~1396B and with white contours representing CO(3-2). In the p-v diagrams the color map is \cii\ (scale shown in the color bar) and the contours are CO(3-2) (with a spacing of 10~K, starting at 5~K). The spatial positions are vertically aligned between the p-v diagrams and the drawn cuts. All images have a resolution of 25\arcsec.}
\label{figure:pvdiagram_IC1396B}
\end{figure}

\begin{figure}
\centering
\includegraphics[width=\hsize]{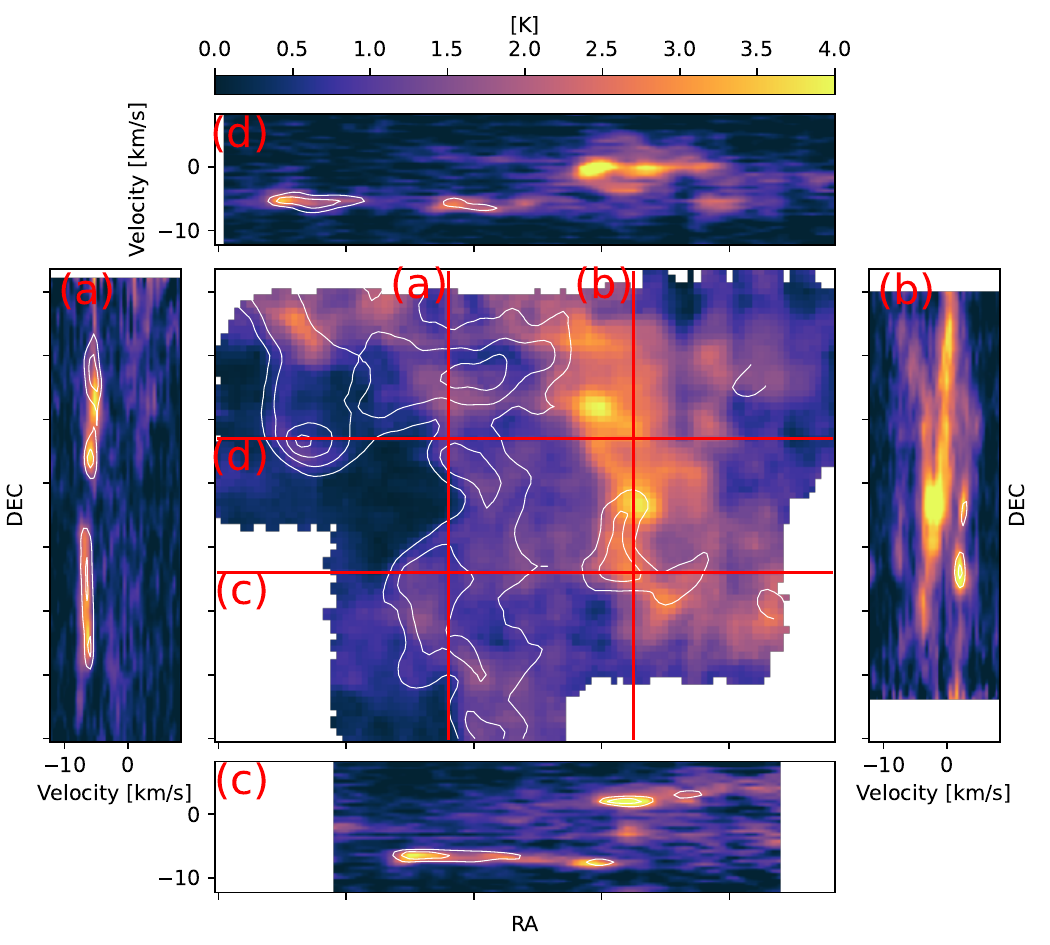}
\caption{Same as Fig.~\ref{figure:pvdiagram_IC1396A} but for IC~1396D and with four cuts, (a), (b), (c), and (d). In the p-v diagrams the color map is \cii\ (scale shown in the color bar) and contours are CO(3-2) (with a spacing of 10~K, starting at 5~K). The spatial positions are horizontally and vertically aligned between the p-v diagrams and the drawn cuts. All images have a resolution of 25\arcsec.}
\label{figure:pvdiagram_IC1396D}
\end{figure}

\begin{figure}
\centering
\includegraphics[width=0.9\hsize]{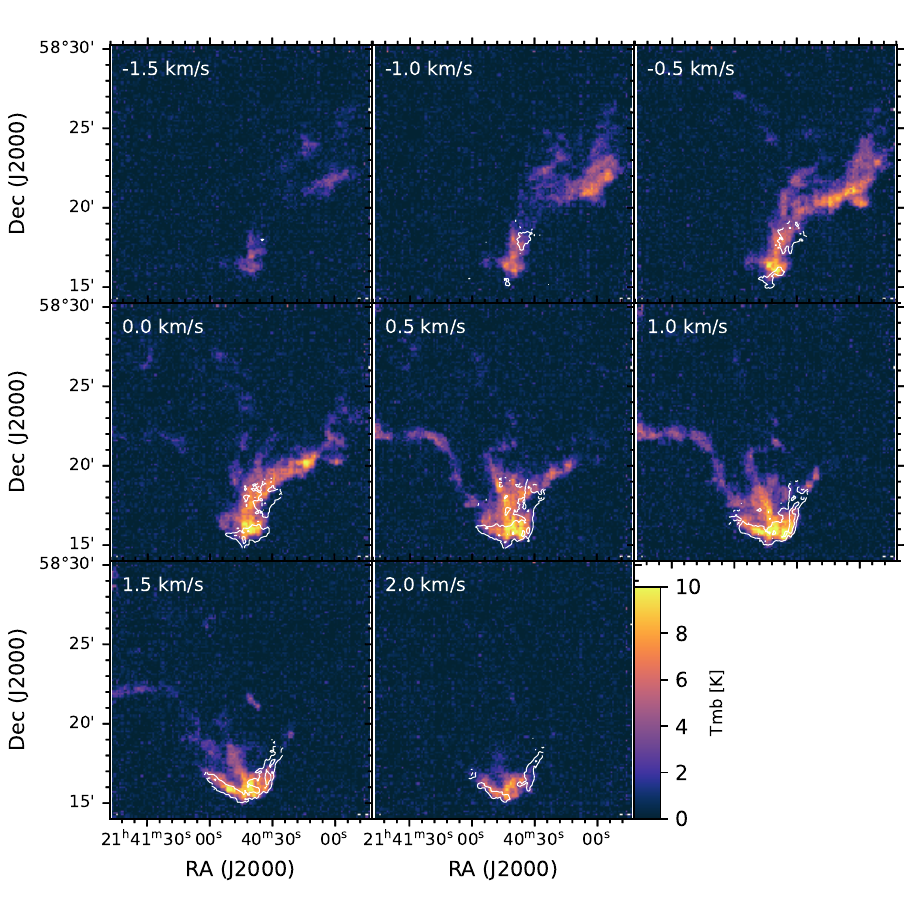}
\caption{Channel maps of \thco(3-2) (prior to resampling to the same resolution and mapping area as \cii), overlaid with contours of \cii\ in IC~1396E.}
\label{figure:channelmap_large_IC1396E}
\end{figure}

\begin{figure}
\centering
\includegraphics[width=\hsize]{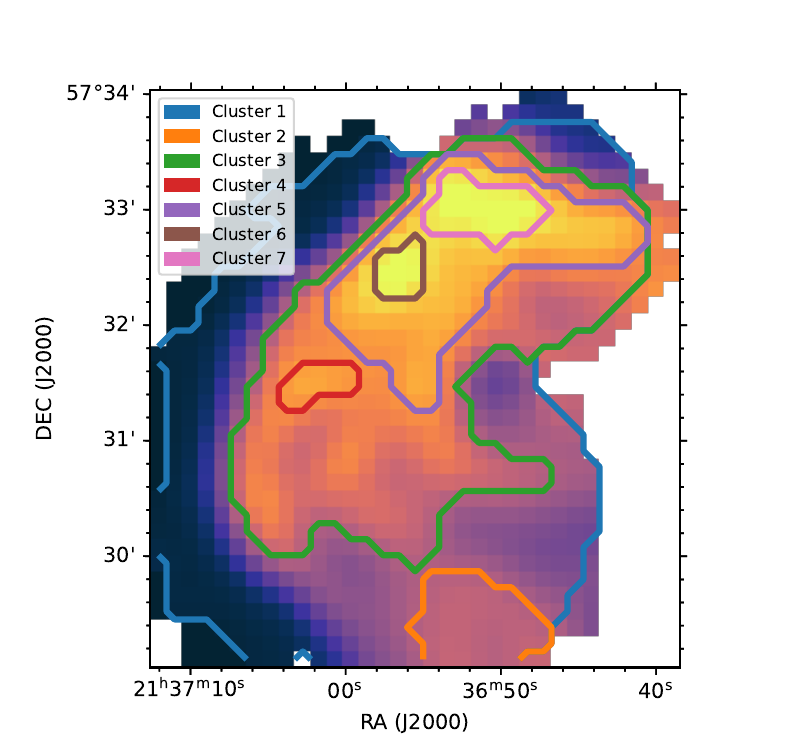}
\caption{2D dendrogram clusters on the \cii\ integrated intensity map in IC~1396A. We used the 25\arcsec\ resolution map because the purpose of the dendrogram analysis is to average over a larger area for the \thcii\ analysis.}
\label{figure:2d_dendro_IC1396A}
\end{figure}

Here we present supplementary figures. Figures~\ref{figure:pvdiagram_IC1396A}--\ref{figure:pvdiagram_IC1396D} are the position-velocity (p-v) diagrams of IC~1396A, B, and D. Figure~\ref{figure:integmap_IC1396A_CI492} is the \ci\ 492 GHz integrated intensity map, shown in \citet{Okada2012}, in order to relate with the positions discussed in this study. Figure~\ref{figure:2d_dendro_IC1396A} shows the clusters detected in the 2D dendrogram analysis in IC~1396A, where the \cii\ optical depth is calculated in Sect.~\ref{subsec:13CII_IC1396A}. Figure~\ref{figure:channelmap_large_IC1396E} show channel maps of \thco(3-2) covering a larger area than the \cii\ observations in this paper.

\end{appendix}

\end{document}